\documentclass[10pt]{article}
\usepackage[english]{babel}
\usepackage{amsmath,amssymb,amsfonts}
\usepackage{subfigure}
\usepackage{multirow}
\usepackage{graphicx}
\usepackage[sectionbib]{natbib}
\usepackage{rotating}
\usepackage{setspace}
\usepackage{enumitem}
\usepackage[hmargin=3cm,vmargin=3.5cm]{geometry}

\setcitestyle{apa}

\title{A computational model of inhibitory control in frontal cortex
and basal ganglia}
\author{Thomas V. Wiecki \& Michael J. Frank}

\begin{document}

\maketitle

\begin{abstract}
  Planning and executing volitional actions in the face of conflicting
habitual responses is a critical aspect of human behavior. At the core
of the interplay between these two control systems lies an override
mechanism that can suppress the habitual action selection process and
allow executive control to take over. Here, we construct a neural
circuit model informed by behavioral and electrophysiological data
collected on various response inhibition paradigms. This model extends
a well established model of action selection in the basal ganglia
by including a frontal executive control network which integrates
information about sensory input and task rules to facilitate
well-informed decision making via the oculomotor system. Our
simulations of the antisaccade, Simon and saccade-override task ensue
in conflict between a prepotent and controlled response which causes
the network to pause action selection via projections to the
subthalamic nucleus. Our model reproduces key behavioral and
electrophysiological patterns and their sensitivity to lesions and
pharmacological manipulations. Finally, we show how this network can
be extended to include the inferior frontal cortex to simulate key
qualitative patterns of global response inhibition demands as required
in the stop-signal task.\\

Download the model at: \texttt{http://ski.clps.brown.edu/BG\_Projects/}
\end{abstract}

\newpage
\tableofcontents
\newpage

\section{Introduction}
``Before you act, listen. Before you react, think. Before you spend, earn. Before your criticize, wait.'' This quote by Ernest Hemingway highlights our basic tendency to act impulsively while reminding us that sometimes it is advisable to inhibit these prepotent response biases and act more thoughtful. Recent scientific advancements have shed light on the neural and cognitive mechanisms that implement inhibitory control of prepotent response biases \citep{Andres03,Aron07,Logan85,MiyakeFriedmanEmersonEtAl00,StuphornSchall06,MunozEverling04}. As part of this effort, a multitude of tasks exist to study response inhibition empirically. Among the tasks thought to require selective response inhibition are the antisaccade task, the Simon task, and the saccade-override task. Each of these tasks induces a prepotent response bias that sometimes needs to be overridden with a controlled response based on executive control. For example, the antisaccade task requires subjects to saccade in the opposite direction of an appearing stimulus. The Simon task requires subjects to respond according to an arbitrary stimulus-response rule (e.g., respond left or right depending on stimulus color), but where the stimulus is presented on one side of the screen, inducing a prepotent response bias to that side. In congruent trials the stimulus is presented on the same side as the correct response indicated by the rule, whereas on incongruent trials it is on the opposite side. Finally, the saccade-override task \citep{IsodaHikosaka07} requires subjects to saccade in the direction of a stimulus of a particular color for several repetitions in a row. On so-called switch-trials the instruction cue indicates that the other colored stimulus is now the target, so that the participant has to override the initial planned response and switch to the other one. While critical differences exist, all of these tasks require subjects to inhibit a prepotent response and replace it with a different response. In contrast, while also requiring response inhibition, the well-studied stop-signal task does not require subsequent initiation of an active response but only outright
inhibition of the planned response \citep{VerbruggenLogan08}.\\

Electrophysiological and functional imaging data implicate key nodes
in frontostriatal circuitry as being active during response inhibition
and executive control. At the cortical level, these include
the right inferior frontal gyrus (rIFG)
\citep{AronFletcherBullmoreEtAl03,ChambersBellgroveGouldEtAl07,SakagamiTsutsuiLauwereynsEtAl01,XueAronPoldrack08}
the dorsolateral prefrontal cortex (DLPFC)
\citep{WegenerJohnstonEverling08,FunahashiChafeeGoldmanRakic93,JohnstonEverling06},
the supplementary eye fields (SEF) \citep{SchlagReySanchezSchlag97},
the presupplementory motor area (pre-SMA)
\citep{CongdonConstableLeschEtAl09,AronBehrensSmithEtAl07,IsodaHikosaka07},
and the frontal eye fields (FEF) \citep{MunozEverling04}. At the
subcortical level,  the striatum
\citep{ZandbeltVink10,WatanabeMunoz11,FordEverling09}, the subthalamic
nucleus (STN)
\citep{EagleBaunezHutchesonEtAl08,IsodaHikosaka08,HikosakaIsoda08,AronPoldrack06,AronBehrensSmithEtAl07}
and the superior colliculus are involved. Manipulations that disrupt
processing in either frontal or subcortical areas cause deficits in
response inhibition
\citep{ChambersBellgroveGouldEtAl07,RayJenkinsonBrittainEtAl09,VerbruggenAronStevensEtAl10}.
Moreover, response inhibition deficits are commonly observed in a wide
range of psychiatric patients with frontostriatal dysregulation,
including attention\-deficit/hyper\-activity disorder (ADHD)
\citep{Nigg01,OosterlaanLoganSergeant98,SchacharLogan90}, obsessive
compulsive disorder (OCD)
\citep{ChamberlainFinebergBlackwellEtAl06,MenziesAchardChamberlainEtAl07,PenadesCatalanRubiaEtAl07,Morein-ZamirFinebergRobbinsEtAl09},
schizophrenia (SZ)
\citep{HuddyAronHarrisonEtAl09,BellgroveChambersVanceEtAl06,BadcockMichieJohnsonEtAl02},
Parkinson's disease (PD) \citep{van_KoningsbruggenPenderMachadoEtAl09}
and substance abuse disorders
\citep{MonterossoAronCordovaEtAl05,NiggWongMartelEtAl06}.\\

Together, the above data suggest that intact functioning of the entire
fronto-basal ganglia network is required to support response
inhibition.  However, it is far from clear that the underlying source
of these deficits is the same.  Inhibitory control is a very dynamic
process, influenced by different interacting cognitive variables and
neuromodulatory systems. Thus, response inhibition can be impacted by
not only dysfunctional stopping {\em per se}, but can also be
influenced by changes in motivational state \citep{LeottiWager10},
attentional saliency \citep{Morein-ZamirKingstone06}, maintenance and
retrieval of task rules
\citep{HuttonEttinger06,NieuwenhuisBroerseNielenEtAl04,ReuterKathmann04,RobertsHagerHeron94},
and separable modulations of selective vs global inhibition mechanisms
\citep{Aron11}, to name a few. Although electrophysiological recording
studies demonstrate neuronal populations that differentiate between
successful and unsuccessful stopping
\citep{IsodaHikosaka08,IsodaHikosaka07}, or inhibition of prepotent
responses in favor of controlled responses
\citep{WatanabeMunoz09,FordEverling09}, there is at present no
coherent framework integrating all of these findings into a single
model attempting to account for patterns of electrophysiological data,
or selective disruptions of component parts and their effects on
behavior.\\

The point of departure for our neural model builds on existing theorizing and data regarding the differential roles of the three main pathways linking frontal cortex with the basal ganglia (BG), often referred to as the direct, indirect and hyperdirect pathways. According to this framework, the corticostriatal direct ``Go'' and indirect ``NoGo'' pathways together implement a selective gating mechanism by computing the evidence for facilitating or suppressing each of the candidate motor actions identified by frontal cortex. Dopamine plays a critical role in this model by differentially modulating the activity levels in the two striatal populations, affecting both learning and choice.  During rewards and punishments, phasic bursts and dips in dopamine neurons convey reward prediction errors \citep{MontagueDayanSejnowski96} that transiently amplify Go or NoGo activity states, and therefore activity-dependent plasticity. In this manner, these striatal populations learn the positive and negative evidence for each cortical action \citep{Frank05}. More chronic increases in tonic dopamine levels also directly affect choice by shifting the overall balance of activity toward the Go pathway over the NoGo pathway, thereby emphasizing learned positive relative to negative associations and speeding responding (and vice-versa for tonic decreases in dopamine). Many of this model's predictions have been validated with behavioral studies involving dopaminergic manipulations and functional imaging in humans and monkeys \citep[e.g.,][]{FrankSeebergerOReilly04,NakamuraHikosaka06,PalminteriLebretonWorbeEtAl09,VoonPessiglioneBrezingEtAl10,JochamKleinUllsperger11}, and synaptic plasticity and opto-genetic and genetic engineering studies in rodents
\citep{KravitzFreezeParkerEtAl10,HikidaKimuraWadaEtAl10,ShenFlajoletGreengardEtAl08,KravitzTyeKreitzer12}.\\

Note that in the above model, responses are {\em selectively}
facilitated or suppressed via separate striatal Go and NoGo
populations modulating the selection of particular cortical actions.
However, more recent models have also incorporated the third
hyperdirect pathway from frontal cortex to the STN to BG output.
Communication along this pathway provides a {\em global} and dynamic
regulation of the gating threshold, by transiently suppressing the
gating of all responses when there is conflict between alternative
actions \citep{Frank06,RatcliffFrank12}. Empirical studies using STN
manipulations
\citep{FrankSamantaMoustafaEtAl07,WylieRidderinkhofEliasEtAl10,CavanaghWieckiCohenEtAl11}
direct recordings
\citep{CavanaghWieckiCohenEtAl11,IsodaHikosaka08,ZaghloulWeidemannLegaEtAl12},
and fMRI/DTI \citep{AronBehrensSmithEtAl07} have similarly supported
this notion.\\

Nevertheless, the existing BG model cannot handle situations in which
an initial prepotent response is activated but then needs to be
suppressed -- either altogether, or in favor of a more controlled
response -- situations typically studied under the rubric of
``response inhibition''. Here, we extend the model by incorporating
additional cortical regions that facilitate executive control and can
inhibit and override the more habitual response selection mechanism.
We consider dynamics of the prepotent response process, the subsequent
detection that this response needs to be inhibited, and the inhibition
process itself -- and how all of these factors are modulated by
biological and cognitive variables. We consider electrophysiological
data in various frontal (DLPFC, FEF, preSMA, ACC) and basal ganglia
(striatum, STN) regions that are well captured by the model, and how
these are linked to functional parameters of a high level decision
making process embodied by a variant of the drift diffusion model.\\

Neural models are complex, in that they involve a number of parameters
interacting to produce nonlinear effects on dynamics and
behavior. There is also a risk of overfitting that could result from
adjusting parameters to precisely match electrophysiological data from
one experiment, which may make it difficult to precisely capture
electrophysiological (or behavioral) data from a different
experiment. Thus our aim was instead to capture qualitative patterns
of data in both electrophysiology at multiple levels of cortical and
subcortical network, and of the effects of their manipulation on
behavior, with a single set of parameters.\footnote{By qualitative we
  mean that we do not attempt to quantitatively fit the precise shape
  of firing of any given cell type, but we do aim to show that a given
  population of cells increases or decreases firing rate at a
  particular point in time relative to some task event or to some
  estimated cognitive process. For example, for an area to be involved
  in inhibition it must show increased activity prior to the time it
  takes to inhibit a response. Or in striatum, particular cell
  populations are active related to biasing the prepotent response,
  suppressing that response, and then activating the controlled
  response - our model recapitulates this qualitative pattern.} In
other work (Wiecki \& Frank, in preparation) we show that systematic
variations in neural model parameters are related in a lawful,
monotonic fashion to more computational level parameters in a modified
drift diffusion framework, providing a principled understanding and
falsifiable experimental predictions. Moreover, despite the
qualitative nature of model fits here, we nevertheless aim to
distinguish our model from others in the literature based on general
principles independent of particular parameterizations. Towards this
goal we extracted a set of qualitative behavioral and neurocognitive
benchmark results (listed in the results section) which we use to
assess the validity of our model and compare to other models.\\

As noted above, despite surface features suggesting a single
integrated response inhibition network, there are actually multiple
dynamic components that can affect inhibition. Our contribution in
this paper is to formalize these separable neural processes, to
explore their interactive dynamics. To summarize and preview the core
aspects of our work:

\begin{itemize}
\item We present a  neural network model of the three main frontal-BG
pathways supporting prepotent action selection, inhibitory control,
conflict-induced slowing, and volitional action generation.

\item We show that behavioral changes in a range of tasks dependent on
these basic processes can result from alterations in brain
connectivity and state and provide testable predictions for effects of
distinct brain disorders.

\item Selective response inhibition involves global conflict-induced
slowing via the hyperdirect pathway, raising the effective decision
threshold to prevent prepotent responding, followed by DLPFC induction
of striatal NoGo activity to inhibit the planned prepotent response.
Subequently, the DLPFC provides top-down facilitation onto striatal Go
populations encoding the controlled response.

\item Response selection and inhibition are further regulated by
neuromodulatory influences including dopamine linked to changes in
motivational and attentional state. Dopamine reflects potential reward
values and facilitates Go actions.  In addition, our model suggests
that while selective response inhibition is influenced by tonic levels
of DA, global response inhibition is not.

\item Our model is challenged in its ability to overcome prepotent
responses and evaluated by its ability to reproduce key qualitative
patterns reported in the literature, including:
\begin{itemize}
\item Behavioral RT distribution patterns in selective response
inhibition tasks.
\item Electrophysiological activity patterns of the FEF
\citep{EverlingMunoz00}, pre-SMA \citep{HikosakaIsoda08}, the STN
\citep{IsodaHikosaka08}, striatum \citep{WatanabeMunoz09}, SC
\citep{PougetLoganPalmeriEtAl11,PareHanes03} and scalp recordings
\citep{YeungBotvinickCohen04}.
\item Psychiatric, developmental, lesion and pharmacological
manipulations of frontal function and DA modulations.
\end{itemize}

\item We show that when our model is extended to include the rIFG it
can recover key electrophysiological and behavioral data from the
stop-signal task literature.
\end{itemize}

In sum, this approach provides a mechanistic account of a major facet
of cognitive control and executive functioning, which we hope will
allow for a richer understanding of the relationship between
behavioral, imaging, and patient findings.

\section{Neural Network Model}
We first introduce the neural circuit model of interacting dynamics
among multiple frontal and basal ganglia nodes and their modulations
by dopamine. We then describe how we vary model parameters to capture
biological and cognitive manipulations.

\paragraph{Overview}
The model is implemented in the Emergent software
\citep{AisaMingusOreilly08} with the neuronal parameters adjusted to
approximate known physiological properties of the different areas
\citep{Frank05,Frank06}. The simulated neurons use a rate-code
approximation of a leaky integrate-and-fire neuron (henceforth
referred to as units) with specific channel conductances (excitatory,
inhibitory and leak). Multiple units (simulated neurons) are grouped
together into layers which correspond to distinct anatomical regions
of the brain. Units within each layer project to those in downstream
areas, and in some cases, when supported by the anatomy, there are
bidirectional projections (e.g., bottom-up superior colliculus
projection to cortex as well as top-down projections from cortex to
colliculus). We summarize the general functionality of the model here
to foster an intuitive understanding; implementational and
mathematical details can be found in the appendix. While a single set
of core parameters (i.e. integration dynamics and overall connection
strength between layers) is used to simulate various
electrophysiological and behavioral data in the intact state, each
reported simulation is tested on 8 networks with randomly initialized
weights between individual neurons. The model can be downloaded from our online-repository
\verb~http://ski.clps.brown.edu/BG_Projects~ .\\

The model represents an extension of our established model of the BG
\citep{Frank05,Frank06,WieckiFrank10}. Because the extended model
involves multiple components, we will progressively introduce each
part, beginning with its core and then describing how each new
component contributes additional functionality.

\paragraph{Basic basal ganglia model}
The architecture of the core model is similar to \citet{Frank06}.
While the original model simulated manual motor responses, our model
features a slightly adapted architecture in accordance to the
neuroanatomy and physiology underlying rapid eye-movements (i.e.
saccades) as reviewed in \citet{Hikosaka07} and
\citet{MunozEverling04}. Stimuli are presented to the network in the
input layer,  corresponding to high level sensory cortical
representations. An arbitrary number of motor responses can be
simulated, but here we include a model with just two candidate
responses. The input layer projects directly to the cortical response
units in the frontal eye fields (FEF) which implements action planning
and monitoring and projects to the superior colliculus (SC), which
acts as an output for saccade generation \citep{Sparks02}. The SC
consists of two units coding for a leftward and a rightward directed
saccade. If the firing rate of one unit crosses a threshold, the
corresponding saccade is initiated \citep{EverlingDorrisKleinEtAl99}.
The time it takes an SC unit to cross its threshold from trial onset
is taken as the network's response time (RT). Stimulus-response
mappings can be prepotently biased by changing projection strengths
(i.e. weights) so that certain input patterns preferentially activate
a set of FEF response units more than the alternative response units.
(These sensory-motor cortical weights can also be learned from
experience, such that they come to reflect the prior probability of
selecting a particular response given the sensory stimulus;
\citep{Frank06}). In fact, with only these three structures our model
would only be capable of prepotent, inflexible responding.\\

\begin{figure}
\center{
\includegraphics[width=\columnwidth]{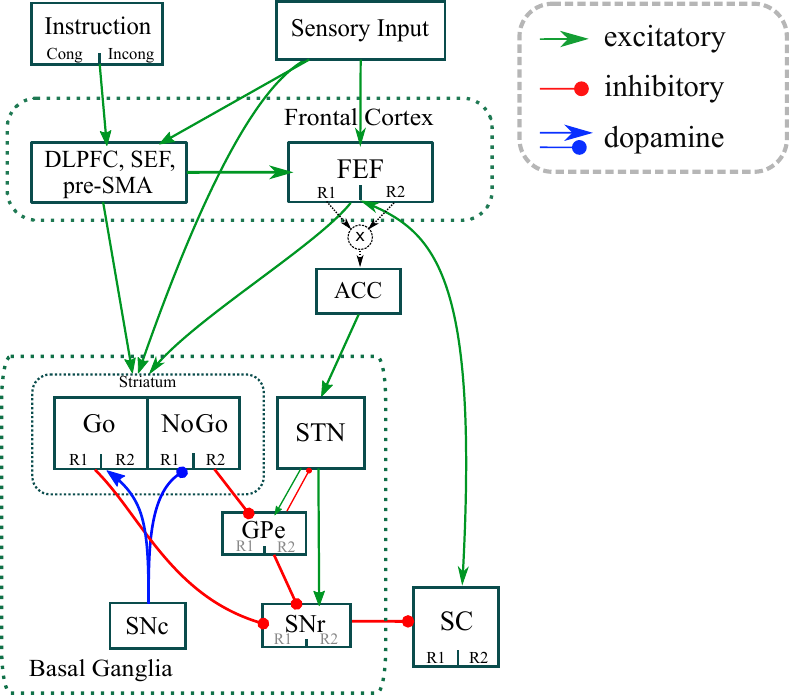}
\caption{Box-and-arrow view of the neural network model. The sensory
input layer projects to the FEF, striatum and executive control (i.e.
DLPFC, SEF and pre-SMA). Via direct projections to FEF (i.e.
cortico-cortical pathway), stimulus-response-mappings can become
ingrained (habitualized). FEF has excitatory projections to the SC
output layer that executes saccades once a threshold is crossed.
However, under baseline conditions, SC is inhibited by tonically
active SNr units. Thus, for SC units to become excited, they have to
be disinhibited via striatal direct pathway Go unit activation and
subsequent inhibition of corresponding SNr units. Conversely,
responses can be selectively suppressed by striatal NoGo activity, via
indirect inhibitory projections from striatum to GP and then to SNr.
Coactivation of mutually incompatible FEF response units leads to dACC
activity (conflict or entropy in choices), which activates STN. This
STN surge makes it more difficult to gate a response until the
conflict is resolved, via excitatory projections to SNr, effectively
raising the gating threshold.  Striatum is innervated by DA from SNc
which amplifies Go relative to NoGo activity in proportion to reward
value and  allows the system to learn which actions to gate and which
to suppress. The instruction layer represents abstract task rule cues
(e.g. antisaccade trial). The DLPFC integrates the task cue together
with the sensory input (i.e. stimulus location) to initiate a
controlled response corresponding to task rules, by activating the
appropriate column of units in FEF and striatum.}
\label{fig.box_arrow}
}
\end{figure}

By itself, FEF activation is not sufficiently strong to initiate
saccade generation because the SC is under tonic inhibition from the
BG output nucleus: the substantia nigra pars reticulata (SNr), whose
neurons fire at high tonic rates. However, the tonic SNr-SC inhibition
is removed following activation of corresponding direct (Go) pathway
striatal units, which inhibit the SNr, and therefore disinhibit the SC
\citep{Hikosaka89,HikosakaTakikawaKawagoe00,GoldbergFarriesFee12}. The
indirect pathway acts in opposition to the direct pathway by further
exciting the SNr (indirectly, via inhibitory projections to the globus
pallidus (GP) which inhibits the SNr). Thus, direct pathway activity
results in gating of a saccade (i.e. Go) while indirect pathway
activity prevents gating (i.e. NoGo). Striking evidence for this
classical model was recently presented by optogenetic stimulation
selectively of direct or indirect pathways cells, showing inhibition
or excitation of SNr respectively, and resulting in increased or
decreased movement \citep{KravitzFreezeParkerEtAl10}.\\

The Go and NoGo striatal populations include multiple units that code for the positive and negative evidence in favor of the FEF candidate actions given the sensory input context.  Relative activity of the striatal pathways is modulated by dopaminergic innervation from the Substantia Nigra pars compacta (SNc) due to differential simulated D1 and D2 receptors present in the two pathways. In particular, dopamine further excites active Go units while inhibiting NoGo units. These effects on activity also produce changes in activity-dependent plasticity, allowing corticostriatal synaptic strength in the Go population to increase following phasic dopamine bursts during rewarding events, and those in the NoGo population to decrease (and vice-versa for negative events; \citep{Frank05}). For simplicity, in the present model we omit learning because the paradigms we simulate do not involve learning, and focus on associations that have already been learned. However, it is now well known that striatal unit activity is modulated by the reward value of the candidate action, such that rewarding saccades are more likely to be disinhibited
\citep{HikosakaNakamuraNakahara06}.\\

Bottom-up projections from SC to FEF allow action-planning to be
modulated according to direct and indirect pathway activity
\citep{SommerWurtz06,SommerWurtz04,SommerWurtz04a,SommerWurtz02}. This
effectively forms a closed loop in which FEF modulates the striatum
which, via gating through SNr and SC, in turn modulates the FEF.
Loosely, FEF considers the candidate responses and "asks" the BG if
the corresponding action should be gated or not. Thus, with these
structures the model can selectively gate responses modulated by DA.\\

In addition to the above gating dynamics, the overall threshold for gating is controlled by the ease with which the SNr units are inhibited by the striatal Go units. The STN sends diffuse excitatory projections to the SNr \citep{ParentHazrati95}, and therefore when STN units are active they increase the gating threshold for all responses, effectively contributing a 'global NoGo' signal \citep{Frank06,RatcliffFrank12}. The STN does not however, act as a static increase in threshold. Rather, the STN receives input directly from frontal cortex, and becomes more active when there is response conflict (or choice entropy) during the early response selection process. In the current model, conflict is computed explicitly by the dorsal anterior cingulate cortex (dACC), which detects when multiple competing FEF response units are activated concurrently, and in turn activates the STN to make it more difficult to gate any response until this conflict is resolved. The full computational role of dACC is far from resolved and likely to be more complex than conflict detection and control \citep[see, e.g.][]{HolroydColes02,BotvinickCohenCarter04,AlexanderBrown11,KollingBehrensMarsEtAl12}. Nevertheless, alternative accounts of dACC function \citep{KollingBehrensMarsEtAl12} are entirely compatible with our model (an issue we return to in the discussion), but for convenience we label the computation as ``conflict''.\\

\paragraph{Frontal Pathway model}
\begin{figure}
\includegraphics[width=\columnwidth]{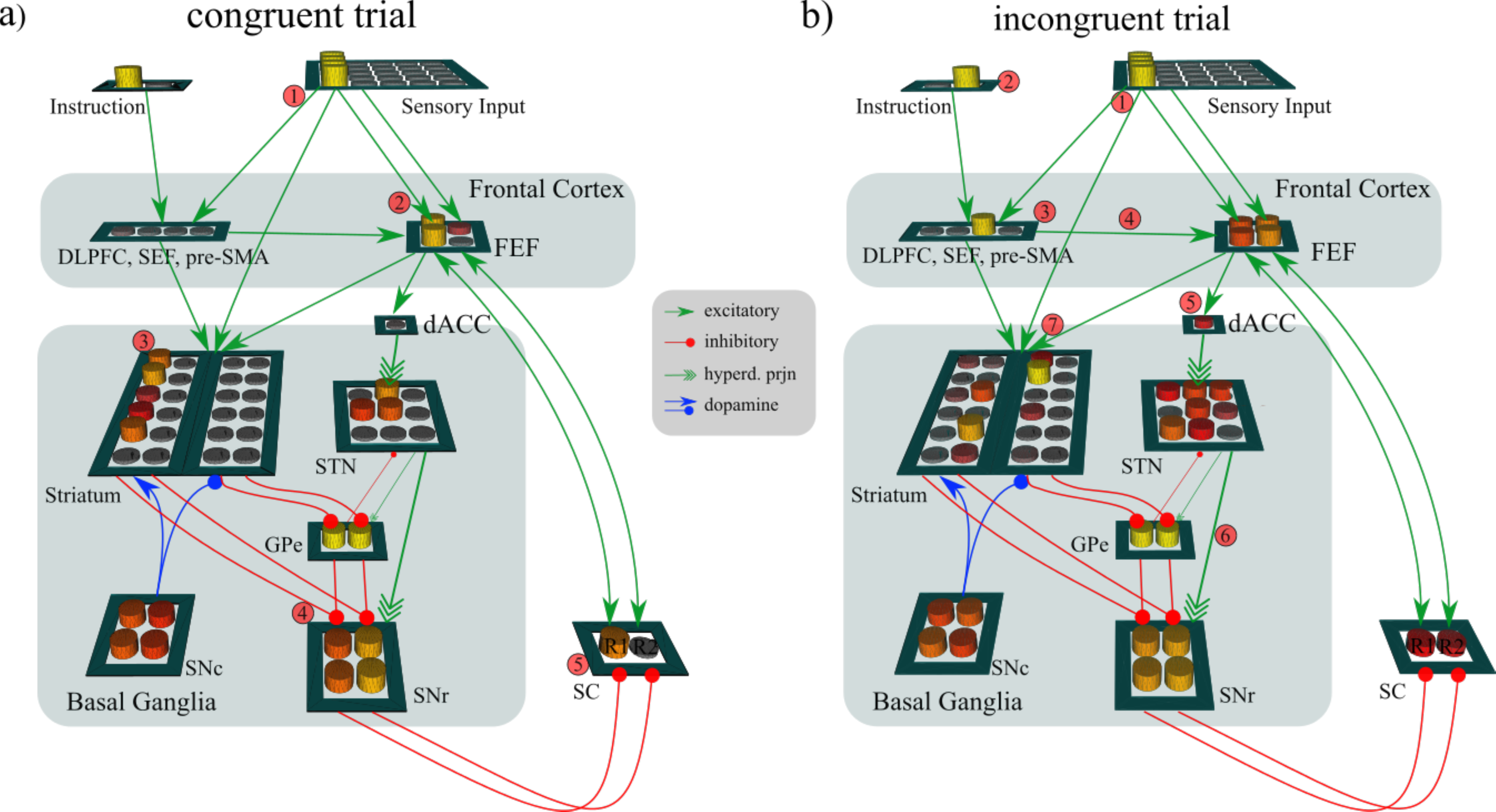}
\caption{Neural network model in different task conditions.
\textbf{a)} Prosaccade condition. (1) Left stimulus is presented in
input layer; (2) Prepotent weights bias left response coding units in
FEF; (3) Left response Go gating neurons in striatum are activated;
(4) Left response coding units in SNr are inhibited; (5) The left
response unit in SC is disinhibited, and due to recurrent excitatory
projections with FEF, is excited and the action is executed.
\textbf{b)} Antisaccade condition. The activity pattern early in the
trial (i.e. before DLPFC comes online) is similar to that in the
prosaccade condition. (1) Left stimulus is presented in input layer
activating prepotent left response in FEF; (2) The unit coding for the
antisaccade condition is externally activated in instruction layer;
(3) DLPFC integrates sensory and instruction input according to task
rules and activates \textit{right} coding units in FEF together with
\textit{right} Go gating units  \textit{left} NoGo units in striatum;
(4) in FEF, right coding units are activated due to DLPFC input in
addition to the prepotent left coding units already active; (5) dACC
detects co-activation of multiple FEF action plans and activates (6)
hyperdirect pathway to excite STN and SNr, globally preventing gating
until conflict is resolved. Eventually, stronger controlled DLPFC
activation of the right coding FEF response results in gating of the
correct antisaccade (7). In some trials, DLPFC activation is too late
and the prepotent left saccade will have already crossed threshold,
resulting in an error.}
\label{fig.BG_inhib_nn_all}
\end{figure}

\paragraph{Volitional response selection}
So far our model is able to select/gate responses and slow down gating
when an alternative response appears to have some value relative to
the initial planned action.\\

However, SRITs require executive control: integration of the sensory
state together with the task rule to not only inhibit the prepotent
response but replace it with a volitional one. Such rule-based
processing is effortful and time-consuming, and hence the controlled
response process lags that of the initial fast response capture. Based
on a variety of evidence, we ascribe the rule-based representations to
the dorso-lateral prefrontal cortex (DLPFC)
\citep[e.g.][]{MillerCohen01,ChambersGaravanBellgrove09}. This
structure is involvedin the active maintenance of stimulus-response
rule representations
\citep{DerrfussBrassCramon04,DerrfussBrassNeumannEtAl05,BrassDerrfussForstmannEtAl05},
is
necessary for correct antisaccade trials
\citep{WegenerJohnstonEverling08,FunahashiChafeeGoldmanRakic93,JohnstonEverling06},
and is involved in selective response inhibition
\citep{GaravanHesterMurphyEtAl06,SimmondsPekarMostofsky08} and
response selection
\citep{BraverBarchGrayEtAl01,RoweFristonFrackowiakEtAl02}.  Moreover,
SEF \citep{SchlagReySanchezSchlag97} and pre-SMA
\citep{IsodaHikosaka07,RidderinkhofForstmannWylieEtAl11} are also
critically involved in correct SRIT performance.\\

We consequently added an abstract executive control layer to summarize
the DLPFC, SEF and pre-SMA complex (in the future referred to as
DLPFC). This layer selects FEF responses and biases BG gating
according to task rules  (see figure \ref{fig.box_arrow}). Although
not explicitly represented separately in the model architecture, we
conceptualize the individual contribution of DLPFC as rule encoding
and abstract action selection whereas SEF and pre-SMA are transforming
this abstract action representation into concrete motor-actions
\citep{Schlag-ReySchlag84,SchlagSchlag-Rey87,CurtisDEsposito03}. In
turn, these planned motor actions can influence the selected response
in FEF and bias gating via projections to striatal Go and NoGo neurons
\citep{MunozEverling04}.\\

Anatomical and functional studies demonstrate projections from both
DLPFC to SEF and pre-SMA
\citep{LuPrestonStrick94,WangIsodaMatsuzakaEtAl05} and to striatum to
affect response gating
\citep{Haber03,DollJacobsSanfeyEtAl09,FrankBadre11}; and from SEF to
FEF \citep{HuertaKrubitzerKaas87}. We explore how these projections
impact dynamics of response selection.  But how does the executive
controller in our model 'know' which rule to activate? We do not
address here how these rule representations arise via learning, which
is the focus of other PFC-BG modeling studies
\citep[see][]{RougierNoelleBraverEtAl05,FrankBadre11,CollinsFrank12}.
Instead, we simulate the state of the network after learning by simply
including an Instruction layer as a second input layer to the model
encoding task condition (e.g. antisaccade trial). In case of the
antisaccade task, the sensory input layer encodes the direction of the
visual stimulus and the instruction layer encodes whether the network
should perform a pro or antisaccade. The DLPFC complex then integrates
these two inputs and activates a (pre-specified) rule unit that (i)
projects to the correct FEF response units supporting the antisaccade;
(ii) activates striatal NoGo units to prevent gating of the active
prepotent pro-saccade response, and (iii) activates striatal Go units
encoding the controlled antisaccade.

Critically, DLPFC units are relatively slow to activate the
appropriate rule unit. This is due to the need to formulate a
conjunctive rule representation between the visual location of the
stimulus and the task instruction (either one of these is not
sufficient to determine the correct response, and indeed, each
individual input provides evidence for multiple potential rules). Time
constants of membrane potential updating is reduced to support this
integration, which also is intended to approximate slower time course
of rule retrieval and subsequent computation to determine the correct
action (via interactions with preSMA and SEF). Moreover, we include
considerable inter-trial noise in DLPFC activation dynamics so that
executive control is available earlier on some trials while later on
others. The slowed integration  and the increase of inter-trial noise
in executive control are necessary for the model to capture the
quantitative benchmark results (demonstrated below). Moreover, the
slower controlled processing is also a core feature of classical dual
process models of cognition \citep[e.g.][]{Sloman96} and the increased
noise accords with the general statistical observation that longer
latencies are typically associated with greater variability.

\paragraph{Competition between the two response selection mechanisms}
As outlined above, our model features two response selection
mechanisms: (i) a fast, prepotent mechanism driven by a biased
projection from sensory input to FEF; and (ii) a slow, volitional
mechanism that originates in the DLPFC which integrates instruction
and sensory input to select and gate the correct response.
Importantly, the volitional mechanism is slower but stronger than the
prepotent one. If, due to noise in the speed of integration, executive
control is slower on some trials, it might be too late to activate the
correct rule representation before the prepotent response is gated. In
contrast, when the executive controller is faster, it activates the
alternative FEF response, leading to conflict-induced slowing, and
then actively suppresses the prepotent response via projections to
striatal NoGo units encoding the prosaccade.  This conceptualization
can be regarded as a biologically plausible implementation of the
cognitive activation-suppression model
\citep{Ridderinkhof02,RidderinkhofWildenbergSegalowitzEtAl04}. Note
however that our implementation involves two suppression mechanisms,
one in which conflict results in global threshold adjustment, and
another in which the prepotent response is selectively inhibited.

\paragraph{Modulations}
To test the influence of different biological manipulations on
executive control paradigms, we modify various parameters in the
network model. Here, we list the different modulations and their
implementation.
\begin{itemize}
\item \textit{Prepotency}: To simulate differences in the strength of
the prepotent response capture of an appearing stimulus (e.g., the
prosaccade stimulus) we modulate the projection strength between
sensory input to the dominant response units in FEF and striatum.
\item \textit{Speed of DLPFC}: To simulate efficacy of prefrontal
function we modulate the speed of DLPFC integration, by adjusting the
time constant of membrane potential updating in these units. Faster
updating implies proactive control.
\item \textit{Connectivity of DLPFC}: To simulate differences in
  intra-cortical connectivity we modulate the DLPFC$\rightarrow$FEF
  projection strength.
\item \textit{Speed-accuracy trade-off}: To simulate strategic
adjustments in the speed-accuracy trade-off, we modulate the
connection strength between frontal cortex and striatum
\citep{ForstmannAnwanderSchaferEtAl10}. In particular, when speed is
emphasized, the FEF more effectively activates striatal Go units so
that it is easier to reach gating threshold. In contrast, accuracy
adjustments are reflected in increased STN baseline ultimately
increasing the response gating threshold.
\item \textit{STN impact}:   STN contributions are simulated by
manipulating the relative synaptic strengths from STN to SNr,
effectively changing the amount of STN activity required to prevent BG
gating \citep{RatcliffFrank12,CavanaghWieckiCohenEtAl11}.
\item \textit{tonic DA}: Pharmacological and disease modulations of DA
levels are simulated by either decreasing (e.g., PD) or increasing
(e.g., SZ) tonic DA activity, which in turn modulates relative
activity of Go vs NoGo units.
\end{itemize}

\subsection{Selective Response Inhibition}
\subsubsection{Methods}
As summarized earlier, all SRITs have a common task structure. (i) A
prepotent response bias is induced by priming an action. In the
antisaccade task this is a result of the appearance of a stimulus that
initiates a 'visual grasping reflex' \citep{HessBurgiBucher46}; in the
Simon task this is the result of placing the target stimuli on either
side of the screen, initiating a response capture
\citep{Ridderinkhof02}; in the saccade-overriding task this is the
result of repeated responding to the same colored stimulus which
renders this response habitual. (ii) In congruent trials, the correct
response is the same as the prepotently biased one. (iii) In
incongruent trials, the correct response is incompatible with the
prepotently biased response, and subjects can use executive control to
suppress the initially predominant action in favor of the
task-appropriate one.

We implemented this common task structure as follows in our neural
network model (alternative task implementations that accommodate the
differences between the tasks lead to similar patterns so we
simplified in order to use a single task representation of this basic
process, but nevertheless simulate patterns of data evident in
specific tasks below). Two stimulus positions, left and right, were
encoded in the input layer as two distinct columns of activated
units. The prepotent bias toward an appearing target was hard-coded by
strong weights from each input stimulus to corresponding response
units in FEF. This prepotent weight facilitates fast responding for
congruent trials, but biases responding in the erroneous direction for
incongruent trials. The DLPFC layer integrates sensory input and
instruction input to activate a conjunctive rule unit encoding the
unique combination of sensory and instruction input, which then
projects to the associated correct response unit in FEF.  Each of the
four DLPFC units project to the appropriate FEF response unit. Note
that weights from the DLPFC to the FEF are stronger than the prepotent
bias connection from the input layer to the FEF so that the DLPFC
would eventually override an erroneous prepotent response. (The same
functionality could be achieved by simply allowing DLPFC units to
reach a higher firing rate or to engage a larger population of units,
instead of adjusting the weights). In addition, DLPFC units activate
corresponding Go and NoGo units in the striatum (e.g. in an
antisaccade trial, Go units coding for the correct response and NoGo
units coding for the incorrect response get activated by top-down PFC
input).

\subsubsection{Results}
\label{sec.results_ast}
We identified a set of key behavioral and neurophysiological
qualitative patterns across SRITs that form desiderata for our model
to capture:

\begin{enumerate}[label=\#\arabic*]
\item \label{bench.accuracy} Incongruent trials are associated with
higher error rates than congruent trials
\citep[e.g][]{ReillyHarrisKeshavanEtAl06,McDowellBrownPaulusEtAl02,IsodaHikosaka08}.
\item \label{bench.RT} Reaction times (RTs) are faster for errors than
correct trials
\citep[e.g][]{ReillyHarrisKeshavanEtAl06,McDowellBrownPaulusEtAl02,IsodaHikosaka08}.
\item \label{bench.RT_accuracy} Strategic adjustments in the
speed-accuracy trade-off (via changes in decision threshold) modulates
functional connection strength between frontal cortex and striatum
\citep{ForstmannAnwanderSchaferEtAl10}. Similarly, STN activity is
associated with modulations of the decision threshold
\citep{RatcliffFrank12,CavanaghWieckiCohenEtAl11}.
\item \label{bench.psych} Various psychiatric diseases associated with
frontostriatal cathecholamine dysregulation lead to increased error
rates and speeded responses
\citep[e.g.][]{ReillyHarrisKeshavanEtAl06,HarrisReillyKeshavanEtAl06,ReillyHarrisKhineEtAl07,McDowellBrownPaulusEtAl02}.
\item \label{bench.early_prepotent} Early activation of prepotent
motor response, e.g. in EMG measurements
\citep{BurlePossamaiVidalEtAl02}.
\item \label{bench.FEF} At least four different types of activation
dynamics in FEF neurons during correct and error incongruent trials
\citep{EverlingMunoz00}. Specifically, neurons coding for the
erroneous (i.e. prepotent) response are fast to activate and their
activity is greater on error trials than correct trials. In contrast,
neurons coding for the correct (i.e. controlled)  response are slower
to activate and their activity is reduced and delayed on error trials.
See figure \ref{fig.AS_FEF_SC_ephys}c for the quantitative data that
forms the basis of this qualitative pattern.
\item \label{bench.striatum} At least four different types of striatal
neurons with dissociable dynamics and direction selectivity in
congruent and incongruent trials
\citep{WatanabeMunoz09,FordEverling09}. Specifically, (i) during
prosaccades, distinct neural populations code for facilitation of the
correct response and suppression of the alternative;   (ii) during
antisaccade trials, (iia) neurons coding for facilitation of the {\it
incorrect} prepotent response initially become active but return to
baseline when (iib) neurons coding for the suppression of that
response become active together with (iic) neurons coding for
facilitation of the correct controlled response  (see figure
\ref{fig.AS_striatum}b).
\item \label{bench.conf} Neurons forming part of the hyperdirect
pathway from frontal cortex (pre-SMA, dACC) to the STN show increased
activity (i) {\em before} correct incongruent responses and (ii) {\em
after} incorrect incongruent responses, but (iii) baseline activity
during congruent response \citep{IsodaHikosaka07,IsodaHikosaka08,
YeungBotvinickCohen04,ZaghloulWeidemannLegaEtAl12}. This pattern of
activity co-occurs with delayed but more accurate incongruent
responding.
\end{enumerate}

In the following, we demonstrate how our model reproduces these
qualitative patterns, before linking its dynamics to a higher level
computational description.

\paragraph{Behavior}
As expected, intact networks make considerably more errors on
incongruent trials (error rate of 15\%) as compared to perfect
performance in congruent trials (error rate close to 0\%, not shown),
thereby capturing qualitative pattern \ref{bench.accuracy}.

\begin{figure}
\subfigure{
 \includegraphics[width=0.5\columnwidth]{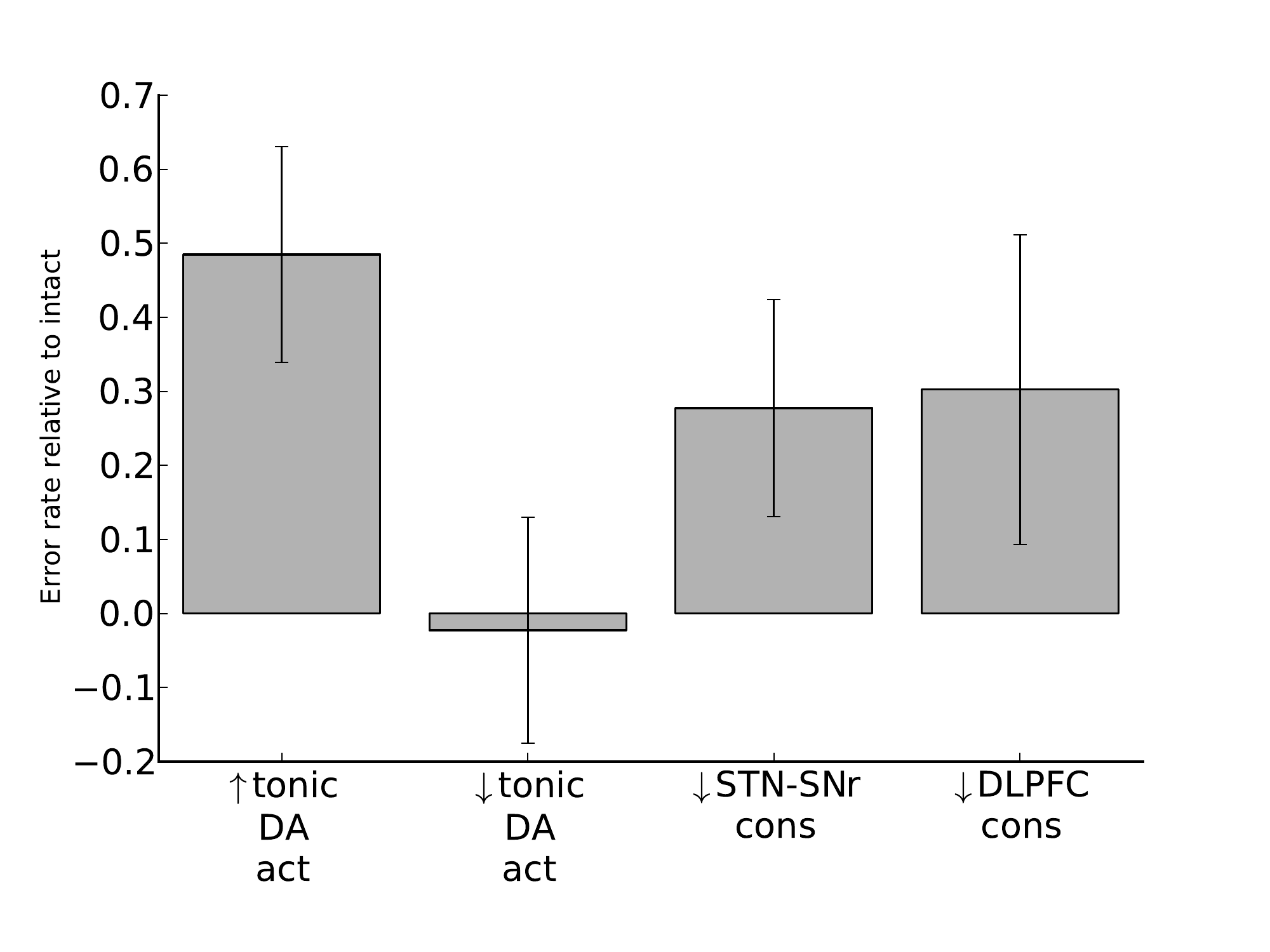}
 \label{fig.AS_error}
}
\subfigure{
 \includegraphics[width=0.5\columnwidth]{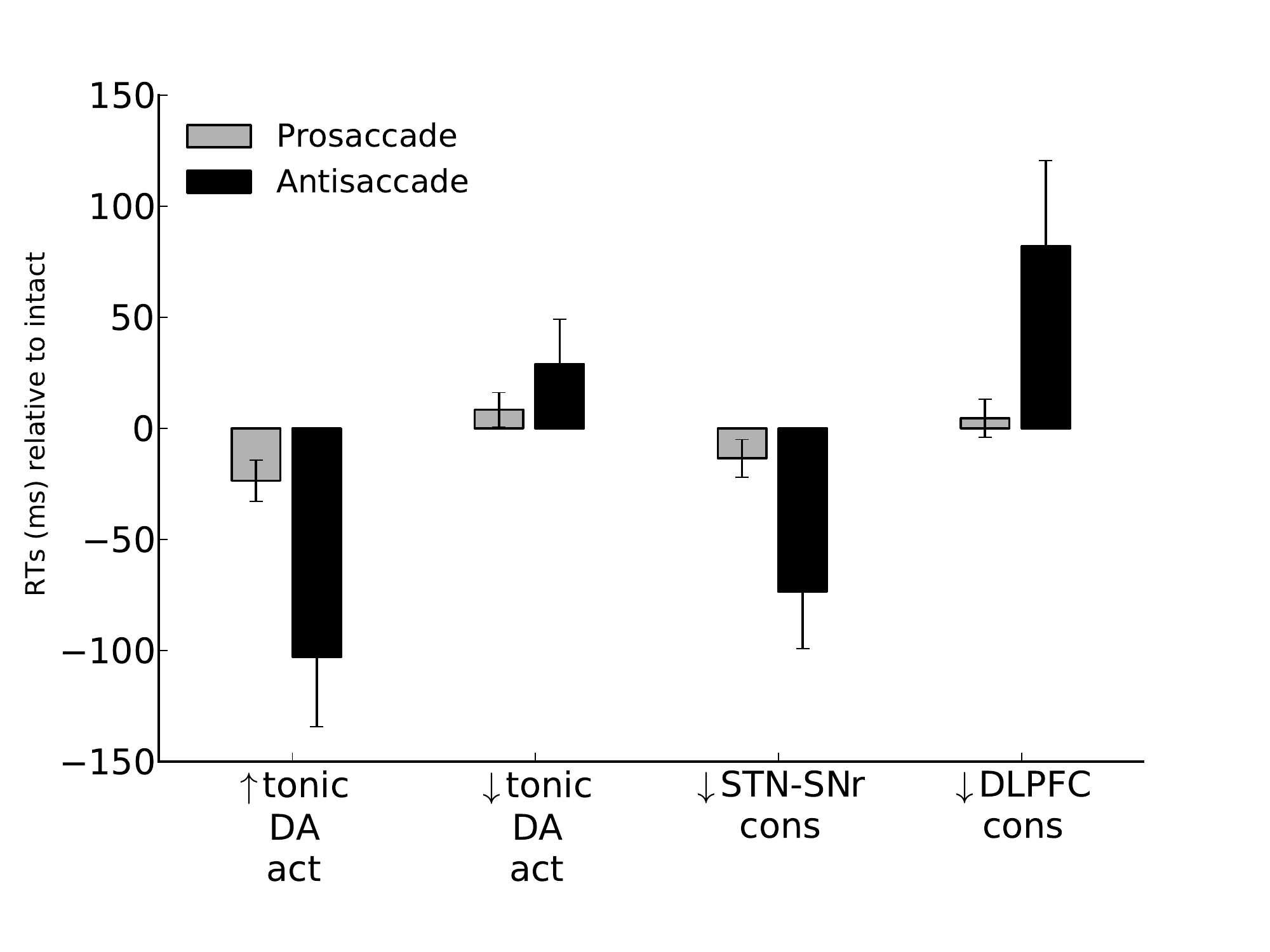}
 \label{fig.AS_RTs}
}
\caption{\textbf{a)} Error rates in incongruent trials $\pm$ SEM
relative to intact networks for different neural   manipulations.
Networks make more errors with increased tonic DA   levels, or STN
dysfunction, compared to intact networks. \textbf{b)}   Response Times
(RTs) $\pm$ SEM relative to intact networks, for pro   and antisaccade
trials as a function of neural manipulations. For   more analysis see
the main text.}
\end{figure}

Further, networks in general have longer response times (RTs) in
incongruent trials (see figure \ref{fig.AS_RTs}) thus capturing
qualitative pattern \ref{bench.RT}.  Incongruent trials are slower for
two reasons:   (i) it takes time for executive control (DLPFC)
computations due  to the requirement to integrate two sources of input
to activate the associated rule; and (ii) once activated, the
controlled response conflicts with the prepotent response, leading to
STN activation and associated increases in BG gating threshold.

\begin{figure}
\includegraphics[width=\columnwidth]{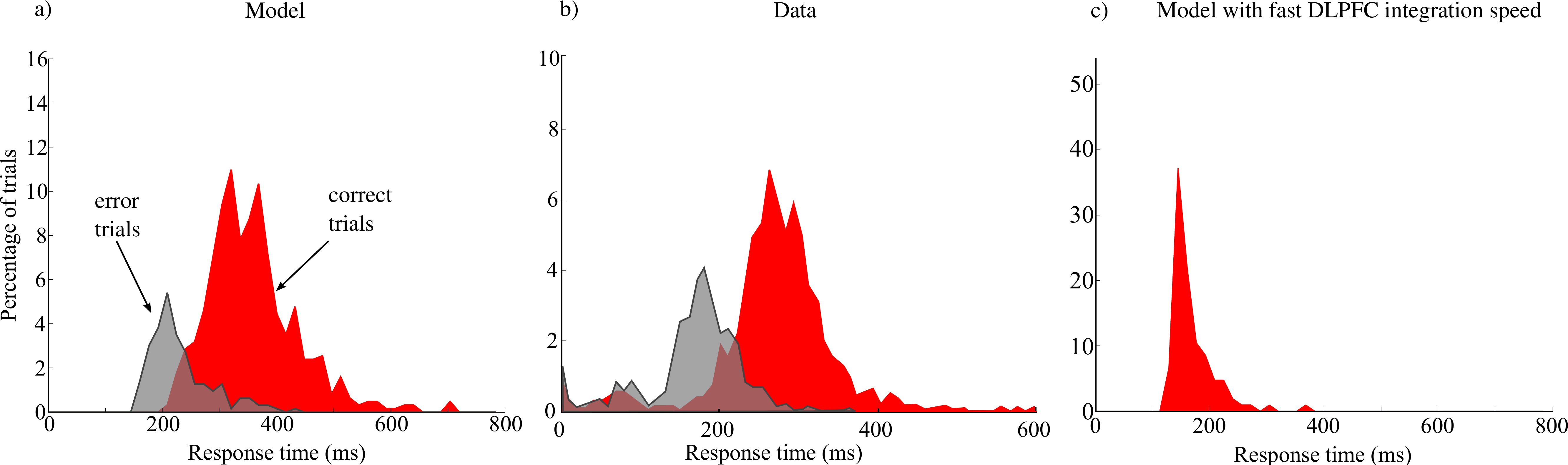}
\caption{\textbf{a)} RT histogram for correct and erroneous
  incongruent trials in the model. Error RT distributions were shifted
  to the left due to prepotent response capture. This pattern is
  exaggerated with increased tonic DA due to lowered effective gating
  threshold. \textbf{b)} RT histograms of a monkey during the
  switch-task (reproduced from \citet{IsodaHikosaka08}). In blocks of
  trials, monkeys are continuously rewarded following saccades to one
  of two targets. On so-called 'switch-trials' a cue indicates that
  the monkey should perform a saccade to the opposite target,
  requiring the monkey to inhibit his planned saccade and perform a
  saccade to the opposite direction. As in the model, errors are
  associated with shorter reaction time. \textbf{c)} Reaction time
  distribution of an alternative model with fast DLPFC integration
  speeds. Correct trials are in red and errors in gray (not
  present). This model cannot account for the behavioral pattern of
  errors and RTs as a function of congruency, in contrast to models
  with slowed DLPFC integration (panel a).}
\label{fig.AS_histo}
\end{figure}

Additional analysis revealed that incongruent error trials are
associated with faster RTs compared to correctly performed incongruent
trials (figure \ref{fig.AS_histo}). In
our model, errors are made when the faster prepotent action reaches
threshold before the inhibitory process can cancel it. This mechanism
allows the model to capture qualitative pattern \ref{bench.RT} and
\ref{bench.RT_accuracy}.\\

We next investigated how these behavioral patterns were affected by
manipulations (see figure \ref{fig.AS_error}). Incongruent error rates
were most exaggerated with increased   tonic DA levels, and by
disrupted STN function to simulate deep brain stimulation. The effect
of increased striatal DA on incongruent error rates captures
corresponding patterns (see \ref{bench.psych}) observed in
non-medicated schizophrenia patients, who have elevated striatal DA
\citep[e.g][]{ReillyHarrisKeshavanEtAl06,HarrisReillyKeshavanEtAl06,ReillyHarrisKhineEtAl07,McDowellBrownPaulusEtAl02}.
Tonic DA elevations are associated with speeded responding in both
congruent and incongruent trials, due to shifted balance toward the Go
pathway facilitating response gating.  This same mechanism explains
the increased antisaccade error rate. Conversely, decreased tonic DA
leads to slowed responding due to increased excitability of the
indirect NoGo pathway. The model also predicts that STN dysfunction
produces increased error rates, due to an inability to raise the
threshold required for striatal facilitation of prepotent responses.
Indeed, STN-DBS induces impulsive (fast but inaccurate) responding in
SRITs \citep{WylieRidderinkhofEliasEtAl10}.\\

\begin{figure}
\subfigure{
\includegraphics[width=\columnwidth]{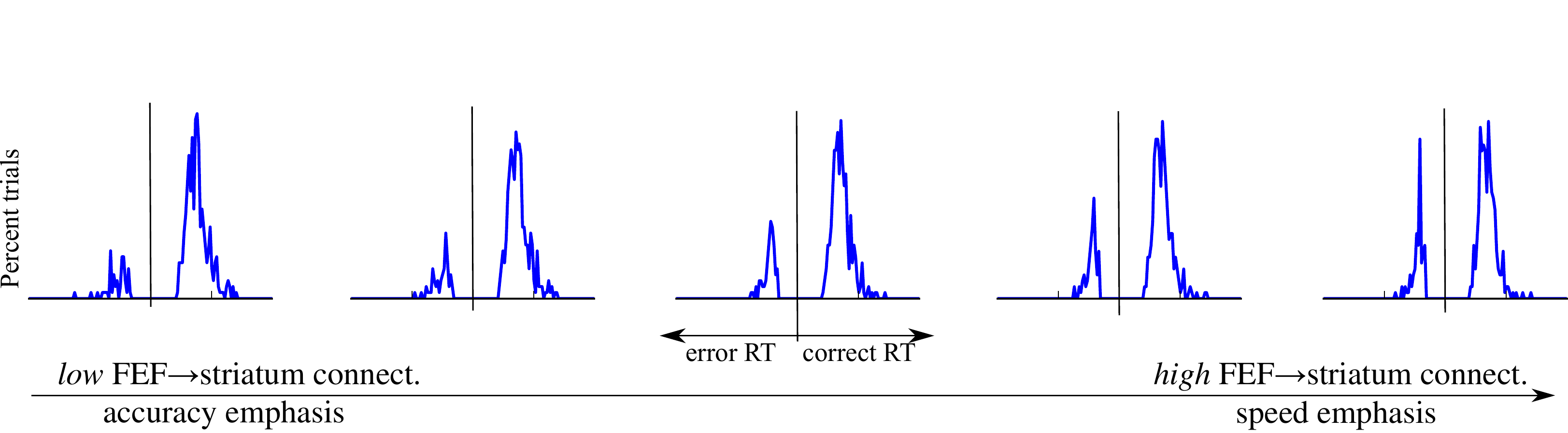}
\label{fig.AS_RT_speed}
}
\subfigure{
  \includegraphics[width=.5\columnwidth]{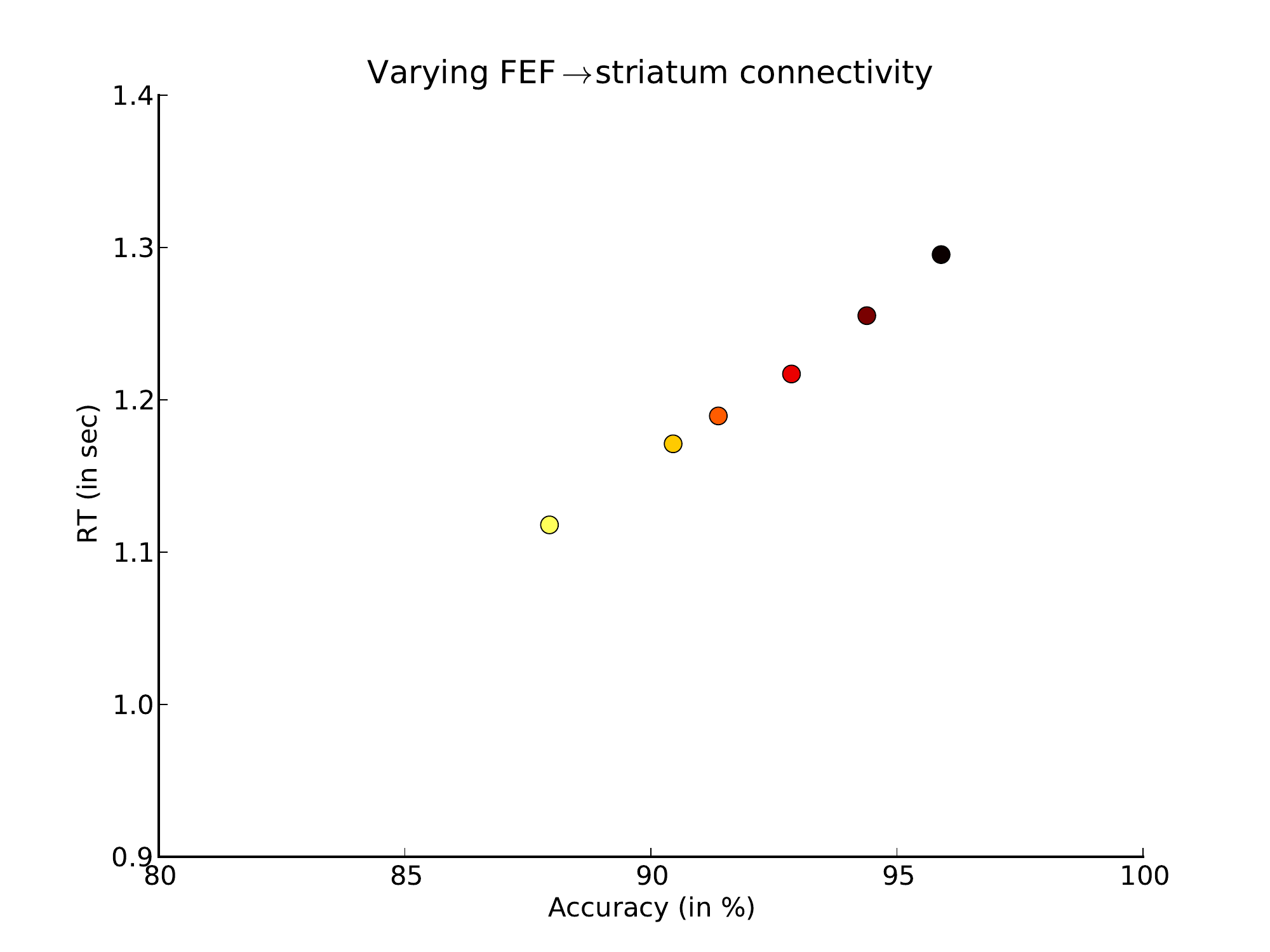}
    \label{fig.AS_FEF_speedacc}
}
\subfigure{
  \includegraphics[width=.5\columnwidth]{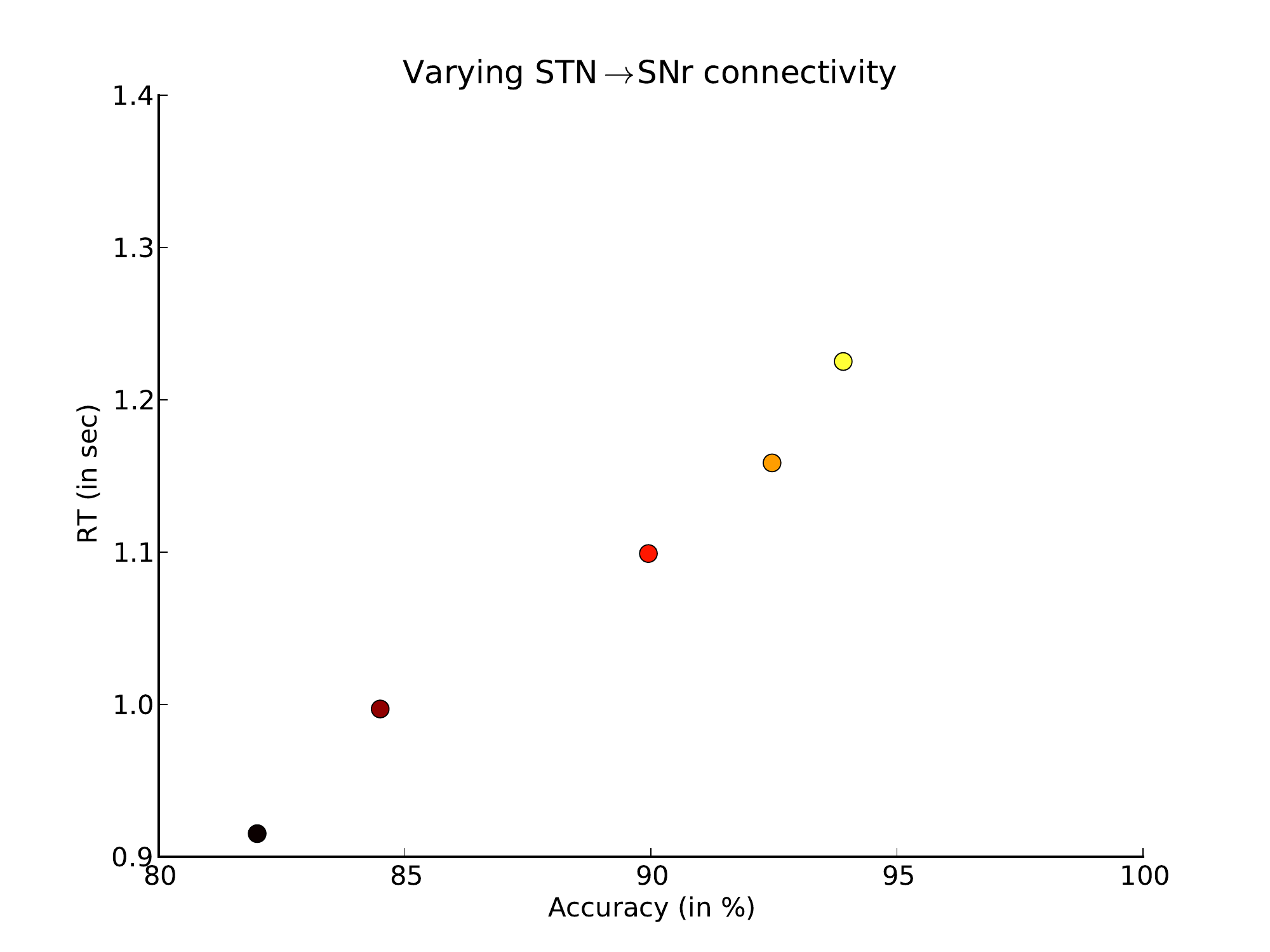}
  \label{fig.AS_STN_speedacc}
}
\caption{\textbf{a)} RT distributions for incongruent trials by
network models. FEF$\rightarrow$striatum projection strengths were
varied along the x-axis. Correct RT distributions are on the right
side of each panel and incorrect RT distributions are on the left
side, mirrored on the y-axis. This manipulation is equivalent to a
speed-accuracy adjustment, as shown empirically to vary with
pre-SMA$\rightarrow$striatal communication (Forstmann et al, 2008;
2010), where here FEF plays the role of pre-SMA for eye movements as
compared to manual movements studied in Forstmann et al.
\textbf{b+c)} Speed-accuracy tradeoff under parametric modulation of
\textbf{(b)} FEF$\rightarrow$striatum connection strength and
\textbf{(c)} STN$\rightarrow$SNr connection strength (color coded).
Black represents low and yellow high connection strength. This
pattern is consistent with decision threshold modulation. The
absolute values of connection strengths in these different routes
are chosen to lie on a sensitive range producing observable effects
for demonstration purposes.}
\end{figure}

\begin{figure}
  \includegraphics[width=\columnwidth]{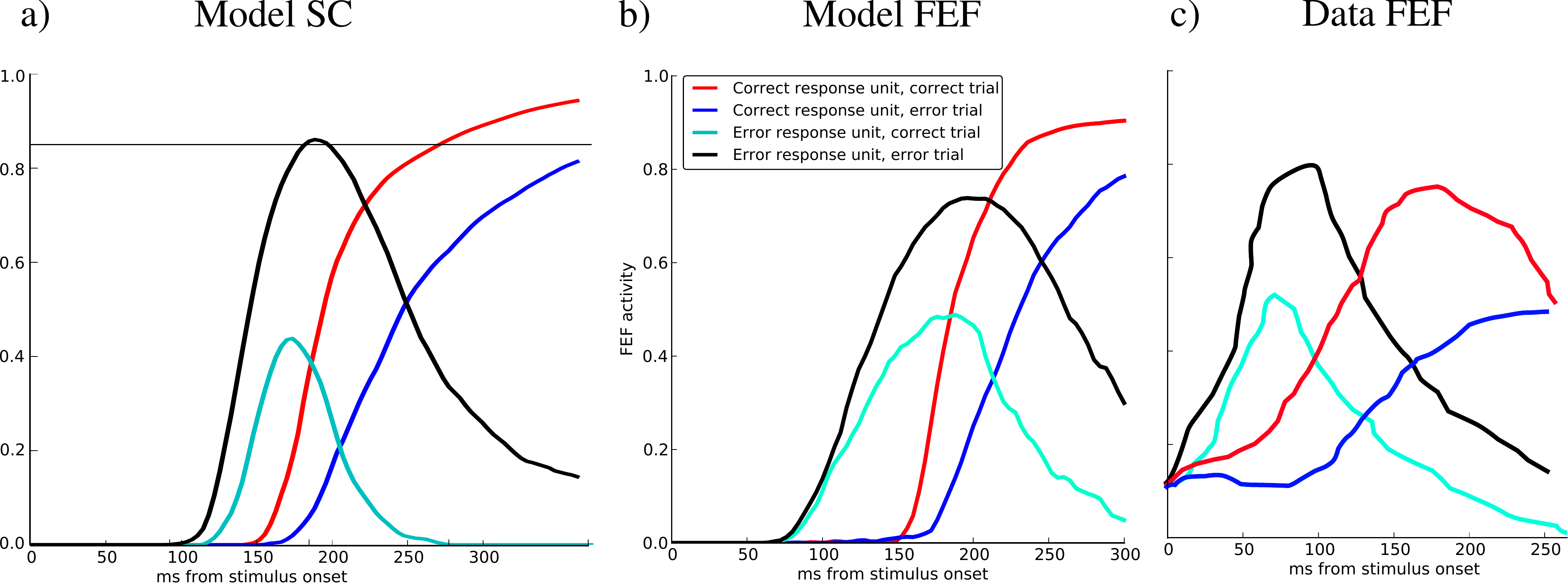}
\caption{\textbf{a)} Average activity of individual superior
colliculus (SC) units coding for the correct and error response in
correct and incorrect trials during incongruent trials aligned to
stimulus onset. The prepotent (i.e. erroneous) response comes on
before the volitional, correct response. In incorrect trials the
error-unit threshold is crossed before the volitional response unit
gets active. In correct trials the error-unit is inhibited in time.
\textbf{b)} Average activity of individual FEF units coding for
prepotent error responses and volitional correct responses during
incongruent trials aligned to stimulus onset (benchmark pattern \ref{bench.FEF}) \textbf{c)}
Electrophsyiological recordings in FEF of monkeys (reproduced from
\citet{EverlingMunoz00}). }
  \label{fig.AS_FEF_SC_ephys}
\end{figure}

Finally, we tested in more detail how systematic parametric changes in
a biological variable affect RT and accuracy. Figure
\ref{fig.AS_RT_speed} shows how RT distributions change under
different settings of FEF$\rightarrow$striatum connection
strength. Figure \ref{fig.AS_FEF_speedacc} shows quantitatively how
increases in FEF$\rightarrow$striatum connectivity leads to faster RT
and decreased accuracy (qualitative pattern
\ref{bench.RT_accuracy}). Loosely, increasing FEF connection strength
onto Go-units in the direct pathway leads to faster gating of
responses. Conversely, increases in STN$\rightarrow$SNr connectivity
lead to slower RT and improved accuracy (figure
\ref{fig.AS_STN_speedacc}). The reason for both of these effects is
that they differentially modulate SNr activity. Recall that the SNr
tonically inhibits the thalamus, unless it is itself inhibited by the
striatal direct pathway. Hence any modulation of the ease with which
SNr units are inhibited -- either via stronger connections from cortex
onto Go units, or by increasing the SNr via the STN -- will change the
threshold required for the BG to gate an action. Indeed,
\citet{RatcliffFrank12} and \citet{LoWang06} have shown that these two
mechanisms are related to changes in the decision threshold in
sequential sampling models. Our model subsumes both of these
mechanisms, and suggests that these different routes are themselves
modulated by distinct cognitive variables, such as volitional
speed-accuracy modulation and conflict/choice entropy
(cortico-striatal and STN). We return to this issue in the Discussion.

In sum, our model captures key qualitative behavioral patterns
described in the literature (see above). Moreover, these patterns hold
over varying biologically plausible parameter ranges leading to
predictable changes in the behavioral patterns. However, given the
complexity of the underlying model, it is also important to establish
whether the internal dynamics of the different nodes of the network
are consistent with  available electrophysiological data in this class
of tasks.

\paragraph{Neurophysiology}

\subparagraph{DLPFC, SEF and pre-SMA activity}
Our model summarizes the computations of the executive control complex
as a single layer corresponding to  DLPFC, SEF and pre-SMA. One of our
central predictions is that DLPFC activation must be delayed relative
to the habitual response mechanism in order to produce the desired
qualitative patterns. To demonstrate the plausibility of this account
we simulated networks with increased DLPFC speed (time constant of
membrane potential updating). Consequently, networks ceased to make
fast errors while correct RTs became much faster and more peaked
(figure \ref{fig.AS_histo}c). The reason for this pattern is that
active executive control now dominates and overrides the prepotent
mechanism during early processing. This result implies that some delay
in executive control is needed to account for empirical findings in
which incongruent RTs are delayed.

\subparagraph{SC and FEF activity}
Comparing single unit activation patterns of SC (see figure
\ref{fig.AS_FEF_SC_ephys}a) to those of FEF (see figure
\ref{fig.AS_FEF_SC_ephys}b) reveals that the activation dynamics are
very similar between those two regions. Our model thus predicts that
FEF can be interpreted as a cortical saccade planning/monitoring area
that directly influences saccade generation via its projections to SC
\citep{MunozEverling04}. Moreover, SC activity reveals that in both,
correct and incorrect incongruent trials, the incorrect prepotent
response unit becomes active before the controlled one, thus matching
qualitative pattern \ref{bench.early_prepotent}.

\begin{figure}
\centering
  \includegraphics[width=\columnwidth]{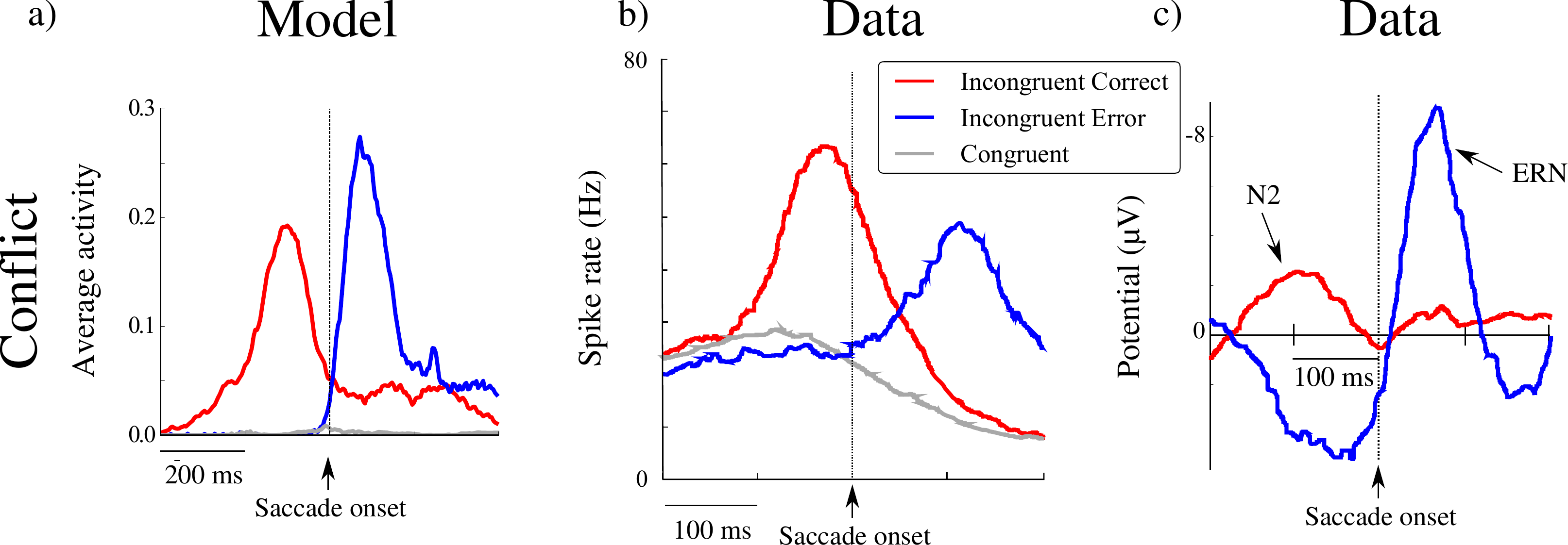}
\caption{\textbf{a)} Averaged dACC activity (corresponding to conflict
in FEF) in prosaccade and correct and incorrect incongruent trials. No
conflict is present in congruent trials. During correct incongruent
trials, conflict is detected and resolved before the response is
gated. During incorrect incongruent trials, an incorrect response is
made before conflict is detected. \textbf{b)} Activity recorded in
monkey pre-SMA during the switch-task (reproduced from
\citet{IsodaHikosaka07}). \textbf{c)} EEG recordings from the central
scalp of humans during the Flanker task (reproduced from
\citet{YeungBotvinickCohen04}), thought to originate from dACC. The N2
and ERN component closely match our modeling results, replicating this
aspect of the Yeung model.}
\label{fig.AS_conflict_ephys}
\end{figure}

\subparagraph{dACC activity}
As described earlier, the dACC computes co-activation of both response
units in FEF (i.e. when average activity is $> 0.5$) -- a direct
measure of conflict (or value of the alternative action to that
initially considered; see above). Consequently, its activity (see
figure \ref{fig.AS_conflict_ephys}a) follows a similar pattern as
average FEF layer activity: conflict is present but resolved prior to
responding in correct trials while conflict is present \textit{after}
responding in error trials. However, dACC does not get active in
congruent trials, because it never shifts from one action to the
other.\\

This qualitative pattern of peak conflict activation {\it before}
correct incongruent trials but {\it after} incorrect incongruent
trials matches event-related potentials (ERPs) commonly observed in
human EEG studies (see figure \ref{fig.AS_conflict_ephys}c). The
so-called \textit{error related negativity} (ERN) which is measured
\textit{after} response errors whereas the so-called \textit{N2}
potential is measured \textit{before} correct high conflict responses
\citep{FalkensteinHohnsbeinHoormannEtAl91,GehringGossColesEtAl93}. The
idea that these two signals could merely represent 'two sides of the
same conflict coin' and reflective of underlying dACC activity was
first presented in the modeling work by Yeung and
colleagues \citep{YeungCohen06,YeungCohenBotvinick04}.\\

\begin{figure}
  \includegraphics[width=\columnwidth]{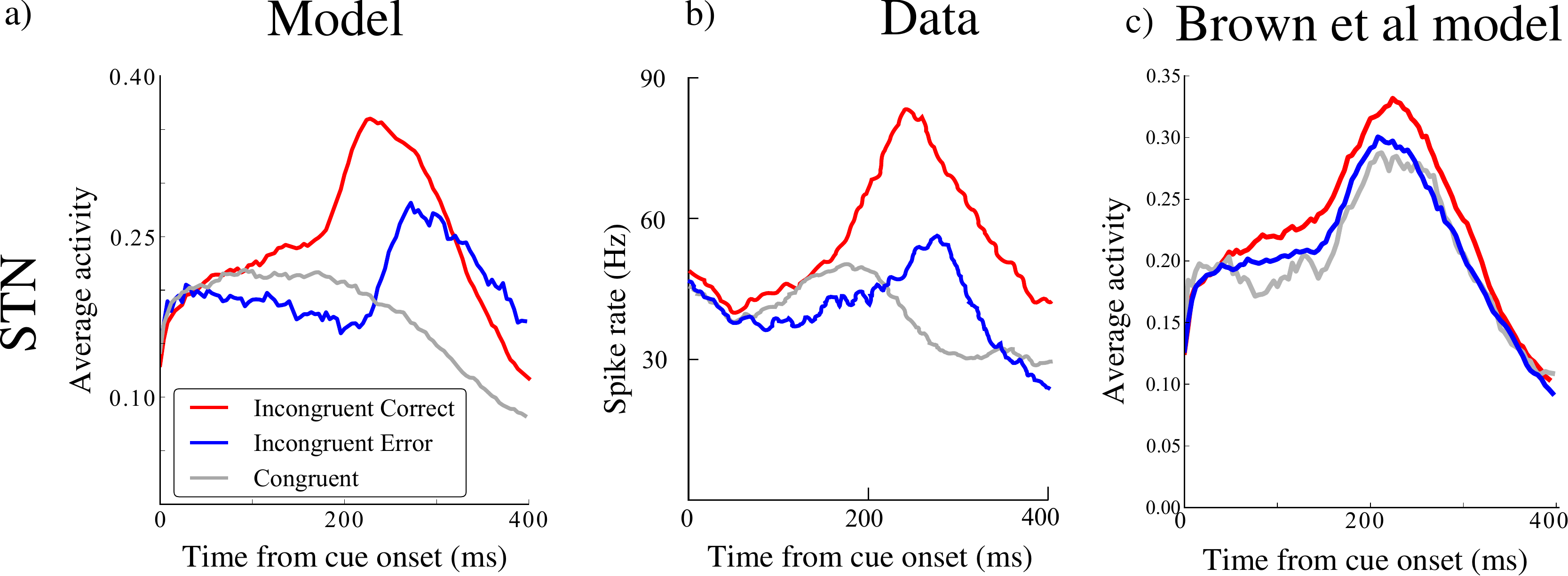}
\caption{\textbf{a)} Averaged activity of the model STN layer during
  prosaccade and correct and incorrect incongruent trials relative to
  response execution. During congruent trials STN units exhibit a
  small early increase in activity that subsides. Correct incongruent
  trials show increased activity early on in the trial which causes
  the conflict-induced slowing and prevents prepotent response
  gating. In error trials, this mechanism is triggered too late and
  the incorrect response gets executed. \textbf{b)}
  Electrophysiological recordings of the monkey STN (reproduced from
  \citet{IsodaHikosaka08}) on correct and incorrect switch trials and
  non-switch trials. \textbf{c)} Average activity of the STN layer of
  an alternative model in which STN is not excited by dACC but instead
  by saccadic output (SC in our model) as proposed by
  \citet{BrownBullockGrossberg04}. This model does not predict
  differences between trial types.}
  \label{fig.AS_STN_ephys}
\end{figure}

\subparagraph{STN activity}
As noted in the model description, conflict detection in the dACC
results in delayed (and more accurate) responding by recruiting the
STN to prevent gating until conflict is resolved. Indeed, this
mechanism is in part responsible for the rightward-shifted RT
distributions in correct incongruent trials. Accordingly, this same
pattern of increased activity before correct responses and increased
activity after error responses can be observed in STN (see figure
\ref{fig.AS_STN_ephys}a). Again, this qualitative pattern was also
found in STN recordings in monkeys by \citet{IsodaHikosaka08} (see
figure \ref{fig.AS_STN_ephys}b), who showed that timing of STN firing
relative to pre-SMA was consistent with communication along this
hyperdirect pathway.

The neurocomputational model of \citep{BrownBullockGrossberg04}
interprets the role of STN differently. In their model, STN is
activated by the output structure (FEF in their case) to lock out the
influence of competing responses {\em after} a response has been
selected.  This is a critical difference to the account presented
herein where STN plays a role in the selection of a response by
raising the threshold prior to response selection, thereby delaying
execution but increasing accuracy.  To show explicitly how our model
predictions can be qualitatively differentiated from this alternative
model of STN function, we disconnected dACC inputs into the STN and
instead allowed only the output structure (SC in our model) to project
to it, so that STN function operates as it does in Brown et al
(2004). As can be seen in figure \ref{fig.AS_STN_ephys}c, the activity
pattern changes dramatically. Specifically, there is no more
differentiation of activation patterns between the different trial
types as is observed in our model and the empirical data
\citep{IsodaHikosaka08}. Because STN only influences processing after
response selection, it also does not lead to delayed responding or
decision threshold adjustment. This qualitative difference in model
predictions is fundamental and not subject to parameter tuning, as it
reflects a distinct computational role for the STN. Although we
focused on the Brown model for demonstration purposes here, other
models of STN function with different connectivity would similarly not
account for these data. For example, the biophysical model of
\citet{RubchinskyKopellSigvardt03} assumes that STN neurons provide
focused selection of a particular action (by disinhibiting SNr, taking
the role of the direct Go pathway) while simultaneously inhibiting
competing actions (by exciting SNr in other columns). This model
cannot explain this activity pattern because co-activation of multiple
cortical inputs does {\it not} result in increased STN activity (see
figure 6b in \citet{RubchinskyKopellSigvardt03}).

\begin{figure}
\end{figure}

\subparagraph{Striatal activity}
Figure \ref{fig.AS_striatum}a shows striatal activity in congruent and incongruent trials (column I and column II, respectively) for direct-path Go and indirect-path NoGo units (upper and lower rows, respectively). In each case, activity selective to the correct and error responses are color coded.  The model closely captures the qualitative pattern across four cell populations (\ref{bench.striatum}) identified in monkey dorsal striatum recordings during the antisaccade task (see figure \ref{fig.AS_striatum}b and \citet{WatanabeMunoz09,FordEverling09}). In particular, for congruent trials, correct-coding Go neurons gate the response while error-coding NoGo units suppress the alternative. In incongruent trials, Go neurons for the error-coding prepotent response are initially activated, but are then followed by increased activity of the corresponding NoGo population which then suppresses the initiated Go activity via NoGo$\rightarrow$Go inhibitory projections \citep{TavernaIlijicSurmeier08}. Finally, the controlled Go-correct units are activated and an incongruent response is executed. Thus our model predicts that the pattern of electrophysiology observed in empirical recordings arises due to top-down cognitive control modulation of direct and indirect pathway neurons.

Note again that we can distinguish our model's predictions from those
of other models that omit the indirect pathway as a distinct source of
computation (there are several) or from models that do include it but
assign a different function.  The neural network model of
\citet{BrownBullockGrossberg04} assumes the indirect pathway
activation defers execution of the correct action plan until the time
is appropriate. This would suggest that the executive control complex
would activate NoGo units coding for the {\it correct} response, not
the incorrect response as in our model. To demonstrate how this leads
to qualitatively different patterns than is observed in our model and
the data (see pattern \ref{bench.striatum} and figure
\ref{fig.AS_striatum}c) in which this alternative account is
simulated in our model. (However, we note that the Brown et al model
could potentially accord with our model in the sense that they also
advocate a mechanism by which negative prediction errors drive
learning in the NoGo cells, which after training on the AST may also
produce the patterns we observe here given that the prepotent response
would be punished). Similarly, the prominent model of
\citet{GurneyPrescottRedgrave01a} suggests that this pathway serves as
a control pathway rather than providing negative evidence against
particular actions as in our model, and it is unclear how this control
function (while not disputed per se) would reproduce the patterns
observed here.

\begin{figure}
\includegraphics[width=\columnwidth]{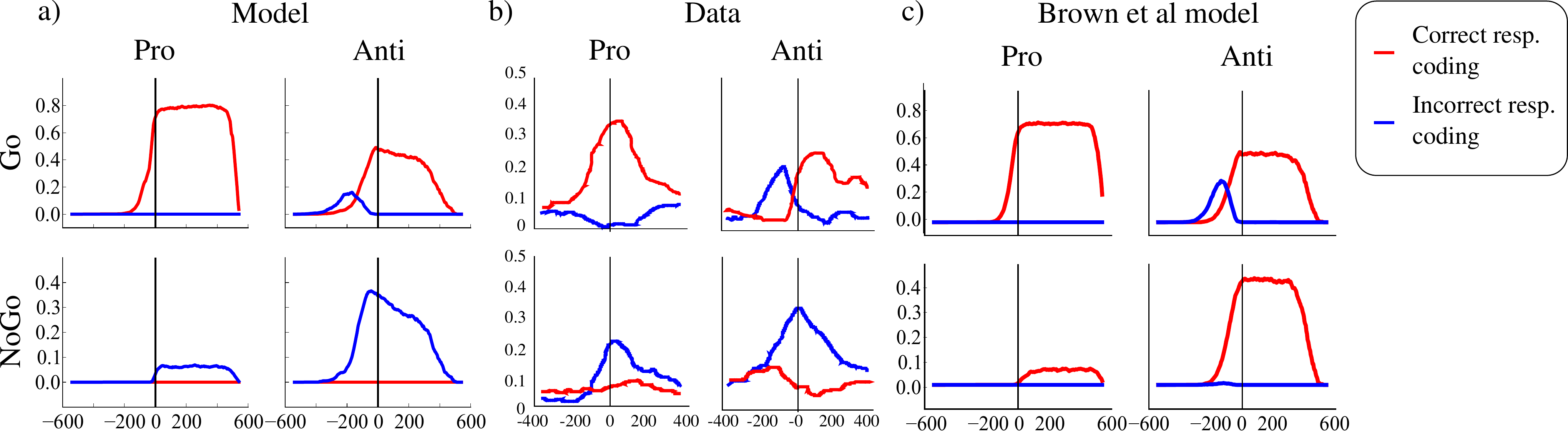}
\caption{\textbf{a)} Averaged striatal activity during correct pro
(first column) and incongruent trials (second column) in Go (first
row) and NoGo (second row) neuronal populations. In each case,
activity for correct (red) and error (blue) response coding units are
shown separately.  As described in the text, the Go units for the
prepotent response become active early in the trial for both trial
types, but in antisaccade trials these are followed by NoGo units
which veto the Go activity and finally Go activity for the controlled
response due to top-down DLPFC activity. \textbf{b)}
Electrophysiological recordings of the monkey striatum (reproduced
from \citet{WatanabeMunoz09}). The first row represents neurons coding
corresponding to the executed response (i.e. Go neurons) and the
second row represents neurons coding that suppress execution of the
corresponding action (i.e. NoGo neurons). \textbf{c)} Alternative
model simulating \citet{BrownBullockGrossberg04} assumption that the
indirect pathway acts to defer the execution of the correct response,
rather than suppress the alternative   response. Note predictions for
Go pathway accord with those of our   model and the data, but
prediction of NoGo neurons differ.}
 \label{fig.AS_striatum}
\end{figure}

\subsection{Global Response Inhibition}
\subsubsection{Methods}
In SRITs the selectively inhibited prepotent response must be replaced
with another, controlled response. Conversely, the stop-signal task
(SST) requires outright response inhibition
\citep[e.g.][]{LoganCowan84,AronPoldrack06,CohenPoldrack08} and is
used to assess global inhibitory control \citep{Aron11}. Specifically,
subjects are required to make press left and right keys in response to
Go-cues appearing on a screen. On a subset of trials {\it after} the
Go-cue has been presented, a stop-signal is presented after variable
delay (i.e. stop-signal delay; SSD) instructing the subject to
withhold responding.

\begin{figure}
\includegraphics[width=\columnwidth]{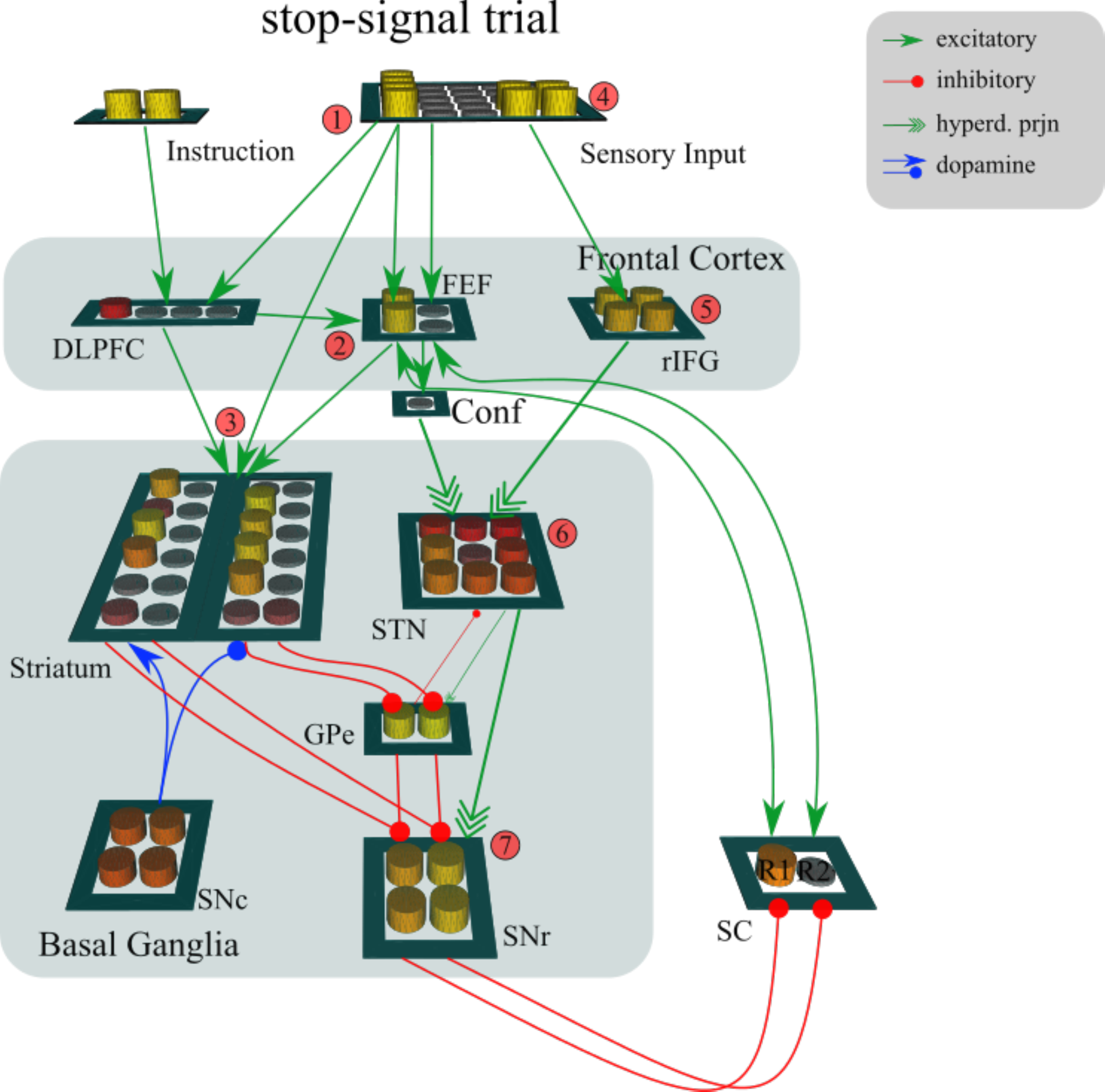}
\caption{Extended neural network model including rFIG  during
stop-signal trials. (1) Left input stimulus activates (2) left-coding
FEF response units and (3) initiates gating via striatum (similar to
pro-saccade trial in a). After a delay, (4) the stop-signal is
presented which activates (5) rIFG, which in turn (6) transiently
activates the STN and finally (7) the whole SNr to globally prevent
gating.  Note, that DLPFC is beginning to get active to initiate
selective response inhibition via striatal NoGo units.}
\label{fig.BG_inhib_nn_SST}
\end{figure}

Here we show that our model can also simulate the SST after we
included the right inferior frontal cortex (rIFG) with direct
projections to STN \citep{AronBehrensSmithEtAl07} see figure
\ref{fig.BG_inhib_nn_SST}. Given the assumptions of the race model
(i.e., a race between Go and Stop processes), one can estimate the
stop-signal reaction time (SSRT) by measuring the probability of
successful inhibition at different SSDs. This inhibition function is
then compared to the distribution of Go reaction times in non-stop
signal trials. There are several extensive reviews of the SST
\citep{VerbruggenLogan09}, so here we focus on how our model captures
the available evidence. Note that the SST typically refers to the task
involving manual movements (and inhibition thereof), but a well
studied equivalent has been used in the oculomotor domain, where it is
referred to as the \textit{'countermanding task'}. While the neuronal
circuitry involved in Go-responding depends on the response modality,
the neuronal circuitry involved in the global mechanism may be
independent of the response modality \citep{LeungCai07}.\\

Networks are presented with one of two input stimuli (left or right),
represented by a column of four units each. As in prior simulations,
prepotent responses are implemented by weights from the input units to
the corresponding FEF response units, such that a left stimulus
suggests a left response. On 25\% of trials, a stop-signal is
presented with variable delay (by activating devoted units in the
sensory input layer). The stop signal units send excitatory
projections directly to the rIFG layer.  rIFG units in the hyperdirect
pathway excite the STN
\citep{AronBehrensSmithEtAl07,NeubertMarsBuchEtAl10} and prevent
striatal response gating, and therefore inhibit responding if the SC
has not already surpassed threshold. In addition to this global
rIFG-STN response suppression mechanism, the DLPFC combines the
stop-signal input and the stimulus location to selectively inhibit the
associated response via activation of the corresponding population of
striatal NoGo units. Critically, this selective mechanism is slower
but remains active after the STN returned to baseline and prevents
subsequent responding. Thus, the model uses a fast, global but
transient response inhibition mechanism and a slower, selective but
lasting mechanism \citep{Aron11}. To estimate the SSRT, we use the
dynamic one-up / one-down \textit{staircase procedure} for adjusting
the SSD \citep[e.g.][]{LoganSchacharTannock97,OsmanKornblumMeyer86}.\\

We tested the influence of rIFG lesions on the SSRT
\citep{AronMonsellSahakianEtAl04} by parametrically reducing the
projection strength of rIFG to the STN.\\

The selective norepinephrine (NE) reuptake inhibitor Atomoxetine
increases NE release and improves stop-signal performance in animals,
healthy adults and adult ADHD patients
\citep{ChamberlainDel_CampoDowsonEtAl07,ChamberlainHampshireMullerEtAl09}.
NE is hypothesized to adaptively change the activation gain of neurons
in frontal cortex \citep{Aston-JonesCohen05}. We consequently tested
the influence of decreasing the gain parameter in units of the frontal
cortex\footnote{Gain modulates how step-like the activation-dynamics
of units are in relation to their input activity. Low gain leads to
linear activation dynamics while high gain levels make a unit respond
in a binary-like fashion.}.\\

Finally, we simulated different motivational influences on stop-signal
accuracy. Evidence for the neural underpinnings of motivational biases
comes from an fMRI study by \citet{LeottiWager10}, who reported that
subjects instructed to focus on speed instead of accuracy exhibited a
greater increase in activations in brain regions associated with
response facilitation, including the FEF and the striatum. Conversely,
when instructed to focus on accuracy, subjects exhibited greater
activity in IFG regions associated with response inhibition. We thus
simulated these activation patterns to account for speed-accuracy
tradeoff in a similar manner as in the antisaccade simulations. In the
speed-condition, we manipulated the strength of FEF to striatum
connections due to evidence that frontostriataal connectivity is
enhanced under speed emphasis
\citep{ForstmannDutilhBrownEtAl08,ForstmannAnwanderSchaferEtAl10,MansfieldKarayanidisJamadarEtAl11}.
Conversely, in the accuracy condition we increased baseline excitatory
input to rIFG, allowing it to be more excitable and hence facilitating
STN recruitment. This simulation approximates the effect of a putative
PFC rule based representation to focus on accuracy. Recent data
supports the notion that the (right) STN, which receives input from
rIFG, shows increased excitability associated with an increased
response caution during accuracy focus
\citep{MansfieldKarayanidisJamadarEtAl11}.

\subsubsection{Results}
\label{sec.results_sst}
As with the SRITs above we extracted a list of key qualitative results
from the literature we use to evaluate the fit of our model.

\begin{enumerate}[label=\#\arabic*]
\item \label{bench.SS_monotonic} The probability of inhibiting a
response decreases monotonically as SSD increases
\citep{VerbruggenLogan08}.
\item \label{bench.SS_compare} Error responses that escape inhibition
  are, on average, faster than  Go responses on no-stop-signal trials.
However, while the distributions begin at the same minimum value, the
responses that escape inhibition have a shorter maximum value
\citep{VerbruggenLogan08}.
\item \label{bench.SS_STN_SNr} STN neurons are excited to stop signals
but show little differentiation between stop-signal inhibition and
stop-respond error trials \citep{AronBehrensSmithEtAl07}. Contrary,
downstream SNr neurons are excited in correct trials but are
disinhibited during error trials
\citep{SchmidtLeventhalPettiboneEtAl12}.
\item \label{bench.SS_PFC} SEF neurons are activated in stop-signal
and stop-response trials {\it after} SSRT and can thus not contribute
to successful stopping \citep{StuphornTaylorSchall00}.
\end{enumerate}

\begin{figure}
\subfigure{
 \includegraphics[width=.5\columnwidth]{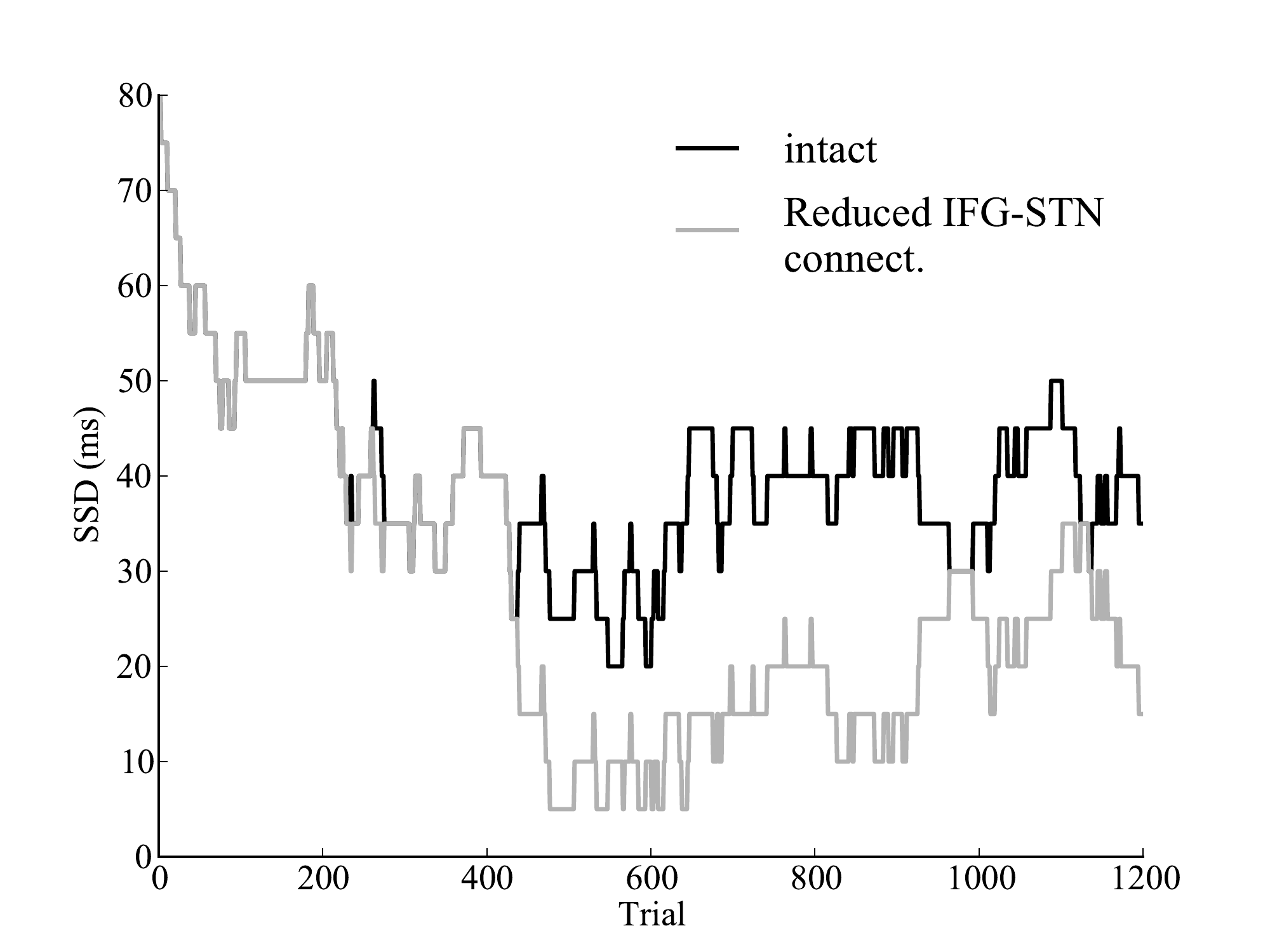}
 \label{fig.SS_staircase}
}
\subfigure{
\includegraphics[width=.5\columnwidth]{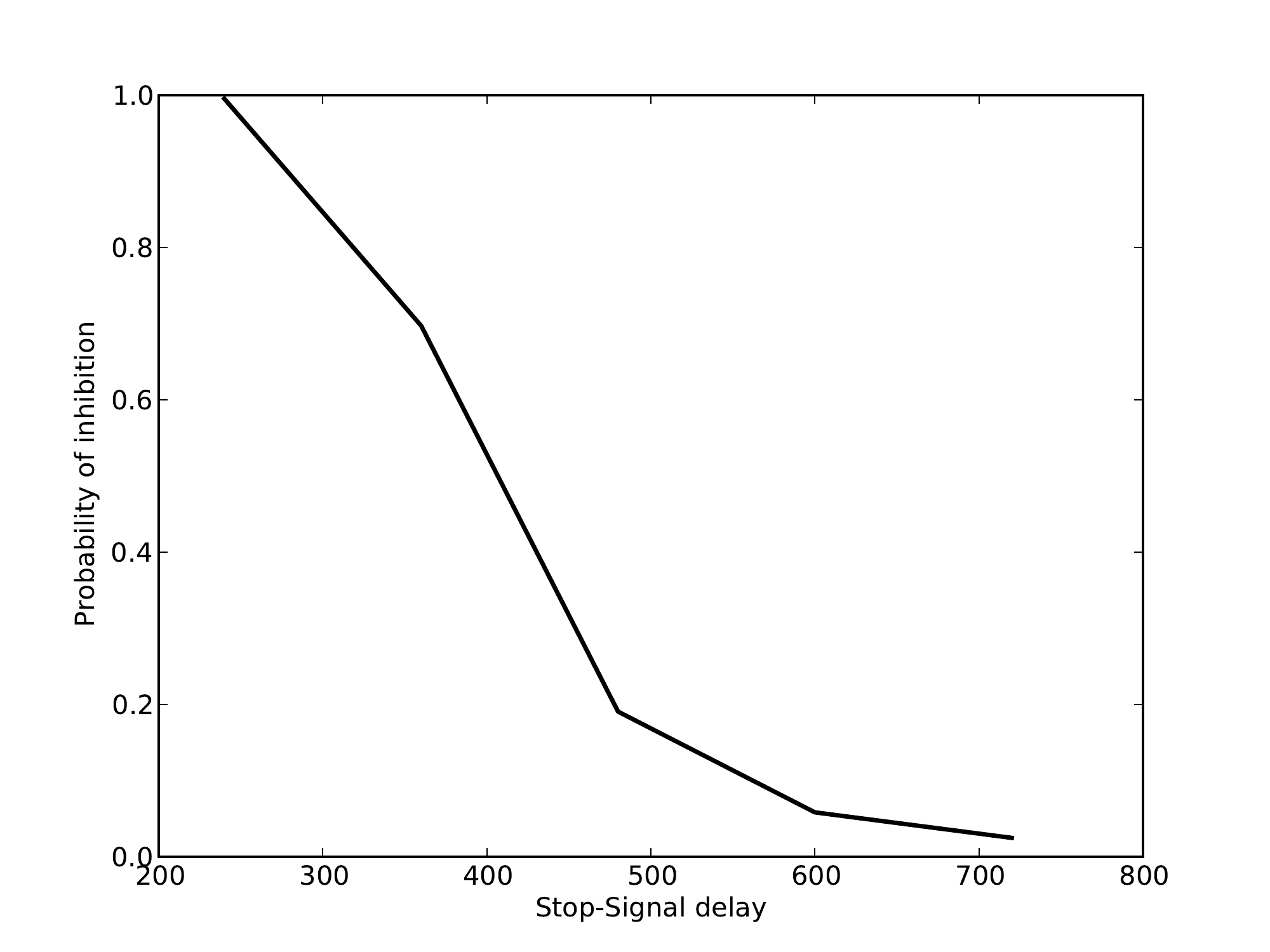}
 \label{fig.SS_SSD_inhib}
}
\caption{\textbf{a)} Progression of the staircase procedure for
manipulating SSD in networks with reduced rIFG-STN connectivity. Trial
number is plotted on the x-axis and the stop-signal delay (SSD) in ms
(converted from simulator time) is plotted on the y-axis. If a
response is successfully inhibited on stop-signal trial, the SSD is
increased by 20 ms to make it harder. If a response is erroneously
made on a stop-signal trial, the SSD is decreased by 20 ms. Networks
without lesion are highest in general representing the most effective
Stop-process that is able to withhold responses even when the SSD is
quite long. \textbf{b)} Inhibition function of the neural network
model in the stop-signal task. The model is tested on systematically
varying levels of stop-signal delay (SSD) in ms and the proportion of
correctly inhibited trials is plotted along the y-axis.}
\end{figure}

\paragraph{Behavior}
To illustrate the staircase procedure, figure \ref{fig.SS_staircase}
shows an example trace of how SSDs are adjusted to assess 50\%
stop-signal accuracy. As can be seen, the network with rIFG lesion is
impaired at stopping and requires shorter SSD on average to inhibit
successfully.\\

As can be seen in figure \ref{fig.SS_SSD_inhib} the inhibition
function resulting from testing the neural network systematically with
different SSDs reveals a monotonically decreasing probability of
correctly stopping (qualitative pattern \ref{bench.SS_monotonic}).\\

\begin{figure}
\includegraphics[width=\columnwidth]{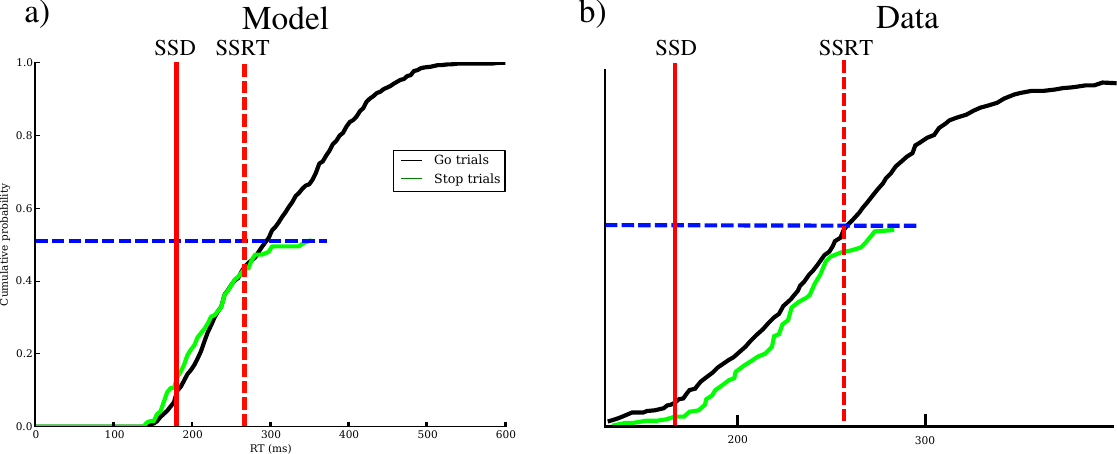}
\caption{\textbf{a)} Cumulative reaction time distributions of the
neural network model and from a monkey experiment. \textbf{b)}
Cumulative reaction time distribution from a monkey experiment for
comparison. Reproduced from \citep{LoBoucherPareEtAl09}. The solid red
line denotes mean stop-signal delay (SSD); the broken red line denotes
stop-signal reaction time (SSRT) offset at SSD. The broken blue
horizontal line represents 50\% stopping accuracy. Note that the
response distribution sums to the response probability -- not
necessarily to 1.}
\label{fig.SS_cum_RT}
\end{figure}

Cumulative RT distributions of Go and non-canceled Stop trials are
presented in figure \ref{fig.SS_cum_RT}. Both distributions match
closely up until SSD+SSRT (qualitative pattern \ref{bench.SS_compare})
suggesting that both are generated by the same process.

\begin{figure}
  \subfigure{
 \includegraphics[width=.5\columnwidth]{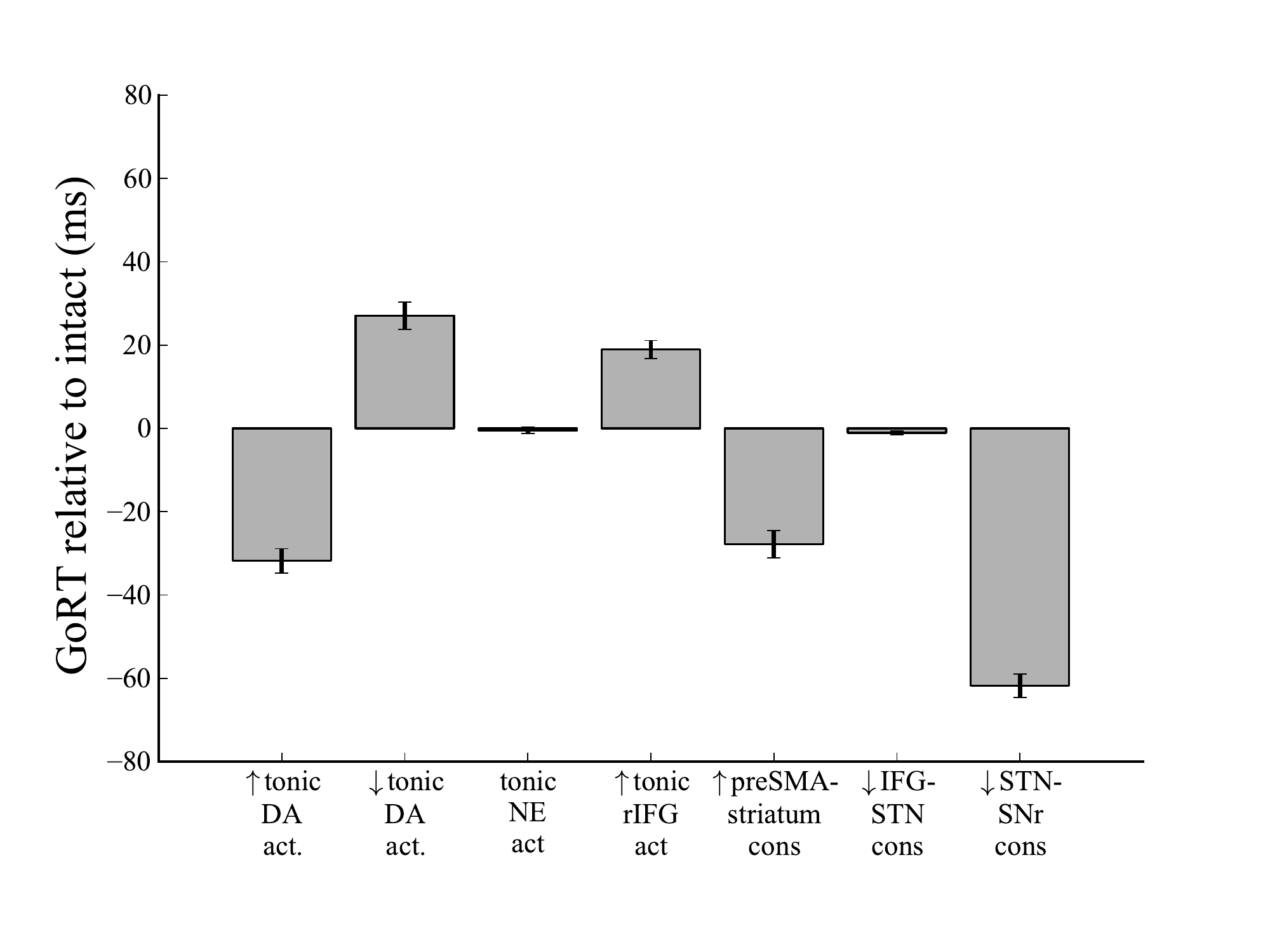}
 \label{fig.SS_GoRT}
}
\subfigure{
 \includegraphics[width=.5\columnwidth]{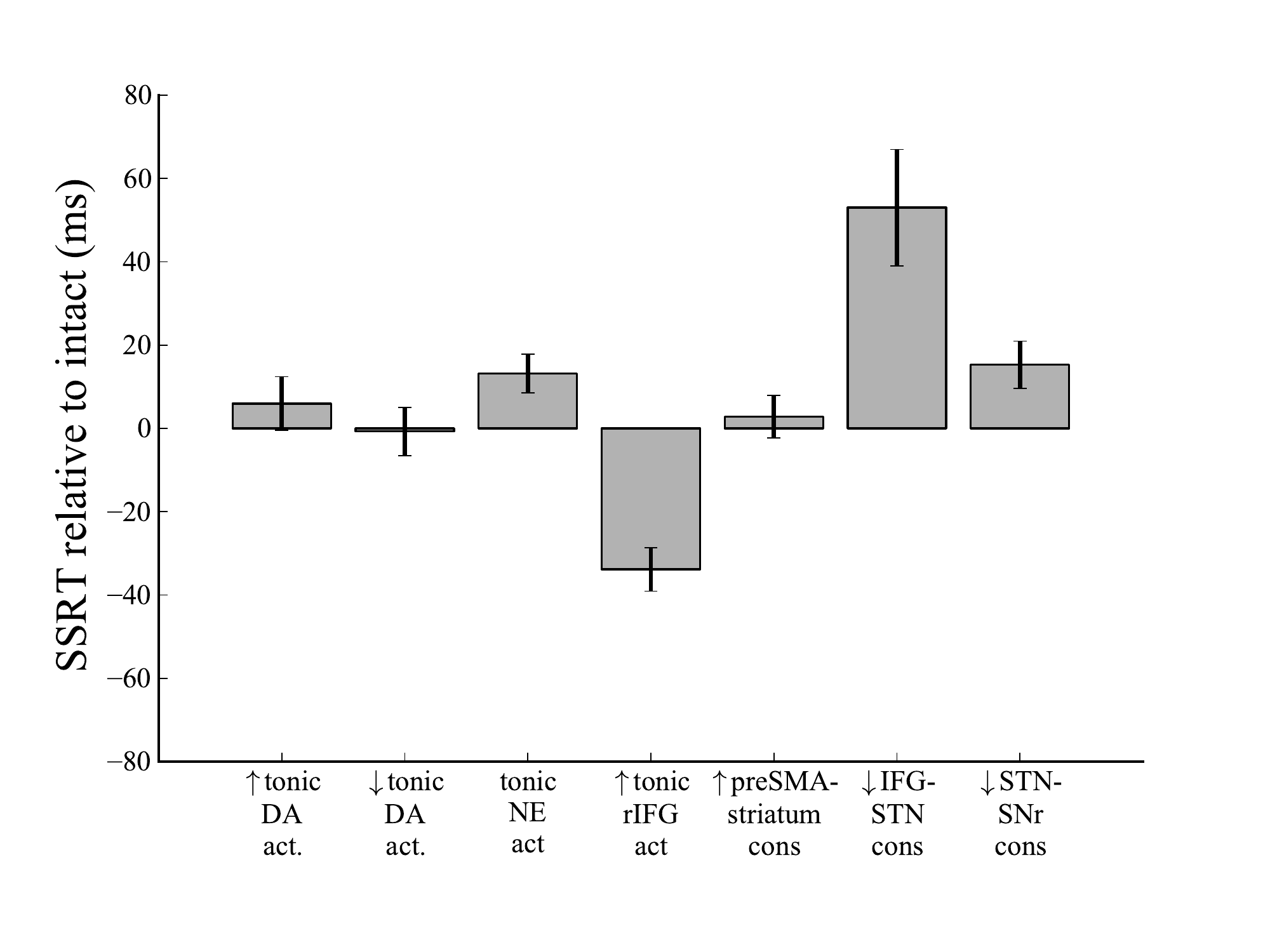}
\label{fig.SS_SSRT}
}
\caption{\textbf{a)} Mean RTs in ms $\pm$ SEM (converted from
simulator time) for Go trials under different modulations (see text).
\textbf{b)} Mean SSRTs in ms $\pm$ SEM (converted from
simulator time)  under different modulations (see text).}
\end{figure}

Different modulations affect GoRT and SSRT in different ways (figures
\ref{fig.SS_GoRT} and \ref{fig.SS_SSRT}). While DA manipulations
certainly speed GoRT, SSRT remains largely unaffected. On the other
hand, when the network is tested with reduced gain (simulating low NE
levels), or has lesions to either STN or rIFG, it exhibits SSRT
deficits (increases). Finally, simulated accuracy emphasis results in
slowed Go RT but faster SSRT (more effective inhibition). The pattern
that emerges from these results is that SSRT is changed by modulations
of parameters that are part of the global inhibitory pathway: rIFG and
STN.\\

\begin{figure}
 \includegraphics[width=\columnwidth]{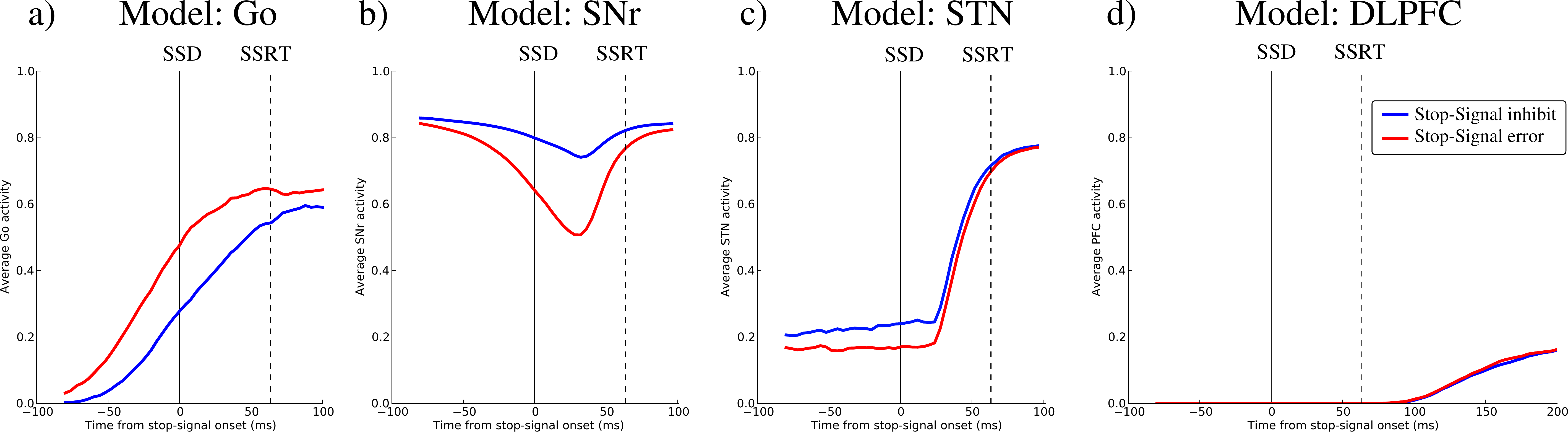}
 \caption{Average activity aligned to stop-signal onset for inhibited
and error stop-signal trials. \textbf{a)} Striatal Go-neuronal
activity. \textbf{b)} Substantia nigra pars reticulata activity.
\textbf{c)} Subthalamic nucleus activity. \textbf{d)} Activity of the
executive control complex consisting of DLPFC, SEF and pre-SMA.}
 \label{fig.SS_STN_SNr}
\end{figure}

\paragraph{Neurophsyiology}
To assess the neural correlates of stopping behavior in our model we
analyzed STN and SNr activity aligned to stop-signal onset. As can be
seen in figure \ref{fig.SS_STN_SNr}, there is little differentiation
between stop-signal inhibition and error trials while SNr units show a
marked dip in error trials that is less pronounced in inhibition
trials (qualitative pattern \ref{bench.SS_STN_SNr}).

We moreover analyzed the activity pattern of our executive control
complex which consists of DLPFC, SEF and pre-SMA. As can be seen in
figure \ref{fig.SS_STN_SNr}, activation is observed in stop-signal
trials (both stop-respond and successful inhibitions) only {\it after}
SSRT and could thus had no influence on the stopping (qualitative
pattern \ref{bench.SS_STN_SNr}). This result implies that global
stopping to salient stop signals is most likely driven by the fast
stop process along the rIFG-STN hyperdirect pathway. We ascertain that
executive control processes are delayed relative to this global
stopping mechanism, and may participate in selective response
inhibition (and in the stop-change task, activation of the correct
response) after the global response pause has passed.

\section{Discussion}
The interaction between executive control and habitual behavior is a
central feature of higher-level brain function, and plays a role in
various domains from cognitive psychology (under the rubric of
``system 1'' vs. ``system 2''; \citep{Evans03}) to machine learning
(model-free vs. model based control \citep{DawNivDayan05}). At the
core of this interaction is a mechanism that allows executive control
to override the habitual system and guide action selection. A
multitude of psychological cognitive tasks have been used to probe the
nature of this interaction. The stop-signal task requires outright
stopping of a response already in the planning stage. The antisaccade
\citep{Hallett78}, Simon \citep{Simon69}, and saccade override
\citep{IsodaHikosaka07,IsodaHikosaka08} tasks all involve inhibition
of a prepotent action together with initiation of an action
incompatible with the prepotent one. Despite the apparent behavioral
simplicity of these tasks, various lines of research have revealed a
highly complex and tightly interconnected brain network underlying
response inhibition consisting of frontal areas including DLPFC, SEF,
pre-SMA, FEF, rIFG, and dACC and basal ganglia structures including
the striatum and STN.\\

We presented a dynamic neural network model of selective and global
response inhibition which provides a description of the distributed
computations carried out by individual brain regions and
neurotransmitters. The complexity of this model is grounded by well
established neuroanatomical and physiological considerations, and
accounts for a wealth of key data including electrophysiology,
psychiatric and pharmacological modulations, behavioral, lesion and
imaging studies. Moreover, this model is constrained (i) by using a
single parameterization across all simulations of intact function and
(ii) by the multitude of qualitative results from different levels of
analysis it is required to reproduce. Although we used one
parameterization across the intact model
simulations, we also generalized the functionality via systematic
manipulations across a range of parameters. In other work (Wiecki \&
Frank, in prep), we have
shown that the emerging fundamental computational properties of this
complex system as a whole are captured by analysis using a modified
drift diffusion model, in which distinct mechanisms within the neural
model (e.g., STN projection strength, DLPFC speed) are monotonically
related to high
level decision parameters (e.g., decision threshold, and drift rate of
the executive process).

\subsection{Selective Response Inhibition}
\label{sec.discussion_ast}
In our SRIT simulations, the model assumes that prepotent, reflexive
actions such as a saccade to an appearing stimulus (e.g. a prosaccade)
are selected via the cortico-cortical route and swiftly gated by the
BG. An abundance of data supports the general involvement of the BG in
saccade generation and inhibition
\citep[e.g.][]{HikosakaTakikawaKawagoe00,Hikosaka89,HikosakaWurtz86}.
Conversely,
the cognitive control system not only represents the task rules needed
to respond appropriately (e.g. in DLPFC), but also incorporates a
downstream mechanism in dACC-STN to detect when these rules indicate
an alternative action than was originally initiated. Thus our model
synthesizes the popular account of dACC in terms of response conflict
\citep{BotvinickBraverBarchEtAl01} with recent studies suggesting that
dACC rather reflects the value of the alternative action
\citep{KollingBehrensMarsEtAl12}. Moreover, via the hyperdirect
pathway to the STN, this mechanism serves to transiently increase the
BG gating threshold to prevent prepotent actions from being
facilitated and allows more time for the controlled PFC-striatal
mechanisms to selectively suppress this response and to facilitate
appropriate alternative courses of action. It has also been shown that
the SEF, FEF \citep{MunozEverling04}, dACC
\citep{BotvinickCohenCarter04}, pre-SMA
\citep{IsodaHikosaka07} and STN
\citep{IsodaHikosaka08} are involved in detecting conflict between a
planned response and the current rule, and for switching from
an automatic to a volitional response (e.g., antisaccades). \\

To detect conflict between reflexive and controlled responses, the
system needs to be able to compute the correct identity of the
controlled response itself. In the model, the DLPFC integrates task
instructions and current stimulus location and forms a conjunctive
rule representation \citep{WallisMiller03a,BungeWallis08} that then
provides evidence for the associated controlled response via its
projection to the FEF, and further biases the gating of this response
(and the selective suppression of the reflexive response) via
striatum. We demonstrated that this is a necessary condition for our
model by showing that a model with faster integration speeds fails to
account for key behavioral patterns.\\

Thus it should be clear that compared to a congruent response, an
incongruent response should (i) be more prone to error because it
depends on successful inhibition of prepotent actions which may be
close to threshold by the time conflict is detected and (ii) take
longer due to (iia) additional computation needed for the DLPFC to
perform the requisite vector inversion (activation of correct rule
representation among multiple competitors based on an integration of
input and instruction), and (iib) the delay in commitment to a
response resulting from the increase in decision threshold along the
hyperdirect pathway.\\

Early cognitive models of interference control assumed a dual-route
mechanism for action selection, including an automatic response route
and a volitional one
\citep{KornblumHasbroucqOsman90,Eimer95,DeLiangKristjanssonNakayama05,Ridderinkhof02}.
This model was extended to include selective suppression of the
automatic response by the volitional response mechanism (i.e. the
activation-suppression model
\citep{Ridderinkhof02,RidderinkhofForstmannWylieEtAl11}). Our model
shares these attributes but makes two crucial contributions to this
discussion: (i) strong predictions on the neural correlates of these
abstract cognitive processes, and (ii) a raise in decision threshold
requiring more evidence to gate \textit{any} response. This latter
mechanism may not only be adaptive as a fast route to prevent gating
of prepotent actions, but could also serve to increase the likelihood
that the alternative action selected is the most accurate
(particularly when there may be more than one, as is often the case in
more realistic executive control scenarios than those typically
studied in simple response inhibition tasks).

\subsubsection{Response time distributions and errors: Neural
underpinnings}
At the behavioral level, our intact model reproduces the same patterns
found empirically -- networks made more errors (see figure
\ref{fig.AS_error}) and were in general slower (see figure
\ref{fig.AS_RTs}) on incongruent trials compared to congruent trials
\citep[e.g.][]{ReillyHarrisKeshavanEtAl06,HarrisReillyKeshavanEtAl06,ReillyHarrisKhineEtAl07,McDowellBrownPaulusEtAl02}.
Incongruent errors were more likely to occur when networks responded
fast (see figure \ref{fig.AS_histo}, \ref{fig.AS_RT_speed} and
\citet{RidderinkhofForstmannWylieEtAl11}). These errors result
primarily from variance in the speed of cognitive control (DLPFC), but
also in the prepotent response (in some trials gating is faster than
others) and in the inhibition process (in some trials the hyperdirect
pathway and/or striatal NoGo process is slower). Moreover, reduced
DLPFC connectivity also degrades accuracy on incongruent trials,
mirroring the empirical performance degradation in antisaccade tasks
during development associated with reduced DLPFC connectivity
\citep{HwangVelanovaLuna10}. A more explicit investigation into the
dynamics of these processes comes from the simulated electrophysiology
across brain regions and trial types.

\subsubsection{Conflict- and error-related activity: relation to
  existing models}
The Error Related Negativity (ERN) is an event-related potential
associated with errors made in forced-choice reaction time tasks
\citep{FalkensteinHohnsbeinHoormannEtAl91,GehringGossColesEtAl93}. The
ERN reaches its peak within 100 ms \textit{after} the erroneous
response. Using a connectionist model, Yeung and collegues hypothesize
the ERN to reflect conflict between the executed, erroneous response
and the still-evolving activation of the correct response
\citep{YeungCohen06,YeungCohenBotvinick04}. Thus, the error detection
mechanism reflects an internal correction of the executed response,
leading to a transient period of response conflict.  According to this
same framework, a similar potential should be observed in high
conflict trials \textit{before} correct responses, when conflict is
resolved prior to responding. These authors indeed reported that the
N2 potential exhibited just this profile and argued that it reflected
the same underlying conflict mechanism in the dACC.\\

Our dACC node exhibits the same qualitative pattern of increased
activity (i) before correct incongruent responses, (ii) after
incorrect incongruent responses and (iii) baseline activity during
congruent responses. However, this pattern is not unique to ERPs
thought to originate from dACC, but is also found in
electrophysiological recordings in pre-SMA, SEF
\citep{EmericLesliePougetEtAl10} and STN \citep{IsodaHikosaka08}. Our
model provides an explicit framework that recapitulates these effects
and explores their influences on behavior. Together, these dynamics
accord with our earlier assertion that our model synthesizes the
conflict model with the notion that the dACC reflects the value of the
alternative action: this network only becomes activated when the
alternative action is deemed to be more correct than the prepotent
one. This process occurs either prior or following response execution
(as in the conflict monitoring account), but must always occur after
the initial activation of an incorrect (often prepotent) response (not
specified by the conflict account but consistent with the alternative
action value account). To more formally describe the computational
dynamic implicated, we devised a modified drift diffusion model which
explicitly incorporates this reversal in evidence.

\subsection{Global Response Inhibition}
\label{sec.discussion_sst}
By adding a single rIFG layer to our model we generalized our model to
capture data from global response inhibition tasks such as the SST.
As we demonstrated above, this model recovers key qualitative
behavioral patterns reported in the literature. Moreover, model
neurophysiology revealed interesting similarities to recent rat
electrophysiological recordings in the SST
\citep{SchmidtLeventhalPettiboneEtAl12}. Specifically,
while STN activity surges in response to the stop signal to the same
extent regardless of whether the response is successfully inhibited or
not, activity in the SNr strongly differentiates between these trial
types. During errors, the striatal Go signals were potent and early
enough to inhibit SNr activity in spite of the STN surge. These
results suggest that the source of response inhibition errors is
variance in the Go process, but that the duration of the stop-process
is rather fixed. This conceptualization matches closely with the
interactive horse-race model \citep{VerbruggenLogan08Rev}. Here, we
hypothesize that the critical point of interaction between the two
processes is the SNr.\\

Why did we add an rIFG layer given that our initial model already
contained an executive control complex including DLPFC? As described
above, rIFG and STN involvement in the SST is well established, and
moreover, simulations showed that the activations in our executive
control complex needed to account for SRITs was too slow to account
for global response inhibition needed in SST. Nevertheless, the nature
of the (potentially separable) mechanisms engaged for detecting when
inhibitory control is necessary, and how it should be implemented,
remains largely elusive. In particular, the role of rIFG is actively
debated. Some studies specifically implicate the rIFG in response
inhibition
\citep{VerbruggenAronStevensEtAl10,AronFletcherBullmoreEtAl03,ChambersBellgroveGouldEtAl07,SakagamiTsutsuiLauwereynsEtAl01,XueAronPoldrack08},
whereas others report rIFG activity in tasks lacking pure response
inhibition demands, suggesting that it is more involved in monitoring
or salience detection
\citep{SharpBonnelleDeBoissezonEtAl10,VerbruggenAronStevensEtAl10,HampshireChamberlainMontiEtAl10,FlemingThomasDolan10,ChathamClausKimEtAl12,MunakataHerdChathamEtAl11}.
Our model unifies these two seemingly opposing views by arguing that
the rIFG in fact detects salient events and, via downstream
processing, engages a stopping mechanism whether or not it is required
by the task rules. In both the stop-signal and stop-change task,
subjects have to detect an infrequent signal which tells them to
update their current action plan. We argue that these signals are
salient events and, via noradrenergic modulation, enhance processing
in the rIFG which, in turn, causes an \textit{orienting} or
\textit{circuit breaker} response by activating the STN
\citep{SwannPoiznerHouserEtAl11} to pause response selection. This
pause enables the volitional DLPFC based response selection mechanism
to take control and either inhibit a specific response (as in the
stop-signal task) or initiate a new response (as in the stop-change
task). This theory of a rIFG triggering a global response-pause is
supported by rIFG involvement in the oddball task
\citep{Stevens00,HuettelMcCarthy04} which requires no behavior
adaptation whatsoever, yet still causes response slowing
\citep{BarceloEsceraCorralEtAl06,ParmentierElfordEsceraEtAl08}.
Indeed, in many of the above-reported studies in which rIFG is
activated under conditions of monitoring or saliency, when they have
been reported, subject RTs were nevertheless delayed despite no overt
inhibitory demands
\citep{SharpBonnelleDeBoissezonEtAl10,FlemingThomasDolan10,ChathamClausKimEtAl12}.\\

\subsection{Different forms of response inhibition}
Inhibitory control can be issued globally or selectively
\citep{AronVerbruggen08,Aron11}. The brain seems to revert to a global
inhibitory mechanism when unexpected events occur that require quick
response adaptation (e.g., stop-signals), and to a selective
inhibitory control mechanism when response inhibition can be prepared
\citep{GreenhouseOldenkampAronEtAl11,HuLi11}. We propose that
selective inhibition of the prepotent response is initiated by the
DLPFC and implemented via the indirect corticostriatal NoGo pathway
\citep{ZandbeltVink10,WatanabeMunoz09,WatanabeMunoz10,HuLi11,JahfariWaldorpWildenbergEtAl11}.
Global response inhibition on the other hand is driven by a salience
detection mechanism implemented in the rIFG which directly projects to
the STN to inhibit responding
\citep{Mink96,NambuTokunoHamadaEtAl00,NambuTokunoTakada02,KuhnWilliamsKupschEtAl04,AronDurstonEagleEtAl07,EagleBaunezHutchesonEtAl08,IsodaHikosaka08,JahfariWaldorpWildenbergEtAl11,JahfariVerbruggenFrankEtAl12}.\\

In addition to the selectivity of inhibitory control, differences
exist between proactive and reactive initiation of response inhibition
\citep{Aron11,GreenhouseOldenkampAronEtAl11,SwannCaiConnerEtAl11,CaiOldenkampAron11}.
Our modeling work suggests multiple possible sources for proactive
control. Speed-accuracy adjustments are implemented by increasing
functional connectivity between frontal motor regions and striatum to
decrease the decision threshold under speed emphasis (see figure
\ref{fig.AS_FEF_speedacc}, \ref{fig.AS_STN_speedacc} and
\citet{LoWang06,ForstmannAnwanderSchaferEtAl10}). The second proactive
mechanism increases response caution by increasing baseline rIFG
activity to prime saliency detection and slow responding via the
rIFG-STN hyperdirect pathway (see figure \ref{fig.SS_SSRT}).
Interestingly, while FEF$\rightarrow$striatum functional connectivity
influence speed and accuracy in our SRIT simulations, SSRT in the
stop-signal task is unaffected by this modulations and is only
improved by an increase in tonic rIFG activity. This suggests that
proactive control in form of mere response slowing is uneffective in
reducing SSRT -- the staircase procedure adapts to slower overall
responding -- but that enhanced attentional monitoring has
preferential influence on global inhibitory control. In other words,
although all these mechanisms can lead to adjustments in decision
threshold, only those associated with active engagement of the stop
process will facilitate inhibitory control per se. If confirmed, this
result may have implications for refining therapy of inhibitory
control disorders like addiction, obesity and OCD. Nevertheless, it
remains important to emphasize that the striatal NoGo pathway is also
thought to help to prevent the proactive selection of maladaptive
responses.

\subsection{Multiple mechanisms of response threshold regulation in
fronto-basal-ganglia circuitry at different time scales}
Different mechanisms in our neural network influence the gating
threshold for initiating motor responses at distinct time scales, and
modulated by distinct cognitive variables. First, the strength of
cortico-striatal projections regulate the ease with which cortical
motor plans can be gated by the BG, allowing for speed emphasis in the
speed-accuracy tradeoff (see figure \ref{fig.AS_STN_speedacc}). This
aspect of our model is quite similar to the model of \citet{LoWang06}
and was subsequently corroborated by
\citet{ForstmannAnwanderSchaferEtAl10}. Our model converges on the
same conclusion but extends this view by showing that gating threshold
is also more dynamically regulated on a shorter time-scale by (i)
motivational state (changes in DA levels, which are modulated by
reinforcement and also facilitate striatal Go signals); and (ii)
response conflict and saliency (via the hyperdirect pathway, making it
more difficult or Go signals to drive BG gating
\citep{JahfariWaldorpWildenbergEtAl11}). Moreover, STN efficacy in the
neural model is positively correlated with increases in estimated
decision threshold \citep{RatcliffFrank12}. Evidence for
conflict-induced decision threshold adjustment via the hyperdirect
pathway has been recently described in a reinforcement-based decision
making task \citep{CavanaghWieckiCohenEtAl11}. Increases in frontal
EEG activity during high conflict decisions were related to increases
in decision threshold estimated by the drift diffusion model.
Intracranial recordings directly within the STN also revealed decision
conflict-related activity during the same time period and frequency
range as observed over frontal electrodes (see also
\citet{ZaghloulWeidemannLegaEtAl12}).  Moreover, disruption of STN
function with deep brain stimulation led to a reversal of the
relationship between frontal EEG and decision threshold, without
altering frontal activity itself. These data thus support the notion
that frontal-STN communication is involved in decision threshold
adjustment as a function of conflict. Similarly, proactive preparation
to increase decision threshold in the stop signal task when stop
signals are likely is associated with hyperdirect pathway activity
\citep{JahfariVerbruggenFrankEtAl12}. \\

In our neural models, conflict-related STN activity subsides with
time (see figure \ref{fig.AS_STN_ephys}), due to resolution of
conflict in FEF/ACC, feedback
inhibition from GPe, and neural accommodation. Thus a more refined
description of this transient STN surge is that it initially increases
the decision threshold (more so with conflict), followed by a dynamic
collapse of the decision threshold over time. Indeed, a recent
multilevel computational modeling and behavioral study by
\citet{RatcliffFrank12} supported this idea by showing that a
collapsing threshold diffusion model provided good fits to both the BG
model and to human participant choices in a reinforcement conflict
task. Moreover, the temporal profile of the best-fitting collapsing
threshold corresponded well to the time course of the collapse in STN
activity across time.\\

\subsection{Psychiatric disorders and differential effects of
  dopamine and norepinephrine}
Abnormal striatal DA signaling is hypothesized to be at the core of
many disorders, including PD
\citep{BernheimerBirkmayerHornykiewiczEtAl73}, SZ
\citep{BreierKestlerAdlerEtAl98} and ADHD
\citep{CaseyNiggDurston07,FrankSantamariaOReillyEtAl07}. Intriguingly,
all of these disorders are linked to response inhibition deficits in
the stop-signal task. Our earlier BG models have successfully
accounted for a wide variety of findings associated with striatal DA
manipulations across reinforcement learning and working memory tasks
\citep[for review,][]{WieckiFrank10}. Yet, we found here that striatal
DA manipulations, while affecting overall RT, had negligible effects
on response inhibition deficits as assessed by SSRT (see figure
\ref{fig.SS_SSRT}).  This prediction converges with recent evidence
\citep[reviewed in,][]{MunakataHerdChathamEtAl11} showing that
levodopa, a drug that increases DA levels in striatum
\citep{HardenGrace95}, had no influence on SSRT in PD patients
\citep{ObesoWilkinsonCasabonaEtAl11,ObesoWilkinsonJahanshahi11}.\\

This lack of DA effect raises the question of the source of the
response inhibition deficits in the aforementioned disorders. One
conspicuous candidate is abnormal NE functioning as suggested by
evidence in both ADHD
\citep{FaraonePerlisDoyleEtAl05,RamosArnsten07,FrankScheresSherman07}
and PD \citep{FarleyPriceMcCulloughEtAl78}. In our simulations, NE
modulation influences SSRT via its gain-modulatory effects in rIFG
\citep{Aston-JonesCohen05}. Additional support for this account comes
from pharmacological experiments using the selective norepinephrine
reuptake inhibitor atomoxetine, which improves response inhibition
performance in animals, healthy adults and ADHD patients
\citep{ChamberlainDel_CampoDowsonEtAl07,ChamberlainHampshireMullerEtAl09}.
Moreover, fMRI analysis revealed that atomoxetine exerted its
beneficial effects via modulation of rIFG
\citep{ChamberlainHampshireMullerEtAl09}, providing additional support
for the model mechanisms. Finally, this highlights an alternative
source for response inhibition deficits observed in PD patients
previously linked to DA dysfunction (see \citet{VazeyAston-Jones12}
for a review highlighting the importance of aberrant NE signaling in
cognitive deficits of PD patients).

\section{Limitations}
Despite our model's success in reproducing and explaining a wide array
of data and offering potential solutions for long standing issues in
the field, we certainly acknowledge that there are many errors of
omission and -- although we did not include any biological features
that are unsupported by data -- perhaps some errors of commission. We
note however note that most of our assumptions and simulations are
largely orthogonal to each other. Thus, each aspect of the model is
falsifiable on its own, without necessarily falsifying other aspects.
We discuss a few salient limitations below; it is by no means
exhaustive.

\subsection{Specificity of PFC regions and function}
While the BG of our neural network model is fairly concrete and
solidly grounded on ample anatomical electrophysiological, and
functional evidence, the individual contributions of frontal regions
including DLPFC, SEF, pre-SMA, FEF and dACC are not as well
established currently. For example, we identified an executive control
network in our model consisting of DLPFC, SEF and pre-SMA. The task
rules and necessary motor commands to follow them are implemented by
hard-coded input and output weight patterns of its extended network
(i.e. sensory input, instruction, FEF and striatum). This
implementation short-circuits a lot of the computational complexities
the biological system has to solve; (i) the executive controller has
to selectively retrieve the appropriate rule for the current trial
from short or long-term working memory; (ii) integrate the sensory
evidence to compute the correct response (e.g. via vector inversion);
(iii) compute the necessary motor sequences to perform the correct
action; and (iv) identify incorrectly activated prepotent responses
and selectively suppress them. While neural network models with a more
detailed representation of PFC exist \citep[e.g.][]{OReillyFrank06} in
which rule-like representations can develop through experience, how
exactly the necessary computations can be implemented dynamically is
as-of-today a still unresolved question.\\

Critically, our focus in this work was on how PFC and BG interact when
inhibitory control is required by extending the detailed BG model by
\citet{Frank06}. We also account for some electrophysiologcal data in
frontal cortex, while acknowledging that there is still some
uncertainty in the respective roles of  these areas and their
interactions which will be open for revision as more data become
available.

\subsection{Learning}
Previous BG models explored the role of DA in feedback driven learning
\citep{WieckiFrank10}.  As humans (but not monkeys) are able to
perform this task without learning, we chose to remain agnostic about
the type of learning that takes place prior to performing the task. We
thus hard-coded task rules into the model. An additional driving
factor is the lack of published reports on specific learning phenomena
in the SST and AST.

\section{Conclusions}
We presented a comprehensive, biologically plausible model of global
and
selective response inhibition which takes known properties of the
neuronal underpinnings into account and tries to link them with
results from cognitive science, electrophysiology, imaging studies and
pharmacological experiments. Here, we showed that augmenting our
previously described BG model with the addition of the FEF, DLPFC, and
rIFG allows us to simulate control over prepotent responses and to
capture a wealth of data in this domain across multiple levels of
analysis. We furthermore provide multiple mechanisms that can lead to
disruptions in inhibitory control processes and which have
implications for interpretation of data from
patients with psychiatric disorders such as SZ and ADHD. Our model
shows that the observed
deficits in inhibitory control paradigms do not necessarily have to
reflect dysfunctional response inhibition per se but could be due to
other factors such as salience, conflict detection and/or
motivation, and related to distinct neural mechanisms.

\section{Acknowledgments}
All modeling was performed by TVW under supervision of MJF. We thank
Jeffrey Schall, Gordon Logan, Christopher H. Chatham, James F.
Cavanagh, and three anonymous reviewers for helpful comments on an
earlier version of the manuscript.\\

This project was supported by NIMH grant R01MH080066-01 and NSF
grant1125788 to MJF, and partially supported by the Intelligence
Advanced Research Projects Activity (IARPA) via Department of the
Interior (DOI) contract number D10PC20023.  The US Government is
authorized to reproduce and distribute reprints for Governmental
purposes notwithstanding any copyright annotation therein. The views
and conclusion contained herein are those of the authors and should
not be interpreted as necessarily representing the official policies
or endorsements, either expressed or implied, of IARPA, DOI, or the US
Government.

\section{Appendix}
\subsection{Software}
The model and the Python scripts are available at \texttt{http://ski.clps.brown.edu/BG\_Projects/}.

\subsection{Implementation details}
Like the original \citet{Frank06} model, this model is implemented in
the Emergent neural modeling software framework
\citep{AisaMingusOreilly08}, which
can be downloaded here:\\
\texttt{http://grey.colorado.edu/emergent/index.php/Main\_Page}.\\
Emergent measures simulator time in cycles. Here, we convert this
time to ms by multiplying cycles by 4 to roughly match behavioral and
electrophysiological data.\\

Emergent uses point neurons with excitatory, inhibitory, and leak
conductances contributing to an integrated membrane potential, which
is then thresholded and transformed via an $\frac{x}{x+1}$ sigmoidal
function to produce a rate code output communicated to other neurons
(discrete spiking can also be used, but produces noisier results).

The membrane potential $V_{m}$ is a function of ionic conductances $g$
with reversal (driving) potentials $E$ as follows:
\begin{equation}
\triangle V_{m}(t) = \tau \sum_{c} g_{c}(t)\overline{g_{c}}(E_{c} -
V_{m}(t))
\end{equation}

with 3 channels (c) corresponding to: e excitatory input; l leak
current; and i inhibitory input. Following electrophysiological
convention, the overall conductance is decomposed into a time-varying
component $g_{c}(t)$ computed as a function of the dynamic state of
the model, and a constant $\overline{g_{c}}$ that controls the
relative influence of the different conductances. The equilibrium
potential can be written in a simplified form by setting the
excitatory driving potential ($E_{e}$) to 1 and the leak and
inhibitory driving potentials ($E_{l}$ and $E_{i}$) of 0:

\begin{equation}
  V_m^\infty = \frac{g_e \overline{g_e}} {g_e
   \overline{g_e} + g_l \overline{g_l} + g_i \overline{g_i}}
\end{equation}

which shows that the neuron is computing a balance between excitation
and the
opposing forces of leak and inhibition. This equilibrium form of the
equation
can be understood in terms of a Bayesian decision making framework
\citep{OReillyMunakata00}.\\

The excitatory net input/conductance $g_{e}(t)$ or $\eta_{j}$ is
computed as the proportion of open excitatory channels as a function
of sending activations times the weight values:

\begin{equation}
\eta_{j} = g_{e}(t) = \langle x_{i}w_{ij} \rangle = \frac{1}{n}
\sum_{i}x_{i}w_{ij}
\label{Net_Input}
\end{equation}

The inhibitory conductance can either be computed by the kWTA function
described in the next section or by modeling inhibitory interneurons.
Leak is a constant.\\
Activation communicated to other cells ($y_{j}$) is a thresholded
($\Theta$) sigmoidal function of the membrane potential with gain
parameter $\gamma$:
\begin{equation}
  y_{j}(t) = \frac{1}{\left(1 + \frac{1}{\gamma [V_{m}(t) -
\Theta]_{+}} \right)}
\end{equation}

where $[x]_{+}$ is a threshold function that returns 0 if $ x « 0$ and
$x$ if $x » 0$. To avoid dividing by 0 we assume $y_{j}(t) = 0$ if it
returns 0.  This activation is subject to scaling factors
(wt\_scale.abs and wt\_scale.rel) which modify how much impact the
projections have on the post-synaptic neurons.

\subsection{Inhibition within and between layers}
Inhibition between layers (i.e. for GABAergic projections between BG
layers and striatal inhibitory interneurons) is achieved via simple
unit inhibition, where the inhibitory current $g_{i}$ for the unit is
determined from the net input of the sending unit.  For
\textit{within} layer lateral inhibition (used here in premotor
cortex), Leabra uses a kWTA (k-Winners-Take-All) function to achieve
inhibitory competition among neurons within each layer (area).  The
kWTA function computes a uniform level of inhibitory current for all
neurons in the layer, such that the k + 1th most excited unit within a
layer is generally below its firing threshold, while the kth is
typically above threshold.  Activation dynamics similar to those
produced by the kWTA function have been shown to result from simulated
inhibitory interneurons that project both feedforward and feedback
inhibition \citep{OReillyMunakata00}.  Thus, although the kWTA
function is somewhat biologically implausible in its implementation
(e.g., requiring global information about activation states and using
sorting mechanisms), it provides a computationally effective
approximation to biologically plausible inhibitory dynamics.  kWTA is
computed via a uniform level of inhibitory current for all neurons in
the layer as follows:

\begin{equation}
g_{i} = g_{k+1}^{\Theta} + q(g_{k}^{\Theta} - g_{k+1}^{\Theta})
\label{kwta}
\end{equation}

where $0 « q « 1$ (0.25 default) is a parameter $\Theta$ for setting
the inhibition between the upper bound of $g_{k}$ and $\Theta$ . These
boundary inhibition values are the lower bound of $g_{k+1}$ computed
as a function of the level of inhibition necessary to keep a unit
right at threshold:

\begin{equation}
  g_i = g^{\Theta}_{k+1} + q (g^{\Theta}_k - g^{\Theta}_{k+1})
  \label{eq.g_i}
\end{equation}

\label{kwta-sec}

In the basic version of the kWTA function, which is relatively rigid
about the kWTA constraint and is therefore used for output layers,
$g_{k}^{\Theta}$ and $g^{\Theta}_{k+1}$ are set to the threshold
inhibition value for the kth and k+1th most excited neurons,
respectively. Thus, the inhibition is placed exactly to allow $k$
neurons to be above threshold, and the remainder below threshold. For
this version, the q parameter is almost always .25, allowing the kth
unit to be sufficiently above the inhibitory threshold. \\

The premotor cortex uses the average-based kWTA version,
$g^{\Theta}_{k}$ is the average $g_{i}^{\Theta}$ value for the top $k$
most excited neurons, and $g^{\Theta}_{k+1}$ is the average of
$g_{i}^{\Theta}$ for the remaining $n-k$ neurons. This version allows
for more flexibility in the actual number of neurons active depending
on the nature of the activation distribution in the layer and the
value of the $q$ parameter (which is typically .6), and is therefore
used for hidden layers.

Hysterisis and Accommodation

\begin{equation}
I_a(t) = g_a(t) \bar{g_a} (V_m(t) - E_a)
 \end{equation}

 \begin{equation}
    I_h(t) = g_h(t) \bar{g_h}(V_m(t) - E_h)
 \end{equation}
$E_h$ is excitatory; $E_a$ inhibitory.

$g_a$ and $g_h$ are time-varying functions that depend on previous
activity,
integrated over different time periods.

\begin{equation}
 g_a(t) = \left\{ \begin{array}{ll}
     g_a(t-1) + dt_{g_a} (1 - g_a(t-1)); & \mbox{if}(b_a(t) =
\Theta_a) \\
     g_a(t-1) + dt_{g_a} (0 - g_a(t-1)); & \mbox{if}(b_a(t) =
\Theta_d)
     \end{array} \right.
\end{equation}

\subsection{Computation of conflict}
dACC activity is the Hopfield energy of pre-SMA:
\begin{equation}
\textrm{dACC}_{\textrm{act}} =
\textrm{FEF}_{\textrm{left}_{\textrm{act}}}*\textrm{FEF}_{\textrm{right}_{\textrm{act}}}
\end{equation}

\bibliographystyle{plainnat}
\bibliography{ccnlab,ccnlab.ip,Whyking_formatted,Whyking,mendeley,pdreview}

\begin{thebibliography}{194}
\providecommand{\natexlab}[1]{#1}
\providecommand{\url}[1]{\texttt{#1}}
\expandafter\ifx\csname urlstyle\endcsname\relax
  \providecommand{\doi}[1]{doi: #1}\else
  \providecommand{\doi}{doi: \begingroup \urlstyle{rm}\Url}\fi

\bibitem[Aisa et~al.(2008)Aisa, Mingus, and O'Reilly]{AisaMingusOreilly08}
Brad Aisa, Brian Mingus, and Randy O'Reilly.
\newblock The emergent neural modeling system.
\newblock \emph{Neural networks}, 21\penalty0 (8):\penalty0 1146--1152, Oct
  2008.
\newblock URL \url{http://www.ncbi.nlm.nih.gov/pubmed/18684591}.

\bibitem[Alexander and Brown(2011)]{AlexanderBrown11}
William~H Alexander and Joshua~W Brown.
\newblock Medial prefrontal cortex as an action-outcome predictor.
\newblock \emph{Nature Neuroscience}, 14\penalty0 (10):\penalty0 1338--1344,
  Oct 2011.
\newblock URL \url{http://www.ncbi.nlm.nih.gov/pubmed/21926982}.

\bibitem[Andr\'{e}s(2003)]{Andres03}
Pilar Andr\'{e}s.
\newblock Frontal cortex as the central executive of working memory: time to
  revise our view.
\newblock \emph{Cortex; a journal devoted to the study of the nervous system
  and behavior}, 39:\penalty0 871--896, 10 2003.
\newblock URL \url{http://www.ncbi.nlm.nih.gov/pubmed/14584557}.

\bibitem[Aron(2007)]{Aron07}
A.~R. Aron.
\newblock The neural basis of inhibition in cognitive control.
\newblock \emph{The Neuroscientist : a review journal bringing neurobiology,
  neurology and psychiatry}, 13\penalty0 (3):\penalty0 214--228, June 2007.
\newblock ISSN 1073-8584.
\newblock \doi{10.1177/1073858407299288}.
\newblock URL \url{http://dx.doi.org/10.1177/1073858407299288}.

\bibitem[Aron and Verbruggen(2008)]{AronVerbruggen08}
A.~R. Aron and F.~Verbruggen.
\newblock Stop the presses: dissociating a selective from a global mechanism
  for stopping.
\newblock \emph{Psychological science : a journal of the American Psychological
  Society / APS}, 19\penalty0 (11):\penalty0 1146--1153, November 2008.
\newblock ISSN 1467-9280.
\newblock \doi{10.1111/j.1467-9280.2008.02216.x}.
\newblock URL \url{http://dx.doi.org/10.1111/j.1467-9280.2008.02216.x}.

\bibitem[Aron(2011)]{Aron11}
Adam~R Aron.
\newblock {From reactive to proactive and selective control: developing a
  richer model for stopping inappropriate responses.}
\newblock \emph{Biological psychiatry}, 69\penalty0 (12):\penalty0 e55--68,
  June 2011.
\newblock ISSN 1873-2402.
\newblock \doi{10.1016/j.biopsych.2010.07.024}.
\newblock URL
  \url{http://www.pubmedcentral.nih.gov/articlerender.fcgi?artid=3039712\&tool%
=pmcentrez\&rendertype=abstract http://www.ncbi.nlm.nih.gov/pubmed/20932513}.

\bibitem[Aron and Poldrack(2006)]{AronPoldrack06}
Adam~R Aron and Russell~A Poldrack.
\newblock Cortical and subcortical contributions to stop signal response
  inhibition: role of the subthalamic nucleus.
\newblock \emph{The Journal of neuroscience : the official journal of the
  Society for Neuroscience}, 26:\penalty0 2424--33, 03 2006.
\newblock URL \url{http://www.ncbi.nlm.nih.gov/pubmed/16510720}.

\bibitem[Aron et~al.(2003)Aron, Fletcher, Bullmore, Sahakian, and
  Robbins]{AronFletcherBullmoreEtAl03}
Adam~R Aron, Paul~C Fletcher, Ed~T Bullmore, Barbara~J Sahakian, and Trevor~W
  Robbins.
\newblock Stop-signal inhibition disrupted by damage to right inferior frontal
  gyrus in humans.
\newblock \emph{Nature neuroscience}, 6:\penalty0 115--116, 01 2003.
\newblock URL \url{http://www.ncbi.nlm.nih.gov/pubmed/12536210}.

\bibitem[Aron et~al.(2004)Aron, Monsell, Sahakian, and
  Robbins]{AronMonsellSahakianEtAl04}
Adam~R Aron, Stephen Monsell, Barbara~J Sahakian, and Trevor~W Robbins.
\newblock A componential analysis of task-switching deficits associated with
  lesions of left and right frontal cortex.
\newblock \emph{Brain : a journal of neurology}, 127\penalty0 (7):\penalty0
  1561--1573, 06 2004.
\newblock URL \url{http://www.ncbi.nlm.nih.gov/pubmed/15090477}.

\bibitem[Aron et~al.(2007{\natexlab{a}})Aron, Behrens, Smith, Frank, and
  Poldrack]{AronBehrensSmithEtAl07}
Adam~R Aron, Tim~E Behrens, Steve Smith, Michael~J Frank, and Russell~A
  Poldrack.
\newblock Triangulating a cognitive control network using diffusion-weighted
  magnetic resonance imaging (mri) and functional mri.
\newblock \emph{The Journal of neuroscience : the official journal of the
  Society for Neuroscience}, 27\penalty0 (14):\penalty0 3743--3752, 04
  2007{\natexlab{a}}.
\newblock URL \url{http://www.ncbi.nlm.nih.gov/pubmed/17409238}.

\bibitem[Aron et~al.(2007{\natexlab{b}})Aron, Durston, Eagle, Logan, Stinear,
  and Stuphorn]{AronDurstonEagleEtAl07}
Adam~R. Aron, Sarah Durston, Dawn~M. Eagle, Gordon~D. Logan, Cathy~M. Stinear,
  and Veit Stuphorn.
\newblock Converging evidence for a fronto-basal-ganglia network for inhibitory
  control of action and cognition.
\newblock \emph{J. Neurosci.}, 27\penalty0 (44):\penalty0 11860--11864, October
  2007{\natexlab{b}}.
\newblock \doi{10.1523/JNEUROSCI.3644-07.2007}.
\newblock URL \url{http://dx.doi.org/10.1523/JNEUROSCI.3644-07.2007}.

\bibitem[Aston-Jones and Cohen(2005)]{Aston-JonesCohen05}
Gary Aston-Jones and Jonathan~D Cohen.
\newblock An integrative theory of locus coeruleus-norepinephrine function:
  adaptive gain and optimal performance.
\newblock \emph{Annual review of neuroscience}, 28:\penalty0 403--450, 07 2005.
\newblock URL \url{http://www.ncbi.nlm.nih.gov/pubmed/16022602}.

\bibitem[Badcock et~al.(2002)Badcock, Michie, Johnson, and
  Combrinck]{BadcockMichieJohnsonEtAl02}
J.~C. Badcock, P.~T. Michie, L.~Johnson, and J.~Combrinck.
\newblock Acts of control in schizophrenia: dissociating the components of
  inhibition.
\newblock \emph{Psychological medicine}, 32\penalty0 (2):\penalty0 287--297,
  February 2002.
\newblock ISSN 0033-2917.
\newblock URL \url{http://view.ncbi.nlm.nih.gov/pubmed/11866323}.

\bibitem[Barcelo et~al.(2006)Barcelo, Escera, Corral, and
  Periáñez]{BarceloEsceraCorralEtAl06}
Francisco Barcelo, Carles Escera, Maria~J Corral, and Jose~A Periáñez.
\newblock Task switching and novelty processing activate a common neural
  network for cognitive control.
\newblock \emph{Journal of cognitive neuroscience}, 18, Oct 2006.
\newblock URL \url{http://www.ncbi.nlm.nih.gov/pubmed/17014377}.

\bibitem[Bellgrove et~al.(2006)Bellgrove, Chambers, Vance, Hall, Karamitsios,
  and Bradshaw]{BellgroveChambersVanceEtAl06}
M.~A. Bellgrove, C.~D. Chambers, A.~Vance, N.~Hall, M.~Karamitsios, and J.~L.
  Bradshaw.
\newblock Lateralized deficit of response inhibition in early-onset
  schizophrenia.
\newblock \emph{Psychological medicine}, 36\penalty0 (4):\penalty0 495--505,
  April 2006.
\newblock ISSN 0033-2917.
\newblock \doi{10.1017/S0033291705006409}.
\newblock URL \url{http://dx.doi.org/10.1017/S0033291705006409}.

\bibitem[Bernheimer et~al.(1973)Bernheimer, Birkmayer, Hornykiewicz, Jellinger,
  and Seitelberger]{BernheimerBirkmayerHornykiewiczEtAl73}
H~Bernheimer, W~Birkmayer, O~Hornykiewicz, K~Jellinger, and F~Seitelberger.
\newblock {Brain dopamine and the syndromes of Parkinson and Huntington
  Clinical, morphological and neurochemical correlations☆}.
\newblock \emph{Journal of the Neurological Sciences}, 20\penalty0
  (4):\penalty0 415--455, December 1973.
\newblock ISSN 0022510X.
\newblock \doi{10.1016/0022-510X(73)90175-5}.
\newblock URL \url{http://dx.doi.org/10.1016/0022-510X(73)90175-5}.

\bibitem[Botvinick et~al.(2001)Botvinick, Braver, Barch, Carter, and
  Cohen]{BotvinickBraverBarchEtAl01}
M.~M. Botvinick, T.~S. Braver, D.~M. Barch, C.~S. Carter, and J.~D. Cohen.
\newblock Conflict monitoring and cognitive control.
\newblock \emph{Psychological Review}, 108:\penalty0 624--652, December 2001.

\bibitem[Botvinick et~al.(2004)Botvinick, Cohen, and
  Carter]{BotvinickCohenCarter04}
Matthew~M Botvinick, Jonathan~D Cohen, and Cameron~S Carter.
\newblock Conflict monitoring and anterior cingulate cortex: an update.
\newblock \emph{Trends in cognitive sciences}, 8\penalty0 (12):\penalty0
  539--546, 11 2004.
\newblock URL \url{http://www.ncbi.nlm.nih.gov/pubmed/15556023}.

\bibitem[Brass et~al.(2005)Brass, Derrfuss, Forstmann, and von
  Cramon]{BrassDerrfussForstmannEtAl05}
Marcel Brass, Jan Derrfuss, Birte Forstmann, and D~Yves von Cramon.
\newblock The role of the inferior frontal junction area in cognitive control.
\newblock \emph{Trends in cognitive sciences}, 9, 2005.
\newblock URL \url{http://www.ncbi.nlm.nih.gov/pubmed/15927520}.

\bibitem[Braver et~al.(2001)Braver, Barch, Gray, Molfese, and
  Snyder]{BraverBarchGrayEtAl01}
T~S Braver, D~M Barch, J~R Gray, D~L Molfese, and A~Snyder.
\newblock Anterior cingulate cortex and response conflict: effects of
  frequency, inhibition and errors.
\newblock \emph{Cerebral cortex (New York, N.Y. : 1991)}, 11:\penalty0
  825--836, 09 2001.
\newblock URL \url{http://www.ncbi.nlm.nih.gov/pubmed/11532888}.

\bibitem[Breier et~al.(1998)Breier, Kestler, Adler, Elman, Wiesenfeld,
  Malhotra, and Pickar]{BreierKestlerAdlerEtAl98}
A~Breier, L~Kestler, C~Adler, I~Elman, N~Wiesenfeld, A~Malhotra, and D~Pickar.
\newblock Dopamine d2 receptor density and personal detachment in healthy
  subjects.
\newblock \emph{The American journal of psychiatry}, 155:\penalty0 1440--1442,
  10 1998.
\newblock URL \url{http://www.ncbi.nlm.nih.gov/pubmed/9766779}.

\bibitem[Brown et~al.(2004)Brown, Bullock, and
  Grossberg]{BrownBullockGrossberg04}
Joshua Brown, Daniel Bullock, and Stephen Grossberg.
\newblock How laminar frontal cortex and basal ganglia circuits interact to
  control planned and reactive saccades.
\newblock \emph{Neural Networks}, 17:\penalty0 471--510, 04 2004.
\newblock URL \url{http://www.ncbi.nlm.nih.gov/pubmed/15109680}.

\bibitem[Bunge and Wallis(2008)]{BungeWallis08}
Silvia~A. Bunge and Jonathan~D. Wallis, editors.
\newblock \emph{Neuroscience of rule-guided behavior}.
\newblock Oxford University Press, January 2008.

\bibitem[Burle et~al.(2002)Burle, Possama\"{\i}, Vidal, Bonnet, and
  Hasbroucq]{BurlePossamaiVidalEtAl02}
Bor\'{\i}s Burle, Camille-Aim\'{e} Possama\"{\i}, Franck Vidal, Michel Bonnet,
  and Thierry Hasbroucq.
\newblock {Executive control in the Simon effect: an electromyographic and
  distributional analysis.}
\newblock \emph{Psychological research}, 66\penalty0 (4):\penalty0 324--36,
  November 2002.
\newblock ISSN 0340-0727.
\newblock \doi{10.1007/s00426-002-0105-6}.
\newblock URL \url{http://www.ncbi.nlm.nih.gov/pubmed/12466929}.

\bibitem[Cai et~al.(2011)Cai, Oldenkamp, and Aron]{CaiOldenkampAron11}
Weidong Cai, Caitlin~L Oldenkamp, and Adam~R Aron.
\newblock {A proactive mechanism for selective suppression of response
  tendencies.}
\newblock \emph{The Journal of neuroscience : the official journal of the
  Society for Neuroscience}, 31\penalty0 (16):\penalty0 5965--9, April 2011.
\newblock ISSN 1529-2401.
\newblock \doi{10.1523/JNEUROSCI.6292-10.2011}.
\newblock URL
  \url{http://www.pubmedcentral.nih.gov/articlerender.fcgi?artid=3111595\&tool%
=pmcentrez\&rendertype=abstract}.

\bibitem[Casey et~al.(2007)Casey, Nigg, and Durston]{CaseyNiggDurston07}
B.~J. Casey, J.~T. Nigg, and S.~Durston.
\newblock New potential leads in the biology and treatment of attention
  deficit-hyperactivity disorder.
\newblock \emph{Current opinion in neurology}, 20\penalty0 (2):\penalty0
  119--124, April 2007.
\newblock ISSN 1350-7540.
\newblock \doi{10.1097/WCO.0b013e3280a02f78}.
\newblock URL \url{http://dx.doi.org/10.1097/WCO.0b013e3280a02f78}.

\bibitem[Cavanagh et~al.(2011)Cavanagh, Wiecki, Cohen, Figueroa, Samanta,
  Sherman, and Frank]{CavanaghWieckiCohenEtAl11}
James~F Cavanagh, Thomas~V Wiecki, Michael~X Cohen, Christina~M Figueroa, Johan
  Samanta, Scott~J Sherman, and Michael~J Frank.
\newblock Subthalamic nucleus stimulation reverses mediofrontal influence over
  decision threshold.
\newblock \emph{Nature neuroscience}, 14:\penalty0 1462--1467, Sep 2011.
\newblock URL \url{http://www.ncbi.nlm.nih.gov/pubmed/21946325}.

\bibitem[Chamberlain et~al.(2007)Chamberlain, Del~Campo, Dowson, M\"{u}ller,
  Clark, Robbins, and Sahakian]{ChamberlainDel_CampoDowsonEtAl07}
S.~R. Chamberlain, N.~Del~Campo, J.~Dowson, U.~M\"{u}ller, L.~Clark, T.~W.
  Robbins, and B.~J. Sahakian.
\newblock Atomoxetine improved response inhibition in adults with attention
  deficit/hyperactivity disorder.
\newblock \emph{Biological psychiatry}, 62\penalty0 (9):\penalty0 977--984,
  November 2007.
\newblock ISSN 0006-3223.
\newblock \doi{10.1016/j.biopsych.2007.03.003}.
\newblock URL \url{http://dx.doi.org/10.1016/j.biopsych.2007.03.003}.

\bibitem[Chamberlain et~al.(2009)Chamberlain, Hampshire, M\"{u}ller, Rubia,
  Del~Campo, Craig, Regenthal, Suckling, Roiser, Grant, Bullmore, Robbins, and
  Sahakian]{ChamberlainHampshireMullerEtAl09}
S.~R. Chamberlain, A.~Hampshire, U.~M\"{u}ller, K.~Rubia, N.~Del~Campo,
  K.~Craig, R.~Regenthal, J.~Suckling, J.~P. Roiser, J.~E. Grant, E.~T.
  Bullmore, T.~W. Robbins, and B.~J. Sahakian.
\newblock Atomoxetine modulates right inferior frontal activation during
  inhibitory control: a pharmacological functional magnetic resonance imaging
  study.
\newblock \emph{Biological psychiatry}, 65\penalty0 (7):\penalty0 550--555,
  April 2009.
\newblock ISSN 1873-2402.
\newblock \doi{10.1016/j.biopsych.2008.10.014}.
\newblock URL \url{http://dx.doi.org/10.1016/j.biopsych.2008.10.014}.

\bibitem[Chamberlain et~al.(2006)Chamberlain, Fineberg, Blackwell, Robbins, and
  Sahakian]{ChamberlainFinebergBlackwellEtAl06}
Samuel~R. Chamberlain, Naomi~A. Fineberg, Andrew~D. Blackwell, Trevor~W.
  Robbins, and Barbara~J. Sahakian.
\newblock Motor inhibition and cognitive flexibility in obsessive-compulsive
  disorder and trichotillomania.
\newblock \emph{Am J Psychiatry}, 163\penalty0 (7):\penalty0 1282--1284, July
  2006.
\newblock \doi{10.1176/appi.ajp.163.7.1282}.
\newblock URL \url{http://dx.doi.org/10.1176/appi.ajp.163.7.1282}.

\bibitem[Chambers et~al.(2007)Chambers, Bellgrove, Gould, English, Garavan,
  Mcnaught, Kamke, and Mattingley]{ChambersBellgroveGouldEtAl07}
Christopher~D. Chambers, Mark~A. Bellgrove, Ian~C. Gould, Therese English, Hugh
  Garavan, Elizabeth Mcnaught, Marc Kamke, and Jason~B. Mattingley.
\newblock Dissociable mechanisms of cognitive control in prefrontal and
  premotor cortex.
\newblock \emph{J Neurophysiol}, 98\penalty0 (6):\penalty0 3638--3647, December
  2007.
\newblock \doi{10.1152/jn.00685.2007}.
\newblock URL \url{http://dx.doi.org/10.1152/jn.00685.2007}.

\bibitem[Chambers et~al.(2009)Chambers, Garavan, and
  Bellgrove]{ChambersGaravanBellgrove09}
Christopher~D. Chambers, Hugh Garavan, and Mark~A. Bellgrove.
\newblock Insights into the neural basis of response inhibition from cognitive
  and clinical neuroscience.
\newblock \emph{Neuroscience \& Biobehavioral Reviews}, 33\penalty0
  (5):\penalty0 631--646, May 2009.
\newblock ISSN 01497634.
\newblock \doi{10.1016/j.neubiorev.2008.08.016}.
\newblock URL \url{http://dx.doi.org/10.1016/j.neubiorev.2008.08.016}.

\bibitem[Chatham et~al.(2012)Chatham, Claus, Kim, Curran, Banich, and
  Munakata]{ChathamClausKimEtAl12}
Christopher~H Chatham, Eric~D Claus, Albert Kim, Tim Curran, Marie~T Banich,
  and Yuko Munakata.
\newblock {Cognitive control reflects context monitoring, not motoric stopping,
  in response inhibition.}
\newblock \emph{PloS one}, 7\penalty0 (2):\penalty0 e31546, January 2012.
\newblock ISSN 1932-6203.
\newblock \doi{10.1371/journal.pone.0031546}.
\newblock URL \url{http://www.ncbi.nlm.nih.gov/pubmed/22384038}.

\bibitem[Cohen and Poldrack(2008)]{CohenPoldrack08}
J.~R. Cohen and R.~A. Poldrack.
\newblock Automaticity in motor sequence learning does not impair response
  inhibition.
\newblock \emph{Psychonomic bulletin \& review}, 15\penalty0 (1):\penalty0
  108--115, February 2008.
\newblock ISSN 1069-9384.
\newblock URL \url{http://view.ncbi.nlm.nih.gov/pubmed/18605489}.

\bibitem[Collins and Frank(2012)]{CollinsFrank12}
Anne G~E Collins and Michael~J Frank.
\newblock {How much of reinforcement learning is working memory, not
  reinforcement learning? A behavioral, computational, and neurogenetic
  analysis.}
\newblock \emph{The European journal of neuroscience}, 35\penalty0
  (7):\penalty0 1024--35, April 2012.
\newblock ISSN 1460-9568.
\newblock \doi{10.1111/j.1460-9568.2011.07980.x}.
\newblock URL
  \url{http://www.pubmedcentral.nih.gov/articlerender.fcgi?artid=3390186\&tool%
=pmcentrez\&rendertype=abstract}.

\bibitem[Congdon et~al.(2009)Congdon, Constable, Lesch, and
  Canli]{CongdonConstableLeschEtAl09}
Eliza Congdon, R.~Todd Constable, Klaus~P. Lesch, and Turhan Canli.
\newblock Influence of slc6a3 and comt variation on neural activation during
  response inhibition.
\newblock \emph{Biological Psychology}, 81\penalty0 (3):\penalty0 144--152,
  July 2009.
\newblock ISSN 03010511.
\newblock \doi{10.1016/j.biopsycho.2009.03.005}.
\newblock URL \url{http://dx.doi.org/10.1016/j.biopsycho.2009.03.005}.

\bibitem[Curtis and DEsposito(2003)]{CurtisDEsposito03}
C.~E. Curtis and M.~DEsposito.
\newblock Persistent activity in the prefrontal cortex during working memory.
\newblock \emph{Trends in Cognitive Sciences}, 7:\penalty0 415--423, January
  2003.

\bibitem[Daw et~al.(2005)Daw, Niv, and Dayan]{DawNivDayan05}
Nathaniel~D Daw, Yael Niv, and Peter Dayan.
\newblock Uncertainty-based competition between prefrontal and dorsolateral
  striatal systems for behavioral control.
\newblock \emph{Nature neuroscience}, 8\penalty0 (12):\penalty0 1704--1711, 11
  2005.
\newblock URL \url{http://www.ncbi.nlm.nih.gov/pubmed/16286932}.

\bibitem[DeLiang et~al.(2005)DeLiang, Kristjansson, and
  Nakayama]{DeLiangKristjanssonNakayama05}
Wang DeLiang, Arnj Kristjansson, and Ken Nakayama.
\newblock Efficient visual search without top-down or bottom-up guidance.
\newblock \emph{Perception \& Psychophysics}, 67\penalty0 (2):\penalty0
  239--253, January 2005.

\bibitem[Derrfuss et~al.(2004)Derrfuss, Brass, and von
  Cramon]{DerrfussBrassCramon04}
Jan Derrfuss, Marcel Brass, and D~Yves von Cramon.
\newblock Cognitive control in the posterior frontolateral cortex: evidence
  from common activations in task coordination, interference control, and
  working memory.
\newblock \emph{NeuroImage}, 23, Oct 2004.
\newblock URL \url{http://www.ncbi.nlm.nih.gov/pubmed/15488410}.

\bibitem[Derrfuss et~al.(2005)Derrfuss, Brass, Neumann, and von
  Cramon]{DerrfussBrassNeumannEtAl05}
Jan Derrfuss, Marcel Brass, Jane Neumann, and D.~Yves von Cramon.
\newblock Involvement of the inferior frontal junction in cognitive control:
  meta-analyses of switching and stroop studies.
\newblock \emph{Human brain mapping}, 25\penalty0 (1):\penalty0 22--34, May
  2005.
\newblock URL \url{http://www.ncbi.nlm.nih.gov/pubmed/15846824?ordinalpos}.

\bibitem[Doll et~al.(2009)Doll, Jacobs, Sanfey, and
  Frank]{DollJacobsSanfeyEtAl09}
Bradley~B. Doll, W.~Jake Jacobs, Alan~G. Sanfey, and Michael~J. Frank.
\newblock Instructional control of reinforcement learning: a behavioral and
  neurocomputational investigation.
\newblock \emph{Brain Research}, 1299:\penalty0 74--94, Nov 2009.
\newblock URL \url{http://www.ncbi.nlm.nih.gov/pubmed/19595993}.

\bibitem[Eagle et~al.(2008)Eagle, Baunez, Hutcheson, Lehmann, Shah, and
  Robbins]{EagleBaunezHutchesonEtAl08}
Dawn~M Eagle, Christelle Baunez, Daniel~M Hutcheson, Olivia Lehmann, Aarti~P
  Shah, and Trevor~W Robbins.
\newblock Stop-signal reaction-time task performance: role of prefrontal cortex
  and subthalamic nucleus.
\newblock \emph{Cerebral cortex}, 18, Jan 2008.
\newblock URL \url{http://www.ncbi.nlm.nih.gov/pubmed/17517682}.

\bibitem[Eimer(1995)]{Eimer95}
M~Eimer.
\newblock {S-R compatibility and response selection}.
\newblock \emph{Acta Psychologica}, 90\penalty0 (1-3):\penalty0 301--313,
  November 1995.
\newblock ISSN 00016918.
\newblock \doi{10.1016/0001-6918(95)00022-M}.
\newblock URL \url{http://dx.doi.org/10.1016/0001-6918(95)00022-M}.

\bibitem[Emeric et~al.(2010)Emeric, Leslie, Pouget, and
  Schall]{EmericLesliePougetEtAl10}
Erik~E Emeric, Melanie Leslie, Pierre Pouget, and Jeffrey~D Schall.
\newblock {Performance monitoring local field potentials in the medial frontal
  cortex of primates: supplementary eye field.}
\newblock \emph{Journal of neurophysiology}, 104\penalty0 (3):\penalty0
  1523--37, September 2010.
\newblock ISSN 1522-1598.
\newblock \doi{10.1152/jn.01001.2009}.
\newblock URL
  \url{http://www.pubmedcentral.nih.gov/articlerender.fcgi?artid=2944693\&tool%
=pmcentrez\&rendertype=abstract}.

\bibitem[Evans(2005)]{Evans03}
Jonathan St B~T Evans.
\newblock In two minds: dual-process accounts of reasoning.
\newblock \emph{Trends in cognitive sciences}, 7\penalty0 (10):\penalty0
  454--459, 01 2005.
\newblock URL \url{http://www.ncbi.nlm.nih.gov/pubmed/14550493}.

\bibitem[Everling and Munoz(2000)]{EverlingMunoz00}
S~Everling and D~P Munoz.
\newblock Neuronal correlates for preparatory set associated with pro-saccades
  and anti-saccades in the primate frontal eye field.
\newblock \emph{The Journal of neuroscience : the official journal of the
  Society for Neuroscience}, 20:\penalty0 387, 01 2000.
\newblock URL \url{http://www.ncbi.nlm.nih.gov/pubmed/10627615}.

\bibitem[Everling et~al.(1999)Everling, Dorris, Klein, and
  Munoz]{EverlingDorrisKleinEtAl99}
Stefan Everling, Michael~C. Dorris, Raymond~M. Klein, and Douglas~P. Munoz.
\newblock Role of primate superior colliculus in preparation and execution of
  anti-saccades and pro-saccades.
\newblock \emph{J. Neurosci.}, 19\penalty0 (7):\penalty0 2740--2754, April
  1999.
\newblock URL \url{http://www.jneurosci.org/cgi/content/abstract/19/7/2740}.

\bibitem[Falkenstein et~al.(1991)Falkenstein, Hohnsbein, Hoormann, and
  Blanke]{FalkensteinHohnsbeinHoormannEtAl91}
M.~Falkenstein, J.~Hohnsbein, J.~Hoormann, and L.~Blanke.
\newblock Effects of cross-modal divided attention on late erp components: Ii.
  error processing in choice reaction tasks.
\newblock \emph{Electroencephalography and Clinical Neurophysiology},
  78:\penalty0 447--55, January 1991.

\bibitem[Faraone et~al.(2005)Faraone, Perlis, Doyle, Smoller, Goralnick,
  Holmgren, and Sklar]{FaraonePerlisDoyleEtAl05}
Stephen~V Faraone, Roy~H Perlis, Alysa~E Doyle, Jordan~W Smoller, Jennifer~J
  Goralnick, Meredith~A Holmgren, and Pamela Sklar.
\newblock Molecular genetics of attention-deficit/hyperactivity disorder.
\newblock \emph{Biol Psychiatry}, 57\penalty0 (11):\penalty0 1313--1323, Jun
  2005.
\newblock URL \url{http://dx.doi.org/10.1016/j.biopsych.2004.11.024}.

\bibitem[Farley et~al.(1978)Farley, Price, McCullough, Deck, Hordynski, and
  Hornykiewicz]{FarleyPriceMcCulloughEtAl78}
I.~Farley, K.~Price, E~McCullough, J.~Deck, W~Hordynski, and O~Hornykiewicz.
\newblock {Norepinephrine in chronic paranoid schizophrenia: above-normal
  levels in limbic forebrain}.
\newblock \emph{Science}, 200\penalty0 (4340):\penalty0 456--458, April 1978.
\newblock ISSN 0036-8075.
\newblock \doi{10.1126/science.644310}.
\newblock URL \url{http://www.sciencemag.org/content/200/4340/456.abstract}.

\bibitem[Fleming et~al.(2010)Fleming, Thomas, and Dolan]{FlemingThomasDolan10}
Stephen~M Fleming, Charlotte~L Thomas, and Raymond~J Dolan.
\newblock Overcoming status quo bias in the human brain.
\newblock \emph{Proceedings of the National Academy of Sciences of the United
  States of America}, 107, Mar 2010.
\newblock URL \url{http://www.ncbi.nlm.nih.gov/pubmed/20231462}.

\bibitem[Ford and Everling(2009)]{FordEverling09}
Kristen~A Ford and Stefan Everling.
\newblock Neural activity in primate caudate nucleus associated with pro- and
  antisaccades.
\newblock \emph{Journal of neurophysiology}, 102\penalty0 (4):\penalty0
  2334--2341, Oct 2009.
\newblock URL \url{http://www.ncbi.nlm.nih.gov/pubmed/19692516}.

\bibitem[Forstmann et~al.(2008)Forstmann, Dutilh, Brown, Neumann, von Cramon,
  Ridderinkhof, and Wagenmakers]{ForstmannDutilhBrownEtAl08}
Birte~U Forstmann, Gilles Dutilh, Scott Brown, Jane Neumann, D~Yves von Cramon,
  K~Richard Ridderinkhof, and Eric-Jan Wagenmakers.
\newblock Striatum and pre-sma facilitate decision-making under time pressure.
\newblock \emph{Proceedings of the National Academy of Sciences of the United
  States of America}, 105, Nov 2008.
\newblock URL \url{http://www.ncbi.nlm.nih.gov/pubmed/18981414}.

\bibitem[Forstmann et~al.(2010)Forstmann, Anwander, Sch채fer, Neumann, Brown,
  Wagenmakers, Bogacz, and Turner]{ForstmannAnwanderSchaferEtAl10}
Birte~U Forstmann, Alfred Anwander, Andreas Sch채fer, Jane Neumann, Scott
  Brown, Eric-Jan Wagenmakers, Rafal Bogacz, and Robert Turner.
\newblock Cortico-striatal connections predict control over speed and accuracy
  in perceptual decision making.
\newblock \emph{Proceedings of the National Academy of Sciences of the United
  States of America}, pages 1--5online, Aug 2010.
\newblock URL \url{http://www.ncbi.nlm.nih.gov/pubmed/20733082}.

\bibitem[Frank(2005)]{Frank05}
M.~J. Frank.
\newblock Dynamic dopamine modulation in the basal ganglia: A
  neurocomputational account of cognitive deficits in medicated and
  non-medicated {P}arkinsonism.
\newblock \emph{Journal of Cognitive Neuroscience}, 17:\penalty0 51--72,
  January 2005.

\bibitem[Frank et~al.(2004)Frank, Seeberger, and
  O'Reilly]{FrankSeebergerOReilly04}
M.~J. Frank, L.~C. Seeberger, and R.~C. O'Reilly.
\newblock By carrot or by stick: Cognitive reinforcement learning in
  {P}arkinsonism.
\newblock \emph{Science}, 306\penalty0 (5703):\penalty0 1940--1943, January
  2004.

\bibitem[Frank and Badre(2011)]{FrankBadre11}
Michael Frank and David Badre.
\newblock Mechanisms of hierarchical reinforcement learning in corticostriatal
  circuits 1: Computational analysis.
\newblock \emph{Cerebral Cortex}, page online, Jun 2011.
\newblock URL
  \url{http://cercor.oxfordjournals.org/content/early/2011/06/21/cercor.bhr114%
.abstract}.

\bibitem[Frank(2006)]{Frank06}
Michael~J Frank.
\newblock Hold your horses: a dynamic computational role for the subthalamic
  nucleus in decision making.
\newblock \emph{Neural networks : the official journal of the International
  Neural Network Society}, 19:\penalty0 1120--1136, 10 2006.
\newblock URL \url{http://www.ncbi.nlm.nih.gov/pubmed/16945502}.

\bibitem[Frank et~al.(2007{\natexlab{a}})Frank, Samanta, Moustafa, and
  Sherman]{FrankSamantaMoustafaEtAl07}
Michael~J Frank, Johan Samanta, Ahmed~A Moustafa, and Scott~J Sherman.
\newblock Hold your horses: impulsivity, deep brain stimulation, and medication
  in parkinsonism.
\newblock \emph{Science (New York, N.Y.)}, 318:\penalty0 1309--1312, 11
  2007{\natexlab{a}}.
\newblock URL \url{http://www.ncbi.nlm.nih.gov/pubmed/17962524}.

\bibitem[Frank et~al.(2007{\natexlab{b}})Frank, Santamaria, O'Reilly, and
  Willcutt]{FrankSantamariaOReillyEtAl07}
Michael~J Frank, Amy Santamaria, Randall~C O'Reilly, and Erik Willcutt.
\newblock Testing computational models of dopamine and noradrenaline
  dysfunction in attention deficit/hyperactivity disorder.
\newblock \emph{Neuropsychopharmacology : official publication of the American
  College of Neuropsychopharmacology}, 32:\penalty0 1583--1599, 06
  2007{\natexlab{b}}.
\newblock URL \url{http://www.ncbi.nlm.nih.gov/pubmed/17164816}.

\bibitem[Frank et~al.(2007{\natexlab{c}})Frank, Scheres, and
  Sherman]{FrankScheresSherman07}
Michael~J Frank, Anouk Scheres, and Scott~J Sherman.
\newblock Understanding decision-making deficits in neurological conditions:
  insights from models of natural action selection.
\newblock \emph{Philosophical transactions of the Royal Society of London.
  Series B, Biological sciences}, 362:\penalty0 1641--1654, 08
  2007{\natexlab{c}}.
\newblock URL \url{http://www.ncbi.nlm.nih.gov/pubmed/17428775}.

\bibitem[Funahashi et~al.(1993)Funahashi, Chafee, and
  Goldman-Rakic]{FunahashiChafeeGoldmanRakic93}
S.~Funahashi, M.~V. Chafee, and P.~S. Goldman-Rakic.
\newblock Prefrontal neuronal activity in rhesus monkeys performing a delayed
  anti-saccade task.
\newblock \emph{Nature}, 365:\penalty0 753--756, 11 1993.
\newblock URL \url{http://www.ncbi.nlm.nih.gov/pubmed/8413653}.

\bibitem[Garavan et~al.(2006)Garavan, Hester, Murphy, Fassbender, and
  Kelly]{GaravanHesterMurphyEtAl06}
H.~Garavan, R.~Hester, K.~Murphy, C.~Fassbender, and C.~Kelly.
\newblock Individual differences in the functional neuroanatomy of inhibitory
  control.
\newblock \emph{Brain Research}, 1105\penalty0 (1):\penalty0 130--142, August
  2006.
\newblock ISSN 00068993.
\newblock \doi{10.1016/j.brainres.2006.03.029}.
\newblock URL \url{http://dx.doi.org/10.1016/j.brainres.2006.03.029}.

\bibitem[Gehring et~al.(1993)Gehring, Goss, Coles, Meyer, and
  Donchin]{GehringGossColesEtAl93}
W.~J. Gehring, B.~Goss, M.~G.~H. Coles, D.~E. Meyer, and E.~Donchin.
\newblock A neural system for error detection and compensation.
\newblock \emph{Psychological Science}, 4\penalty0 (6):\penalty0 385--390,
  January 1993.

\bibitem[Goldberg et~al.(2012)Goldberg, Farries, and Fee]{GoldbergFarriesFee12}
Jesse~Heymann Goldberg, Michael~Alan Farries, and Michale~S Fee.
\newblock {Integration of cortical and pallidal inputs in the basal
  ganglia-recipient thalamus of singing birds.}
\newblock \emph{Journal of neurophysiology}, 108\penalty0 (5):\penalty0
  1403--29, June 2012.
\newblock ISSN 1522-1598.
\newblock \doi{10.1152/jn.00056.2012}.
\newblock URL \url{http://www.ncbi.nlm.nih.gov/pubmed/22673333}.

\bibitem[Greenhouse et~al.(2011)Greenhouse, Oldenkamp, Aron, and
  Diego]{GreenhouseOldenkampAronEtAl11}
Ian Greenhouse, Caitlin~L Oldenkamp, Adam~R Aron, and San Diego.
\newblock {Stopping a response has global or non-global effects on the motor
  system depending on preparation.}
\newblock \emph{Journal of neurophysiology}, October 2011.
\newblock ISSN 1522-1598.
\newblock \doi{10.1152/jn.00704.2011}.
\newblock URL \url{http://www.ncbi.nlm.nih.gov/pubmed/22013239}.

\bibitem[Gurney et~al.(2001)Gurney, Prescott, and
  Redgrave]{GurneyPrescottRedgrave01a}
K~Gurney, T~J Prescott, and P~Redgrave.
\newblock A computational model of action selection in the basal ganglia. i. a
  new functional anatomy.
\newblock \emph{Biological cybernetics}, 84:\penalty0 401--410, 06 2001.
\newblock URL \url{http://www.ncbi.nlm.nih.gov/pubmed/11417052}.

\bibitem[Haber(2003)]{Haber03}
Suzanne~N Haber.
\newblock The primate basal ganglia: parallel and integrative networks.
\newblock \emph{Journal of chemical neuroanatomy}, 26\penalty0 (4):\penalty0
  317--330, 01 2003.
\newblock URL \url{http://www.ncbi.nlm.nih.gov/pubmed/14729134}.

\bibitem[Hallett(1979)]{Hallett78}
P~E Hallett.
\newblock Primary and secondary saccades to goals defined by instructions.
\newblock \emph{Vision research}, 18:\penalty0 1270--1296, 02 1979.
\newblock URL \url{http://www.ncbi.nlm.nih.gov/pubmed/726270}.

\bibitem[Hampshire et~al.(2010)Hampshire, Chamberlain, Monti, Duncan, and
  Owen]{HampshireChamberlainMontiEtAl10}
Adam Hampshire, Samuel~R Chamberlain, Martin~M Monti, John Duncan, and Adrian~M
  Owen.
\newblock The role of the right inferior frontal gyrus: inhibition and
  attentional control.
\newblock \emph{NeuroImage}, 50:\penalty0 1313--1319, Jan 2010.
\newblock URL \url{http://www.ncbi.nlm.nih.gov/pubmed/20056157}.

\bibitem[Harden and Grace(1995)]{HardenGrace95}
D~G Harden and A~A Grace.
\newblock Activation of dopamine cell firing by repeated l-dopa administration
  to dopamine-depleted rats: its potential role in mediating the therapeutic
  response to l-dopa treatment.
\newblock \emph{J Neurosci}, 15\penalty0 (9):\penalty0 6157--66, January 1995.

\bibitem[Harris et~al.(2006)Harris, Reilly, Keshavan, and
  Sweeney]{HarrisReillyKeshavanEtAl06}
M.~S. Harris, J.~L. Reilly, M.~S. Keshavan, and J.~A. Sweeney.
\newblock Longitudinal studies of antisaccades in antipsychotic-naive
  first-episode schizophrenia.
\newblock \emph{Psychological medicine}, 36\penalty0 (4):\penalty0 485--494,
  April 2006.
\newblock ISSN 0033-2917.
\newblock \doi{10.1017/S0033291705006756}.
\newblock URL \url{http://dx.doi.org/10.1017/S0033291705006756}.

\bibitem[Hess et~al.(1946)Hess, B\"{u}rgi, and Bucher]{HessBurgiBucher46}
WR~Hess, S~B\"{u}rgi, and V~Bucher.
\newblock {Motorische Funktion des Tectal- und Tegmentalgebietes.}
\newblock \emph{Mschr Psychiat Neurol}, 112:\penalty0 1--52, 1946.

\bibitem[Hikida et~al.(2010)Hikida, Kimura, Wada, Funabiki, and
  Nakanishi]{HikidaKimuraWadaEtAl10}
Takatoshi Hikida, Kensuke Kimura, Norio Wada, Kazuo Funabiki, and Shigetada
  Nakanishi.
\newblock Distinct roles of synaptic transmission in direct and indirect
  striatal pathways to reward and aversive behavior.
\newblock \emph{Neuron}, 66:\penalty0 896--907, 2010.

\bibitem[Hikosaka(1989)]{Hikosaka89}
O.~Hikosaka.
\newblock Role of basal ganglia in initiation of voluntary movements.
\newblock In M.~A. Arbib and S.~Amari, editors, \emph{Dynamic {Interactions in
  Neural Networks: Models and Data}}, pages 153--167. Springer-Verlag, Berlin,
  January 1989.

\bibitem[Hikosaka(2007)]{Hikosaka07}
O.~Hikosaka.
\newblock \emph{GABAergic output of the basal ganglia}, volume 160 of
  \emph{Progress in Brain Research}, pages 209--226.
\newblock 2007.
\newblock \doi{10.1016/S0079-6123(06)60012-5}.
\newblock URL \url{http://dx.doi.org/10.1016/S0079-6123(06)60012-5}.

\bibitem[Hikosaka and Isoda(2008)]{HikosakaIsoda08}
O.~Hikosaka and M.~Isoda.
\newblock Brain mechanisms for switching from automatic to controlled eye
  movements.
\newblock \emph{Progress in brain research}, 171:\penalty0 375--382, 2008.
\newblock ISSN 1875-7855.
\newblock \doi{10.1016/S0079-6123(08)00655-9}.
\newblock URL \url{http://dx.doi.org/10.1016/S0079-6123(08)00655-9}.

\bibitem[Hikosaka and Wurtz(1986)]{HikosakaWurtz86}
O.~Hikosaka and R.~H. Wurtz.
\newblock Saccadic eye movements following injection of lidocaine into the
  superior colliculus.
\newblock \emph{Experimental brain research. Experimentelle Hirnforschung.
  Exp\'{e}rimentation c\'{e}r\'{e}brale}, 61\penalty0 (3):\penalty0 531--539,
  1986.
\newblock ISSN 0014-4819.
\newblock URL \url{http://view.ncbi.nlm.nih.gov/pubmed/3082658}.

\bibitem[Hikosaka et~al.(2000)Hikosaka, Takikawa, and
  Kawagoe]{HikosakaTakikawaKawagoe00}
O.~Hikosaka, Y.~Takikawa, and R.~Kawagoe.
\newblock Role of the basal ganglia in the control of purposive saccadic eye
  movements.
\newblock \emph{Physiological Reviews}, 80\penalty0 (3):\penalty0 953--978, Jul
  2000.
\newblock URL \url{http://www.ncbi.nlm.nih.gov/pubmed/10893428}.

\bibitem[Hikosaka et~al.(2006)Hikosaka, Nakamura, and
  Nakahara]{HikosakaNakamuraNakahara06}
Okihide Hikosaka, Kae Nakamura, and Hiroyuki Nakahara.
\newblock Basal ganglia orient eyes to reward.
\newblock \emph{Journal of neurophysiology}, 95\penalty0 (2):\penalty0
  567--584, Feb 2006.
\newblock URL \url{http://www.ncbi.nlm.nih.gov/pubmed/16424448}.

\bibitem[Holroyd and Coles(2002)]{HolroydColes02}
Clay~B Holroyd and Michael G~H Coles.
\newblock The neural basis of human error processing: reinforcement learning,
  dopamine, and the error-related negativity.
\newblock \emph{Psychological review}, 109:\penalty0 679--709, 10 2002.
\newblock URL \url{http://www.ncbi.nlm.nih.gov/pubmed/12374324}.

\bibitem[Hu and Li(2011)]{HuLi11}
Sien Hu and Chiang-Shan~R Li.
\newblock {Neural processes of preparatory control for stop signal inhibition.}
\newblock \emph{Human brain mapping}, 000, October 2011.
\newblock ISSN 1097-0193.
\newblock \doi{10.1002/hbm.21399}.
\newblock URL \url{http://www.ncbi.nlm.nih.gov/pubmed/21976392}.

\bibitem[Huddy et~al.(2009)Huddy, Aron, Harrison, Barnes, Robbins, and
  Joyce]{HuddyAronHarrisonEtAl09}
V.~C. Huddy, A.~R. Aron, M.~Harrison, T.~R.~E. Barnes, T.~W. Robbins, and E.~M.
  Joyce.
\newblock Impaired conscious and preserved unconscious inhibitory processing in
  recent onset schizophrenia.
\newblock \emph{Psychological Medicine}, 39\penalty0 (06):\penalty0 907--916,
  2009.
\newblock \doi{10.1017/S0033291708004340}.
\newblock URL \url{http://dx.doi.org/10.1017/S0033291708004340}.

\bibitem[Huerta et~al.(1987)Huerta, Krubitzer, and Kaas]{HuertaKrubitzerKaas87}
M~F Huerta, L~A Krubitzer, and J~H Kaas.
\newblock {Frontal eye field as defined by intracortical microstimulation in
  squirrel monkeys, owl monkeys, and macaque monkeys. II. Cortical
  connections.}
\newblock \emph{The Journal of comparative neurology}, 265\penalty0
  (3):\penalty0 332--61, November 1987.
\newblock ISSN 0021-9967.
\newblock URL \url{http://www.ncbi.nlm.nih.gov/pubmed/2447132}.

\bibitem[Huettel and McCarthy(2004)]{HuettelMcCarthy04}
Scott~A Huettel and Gregory McCarthy.
\newblock {What is odd in the oddball task? Prefrontal cortex is activated by
  dynamic changes in response strategy.}
\newblock \emph{Neuropsychologia}, 42\penalty0 (3):\penalty0 379--86, January
  2004.
\newblock ISSN 0028-3932.
\newblock URL \url{http://www.ncbi.nlm.nih.gov/pubmed/14670576}.

\bibitem[Hutton and Ettinger(2006)]{HuttonEttinger06}
Samuel~B Hutton and Ulrich Ettinger.
\newblock {The antisaccade task as a research tool in psychopathology: a
  critical review.}
\newblock \emph{Psychophysiology}, 43\penalty0 (3):\penalty0 302--13, May 2006.
\newblock ISSN 0048-5772.
\newblock \doi{10.1111/j.1469-8986.2006.00403.x}.
\newblock URL \url{http://www.ncbi.nlm.nih.gov/pubmed/16805870}.

\bibitem[Hwang et~al.(2010)Hwang, Velanova, and Luna]{HwangVelanovaLuna10}
Kai Hwang, Katerina Velanova, and Beatriz Luna.
\newblock {Strengthening of top-down frontal cognitive control networks
  underlying the development of inhibitory control: a functional magnetic
  resonance imaging effective connectivity study.}
\newblock \emph{The Journal of neuroscience : the official journal of the
  Society for Neuroscience}, 30\penalty0 (46):\penalty0 15535--45, November
  2010.
\newblock ISSN 1529-2401.
\newblock \doi{10.1523/JNEUROSCI.2825-10.2010}.
\newblock URL
  \url{http://www.pubmedcentral.nih.gov/articlerender.fcgi?artid=2995693\&tool%
=pmcentrez\&rendertype=abstract}.

\bibitem[Isoda and Hikosaka(2007)]{IsodaHikosaka07}
Masaki Isoda and Okihide Hikosaka.
\newblock Switching from automatic to controlled action by monkey medial
  frontal cortex.
\newblock \emph{Nature neuroscience}, 10\penalty0 (2):\penalty0 240--248, 01
  2007.
\newblock URL \url{http://www.ncbi.nlm.nih.gov/pubmed/17237780}.

\bibitem[Isoda and Hikosaka(2008)]{IsodaHikosaka08}
Masaki Isoda and Okihide Hikosaka.
\newblock Role for subthalamic nucleus neurons in switching from automatic to
  controlled eye movement.
\newblock \emph{The Journal of neuroscience : the official journal of the
  Society for Neuroscience}, 28:\penalty0 7209--7218, 07 2008.
\newblock URL \url{http://www.ncbi.nlm.nih.gov/pubmed/18614691}.

\bibitem[Jahfari et~al.(2012)Jahfari, Verbruggen, Frank, Waldorp, Colzato,
  Ridderinkhof, and Forstmann]{JahfariVerbruggenFrankEtAl12}
S.~Jahfari, F.~Verbruggen, M.~J. Frank, L.~J. Waldorp, L.~Colzato, K.~R.
  Ridderinkhof, and B.~U. Forstmann.
\newblock {How Preparation Changes the Need for Top-Down Control of the Basal
  Ganglia When Inhibiting Premature Actions}.
\newblock \emph{Journal of Neuroscience}, 32\penalty0 (32):\penalty0
  10870--10878, August 2012.
\newblock ISSN 0270-6474.
\newblock \doi{10.1523/JNEUROSCI.0902-12.2012}.
\newblock URL \url{http://www.jneurosci.org/cgi/content/abstract/32/32/10870}.

\bibitem[Jahfari et~al.(2011)Jahfari, Waldorp, van~den Wildenberg, Scholte,
  Ridderinkhof, and Forstmann]{JahfariWaldorpWildenbergEtAl11}
Sara Jahfari, Lourens Waldorp, Wery P~M van~den Wildenberg, H~Steven Scholte,
  K~Richard Ridderinkhof, and Birte~U Forstmann.
\newblock {Effective connectivity reveals important roles for both the
  hyperdirect (fronto-subthalamic) and the indirect (fronto-striatal-pallidal)
  fronto-basal ganglia pathways during response inhibition.}
\newblock \emph{The Journal of neuroscience : the official journal of the
  Society for Neuroscience}, 31\penalty0 (18):\penalty0 6891--9, May 2011.
\newblock ISSN 1529-2401.
\newblock \doi{10.1523/JNEUROSCI.5253-10.2011}.
\newblock URL \url{http://www.ncbi.nlm.nih.gov/pubmed/21543619}.

\bibitem[Jocham et~al.(2011)Jocham, Klein, and
  Ullsperger]{JochamKleinUllsperger11}
Gerhard Jocham, Tilmann~A Klein, and Markus Ullsperger.
\newblock Dopamine-mediated reinforcement learning signals in the striatum and
  ventromedial prefrontal cortex underlie value-based choices.
\newblock \emph{The Journal of neuroscience}, 31, Feb 2011.
\newblock URL \url{http://www.ncbi.nlm.nih.gov/pubmed/21289169}.

\bibitem[Johnston and Everling(2006)]{JohnstonEverling06}
Kevin Johnston and Stefan Everling.
\newblock {Monkey dorsolateral prefrontal cortex sends task-selective signals
  directly to the superior colliculus.}
\newblock \emph{The Journal of neuroscience : the official journal of the
  Society for Neuroscience}, 26\penalty0 (48):\penalty0 12471--8, November
  2006.
\newblock ISSN 1529-2401.
\newblock \doi{10.1523/JNEUROSCI.4101-06.2006}.
\newblock URL \url{http://www.ncbi.nlm.nih.gov/pubmed/17135409}.

\bibitem[Kolling et~al.(2012)Kolling, Behrens, Mars, and
  Rushworth]{KollingBehrensMarsEtAl12}
Nils Kolling, Timothy Behrens, Rogier Mars, and Matthew Rushworth.
\newblock Neural mechanisms of foraging.
\newblock \emph{Science}, 336\penalty0 (6077):\penalty0 95--98, 4 2012.
\newblock URL \url{http://www.sciencemag.org/content/336/6077/95.abstract}.

\bibitem[Kornblum et~al.(1990)Kornblum, Hasbroucq, and
  Osman]{KornblumHasbroucqOsman90}
S~Kornblum, T~Hasbroucq, and A~Osman.
\newblock Dimensional overlap: cognitive basis for stimulus-response
  compatibility--a model and taxonomy.
\newblock \emph{Psychological review}, 97\penalty0 (2):\penalty0 253--270, 06
  1990.
\newblock URL \url{http://www.ncbi.nlm.nih.gov/pubmed/2186425}.

\bibitem[Kravitz et~al.(2010)Kravitz, Freeze, Parker, Kay, Thwin, Deisseroth,
  and Kreitzer]{KravitzFreezeParkerEtAl10}
Alexxai Kravitz, Benjamin Freeze, Philip Parker, Kenneth Kay, Myo Thwin, Karl
  Deisseroth, and Anatol Kreitzer.
\newblock Regulation of{ P}arkinsonian motor behaviours by optogenetic control
  of basal ganglia circuitry.
\newblock \emph{Nature}, 466:\penalty0 622--626, Jul 2010.
\newblock URL
  \url{http://www.nature.com/nature/journal/vaop/ncurrent/full/nature09159.htm%
l}.

\bibitem[Kravitz et~al.(2012)Kravitz, Tye, and Kreitzer]{KravitzTyeKreitzer12}
Alexxai~V Kravitz, Lynne~D Tye, and Anatol~C Kreitzer.
\newblock Distinct roles for direct and indirect pathway striatal neurons in
  reinforcement.
\newblock \emph{Nature neuroscience}, Apr 2012.
\newblock URL \url{http://www.ncbi.nlm.nih.gov/pubmed/22544310}.

\bibitem[Kuhn et~al.(2004)Kuhn, Williams, Kupsch, Limousin, Hariz, Schneider,
  Yarrow, and Brown]{KuhnWilliamsKupschEtAl04}
Andrea~A. Kuhn, David Williams, Andreas Kupsch, Patricia Limousin, Marwan
  Hariz, Gerd-Helge Schneider, Kielan Yarrow, and Peter Brown.
\newblock Event-related beta desynchronization in human subthalamic nucleus
  correlates with motor performance.
\newblock \emph{Brain}, 127\penalty0 (4):\penalty0 735--746, April 2004.
\newblock \doi{10.1093/brain/awh106}.
\newblock URL \url{http://dx.doi.org/10.1093/brain/awh106}.

\bibitem[Leotti and Wager(2010)]{LeottiWager10}
Lauren~a Leotti and Tor~D Wager.
\newblock {Motivational influences on response inhibition measures.}
\newblock \emph{Journal of experimental psychology. Human perception and
  performance}, 36\penalty0 (2):\penalty0 430--47, April 2010.
\newblock ISSN 1939-1277.
\newblock \doi{10.1037/a0016802}.
\newblock URL \url{http://www.ncbi.nlm.nih.gov/pubmed/20364928}.

\bibitem[Leung and Cai(2007)]{LeungCai07}
H.~C. Leung and W.~Cai.
\newblock Common and differential ventrolateral prefrontal activity during
  inhibition of hand and eye movements.
\newblock \emph{The Journal of neuroscience : the official journal of the
  Society for Neuroscience}, 27\penalty0 (37):\penalty0 9893--9900, September
  2007.
\newblock ISSN 1529-2401.
\newblock \doi{10.1523/JNEUROSCI.2837-07.2007}.
\newblock URL \url{http://dx.doi.org/10.1523/JNEUROSCI.2837-07.2007}.

\bibitem[Lo and Wang(2006)]{LoWang06}
Chung-Chuan Lo and Xiao-Jing Wang.
\newblock Cortico-basal ganglia circuit mechanism for a decision threshold in
  reaction time tasks.
\newblock \emph{Nature neuroscience}, 9\penalty0 (7):\penalty0 956--963, 06
  2006.
\newblock URL \url{http://www.ncbi.nlm.nih.gov/pubmed/16767089}.

\bibitem[Lo et~al.(2009)Lo, Boucher, Pare, Schall, and
  Wang]{LoBoucherPareEtAl09}
Chung-Chuan Lo, Leanne Boucher, Martin Pare, Jeffrey~D. Schall, and Xiao-Jing
  Wang.
\newblock Proactive inhibitory control and attractor dynamics in countermanding
  action: a spiking neural circuit model.
\newblock \emph{The Journal of Neuroscience}, 29\penalty0 (28):\penalty0
  9059--9071, Jul 2009.
\newblock URL \url{http://www.ncbi.nlm.nih.gov/pubmed/19605643}.

\bibitem[Logan(1985)]{Logan85}
G.~D. Logan.
\newblock On the ability to inhibit simple thoughts and actions: Ii.
  stop-signal studies of repetition priming.
\newblock \emph{Journal of Experimental Psychology}, 11\penalty0 (4):\penalty0
  675--691, January 1985.

\bibitem[Logan and Cowan(1984)]{LoganCowan84}
G.~D. Logan and W.~B. Cowan.
\newblock On the ability to inhibit thought and action: A theory of an act of
  control.
\newblock \emph{Psychological Review}, 91\penalty0 (3):\penalty0 295--327,
  January 1984.

\bibitem[Logan et~al.(1997)Logan, Schachar, and
  Tannock]{LoganSchacharTannock97}
Gordon~D. Logan, Russell~J. Schachar, and Rosemary Tannock.
\newblock Impulsivity and inhibitory control.
\newblock \emph{Psychological Science}, 8\penalty0 (1):\penalty0 60--64, 1997.
\newblock \doi{10.1111/j.1467-9280.1997.tb00545.x}.
\newblock URL \url{http://dx.doi.org/10.1111/j.1467-9280.1997.tb00545.x}.

\bibitem[Lu et~al.(1994)Lu, Preston, and Strick]{LuPrestonStrick94}
M~T Lu, J~B Preston, and P~L Strick.
\newblock {Interconnections between the prefrontal cortex and the premotor
  areas in the frontal lobe.}
\newblock \emph{The Journal of comparative neurology}, 341\penalty0
  (3):\penalty0 375--92, March 1994.
\newblock ISSN 0021-9967.
\newblock \doi{10.1002/cne.903410308}.
\newblock URL \url{http://www.ncbi.nlm.nih.gov/pubmed/7515081}.

\bibitem[Mansfield et~al.(2011)Mansfield, Karayanidis, Jamadar, Heathcote, and
  Forstmann]{MansfieldKarayanidisJamadarEtAl11}
Elise~L Mansfield, Frini Karayanidis, Sharna Jamadar, Andrew Heathcote, and
  Birte~U Forstmann.
\newblock Adjustments of response threshold during task switching: A
  model-based functional magnetic resonance imaging study.
\newblock \emph{The Journal of neuroscience}, 31\penalty0 (41), Oct 2011.
\newblock URL \url{http://www.ncbi.nlm.nih.gov/pubmed/21994385}.

\bibitem[McDowell et~al.(2002)McDowell, Brown, Paulus, Martinez, Stewart,
  Dubowitz, and Braff]{McDowellBrownPaulusEtAl02}
Jennifer~E McDowell, Gregory~G Brown, Martin Paulus, Antigona Martinez, Sara~E
  Stewart, David~J Dubowitz, and David~L Braff.
\newblock Neural correlates of refixation saccades and antisaccades in normal
  and schizophrenia subjects.
\newblock \emph{Biological psychiatry}, 51\penalty0 (3):\penalty0 216--223, 02
  2002.
\newblock URL \url{http://www.ncbi.nlm.nih.gov/pubmed/11839364}.

\bibitem[Menzies et~al.(2007)Menzies, Achard, Chamberlain, Fineberg, Chen, del
  Campo, Sahakian, Robbins, and Bullmore]{MenziesAchardChamberlainEtAl07}
L.~Menzies, S.~Achard, S.~R. Chamberlain, N.~Fineberg, C.~H. Chen, N.~del
  Campo, B.~J. Sahakian, T.~W. Robbins, and E.~Bullmore.
\newblock Neurocognitive endophenotypes of obsessive-compulsive disorder.
\newblock \emph{Brain : a journal of neurology}, 130\penalty0 (Pt 12):\penalty0
  3223--3236, December 2007.
\newblock ISSN 1460-2156.
\newblock \doi{10.1093/brain/awm205}.
\newblock URL \url{http://dx.doi.org/10.1093/brain/awm205}.

\bibitem[Miller and Cohen(2001)]{MillerCohen01}
E~K Miller and J~D Cohen.
\newblock An integrative theory of prefrontal cortex function.
\newblock \emph{Annual Review of Neuroscience}, 24:\penalty0 167--202, 2001.
\newblock URL \url{http://www.ncbi.nlm.nih.gov/pubmed/11283309}.

\bibitem[Mink(1996)]{Mink96}
J~W Mink.
\newblock The basal ganglia: Focused selection and inhibition of competing
  motor programs.
\newblock \emph{Progress in Neurobiology}, 50:\penalty0 381--425, 03 1996.
\newblock URL \url{http://www.ncbi.nlm.nih.gov/pubmed/9004351}.

\bibitem[Miyake et~al.(2000)Miyake, Friedman, Emerson, Witzki, Howerter, and
  Wager]{MiyakeFriedmanEmersonEtAl00}
A~Miyake, N~P Friedman, M~J Emerson, A~H Witzki, A~Howerter, and T~D Wager.
\newblock The unity and diversity of executive functions and their
  contributions to complex "frontal lobe" tasks: a latent variable analysis.
\newblock \emph{Cognitive psychology}, 41:\penalty0 49--100, 09 2000.
\newblock URL \url{http://www.ncbi.nlm.nih.gov/pubmed/10945922}.

\bibitem[Montague et~al.(1997)Montague, Dayan, and
  Sejnowski]{MontagueDayanSejnowski96}
P~R Montague, P~Dayan, and T~J Sejnowski.
\newblock A framework for mesencephalic dopamine systems based on predictive
  hebbian learning.
\newblock \emph{The Journal of Neuroscience}, 16:\penalty0 1936--1947, 01 1997.
\newblock URL \url{http://www.ncbi.nlm.nih.gov/pubmed/8774460}.

\bibitem[Monterosso et~al.(2005)Monterosso, Aron, Cordova, Xu, and
  London]{MonterossoAronCordovaEtAl05}
J.~R. Monterosso, A.~R. Aron, X.~Cordova, J.~Xu, and E.~D. London.
\newblock Deficits in response inhibition associated with chronic
  methamphetamine abuse.
\newblock \emph{Drug and alcohol dependence}, 79\penalty0 (2):\penalty0
  273--277, August 2005.
\newblock ISSN 0376-8716.
\newblock \doi{10.1016/j.drugalcdep.2005.02.002}.
\newblock URL \url{http://dx.doi.org/10.1016/j.drugalcdep.2005.02.002}.

\bibitem[Morein-Zamir et~al.(2009)Morein-Zamir, Fineberg, Robbins, and
  Sahakian]{Morein-ZamirFinebergRobbinsEtAl09}
S.~Morein-Zamir, N.~A. Fineberg, T.~W. Robbins, and B.~J. Sahakian.
\newblock Inhibition of thoughts and actions in obsessive-compulsive disorder:
  extending the endophenotype?
\newblock \emph{Psychological medicine}, pages 1--10, July 2009.
\newblock ISSN 1469-8978.
\newblock \doi{10.1017/S003329170999033X}.
\newblock URL \url{http://dx.doi.org/10.1017/S003329170999033X}.

\bibitem[Morein-Zamir and Kingstone(2006)]{Morein-ZamirKingstone06}
Sharon Morein-Zamir and Alan Kingstone.
\newblock {Fixation offset and stop signal intensity effects on saccadic
  countermanding: a crossmodal investigation.}
\newblock \emph{Experimental brain research. Experimentelle Hirnforschung.
  Exp\'{e}rimentation c\'{e}r\'{e}brale}, 175\penalty0 (3):\penalty0 453--62,
  November 2006.
\newblock ISSN 0014-4819.
\newblock \doi{10.1007/s00221-006-0564-x}.
\newblock URL \url{http://www.ncbi.nlm.nih.gov/pubmed/16783558}.

\bibitem[Munakata et~al.(2011)Munakata, Herd, Chatham, Depue, Banich, and
  O'Reilly]{MunakataHerdChathamEtAl11}
Yuko Munakata, Seth~A. Herd, Christopher~H. Chatham, Brendan~E. Depue, Marie~T.
  Banich, and Randall~C. O'Reilly.
\newblock A unified framework for inhibitory control.
\newblock \emph{Trends in Cognitive Sciences}, 15\penalty0 (10):\penalty0
  453--459, Oct 2011.
\newblock URL \url{http://www.ncbi.nlm.nih.gov/pubmed/21889391}.

\bibitem[Munoz and Everling(2004)]{MunozEverling04}
Douglas~P. Munoz and Stefan Everling.
\newblock Look away: the anti-saccade task and the voluntary control of eye
  movement.
\newblock \emph{Nature Reviews Neuroscience}, 5\penalty0 (3):\penalty0
  218--228, Mar 2004.
\newblock URL \url{http://www.ncbi.nlm.nih.gov/pubmed/14976521}.

\bibitem[Nakamura and Hikosaka(2006)]{NakamuraHikosaka06}
Kae Nakamura and Okihide Hikosaka.
\newblock Role of dopamine in the primate caudate nucleus in reward modulation
  of saccades.
\newblock \emph{The Journal of neuroscience : the official journal of the
  Society for Neuroscience}, 26\penalty0 (20):\penalty0 5360--5369, 05 2006.
\newblock URL \url{http://www.ncbi.nlm.nih.gov/pubmed/16707788}.

\bibitem[Nambu et~al.(2000)Nambu, Tokuno, Hamada, Kita, Imanishi, Akazawa,
  Ikeuchi, and Hasegawa]{NambuTokunoHamadaEtAl00}
A~Nambu, H~Tokuno, I~Hamada, H~Kita, M~Imanishi, T~Akazawa, Y~Ikeuchi, and
  N~Hasegawa.
\newblock Excitatory cortical inputs to pallidal neurons via the subthalamic
  nucleus in the monkey.
\newblock \emph{Journal of Neurophysiology}, 84\penalty0 (1):\penalty0
  289--300, 09 2000.
\newblock URL \url{http://www.ncbi.nlm.nih.gov/pubmed/10899204}.

\bibitem[Nambu et~al.(2002)Nambu, Tokuno, and Takada]{NambuTokunoTakada02}
Atsushi Nambu, Hironobu Tokuno, and Masahiko Takada.
\newblock Functional significance of the cortico-subthalamo-pallidal
  'hyperdirect' pathway.
\newblock \emph{Neuroscience research}, 43:\penalty0 111--7, 06 2002.
\newblock URL \url{http://www.ncbi.nlm.nih.gov/pubmed/12067746}.

\bibitem[Neubert et~al.(2010)Neubert, Mars, Buch, Olivier, and
  Rushworth]{NeubertMarsBuchEtAl10}
Franz-Xaver F.-X. Neubert, Rogier~B. Mars, Ethan~R. Buch, Etienne Olivier, and
  Matthew F.~S. Rushworth.
\newblock {Cortical and subcortical interactions during action reprogramming
  and their related white matter pathways.}
\newblock \emph{Proceedings of the National Academy of Sciences of the United
  States of America}, 107\penalty0 (30):\penalty0 13240--5, July 2010.
\newblock ISSN 1091-6490.
\newblock \doi{10.1073/pnas.1000674107}.
\newblock URL
  \url{http://www.pubmedcentral.nih.gov/articlerender.fcgi?artid=2922153\&tool%
=pmcentrez\&rendertype=abstract
  http://www.pnas.org/cgi/doi/10.1073/pnas.1000674107}.

\bibitem[Nieuwenhuis et~al.(2004)Nieuwenhuis, Broerse, Nielen, and
  de~Jong]{NieuwenhuisBroerseNielenEtAl04}
Sander Nieuwenhuis, Annelies Broerse, Marjan M~a Nielen, and Ritske de~Jong.
\newblock {A goal activation approach to the study of executive function: an
  application to antisaccade tasks.}
\newblock \emph{Brain and cognition}, 56\penalty0 (2):\penalty0 198--214,
  November 2004.
\newblock ISSN 0278-2626.
\newblock \doi{10.1016/j.bandc.2003.12.002}.
\newblock URL \url{http://www.ncbi.nlm.nih.gov/pubmed/15518936}.

\bibitem[Nigg(2001)]{Nigg01}
J.~T. Nigg.
\newblock Is adhd a disinhibitory disorder?
\newblock \emph{Psychological bulletin}, 127\penalty0 (5):\penalty0 571--598,
  September 2001.
\newblock ISSN 0033-2909.
\newblock URL \url{http://view.ncbi.nlm.nih.gov/pubmed/11548968}.

\bibitem[Nigg et~al.(2006)Nigg, Wong, Martel, Jester, Puttler, Glass, Adams,
  Fitzgerald, and Zucker]{NiggWongMartelEtAl06}
J.~T. Nigg, M.~M. Wong, M.~M. Martel, J.~M. Jester, L.~I. Puttler, J.~M. Glass,
  K.~M. Adams, H.~E. Fitzgerald, and R.~A. Zucker.
\newblock Poor response inhibition as a predictor of problem drinking and
  illicit drug use in adolescents at risk for alcoholism and other substance
  use disorders.
\newblock \emph{Journal of the American Academy of Child and Adolescent
  Psychiatry}, 45\penalty0 (4):\penalty0 468--475, April 2006.
\newblock ISSN 0890-8567.
\newblock \doi{10.1097/01.chi.0000199028.76452.a9}.
\newblock URL \url{http://dx.doi.org/10.1097/01.chi.0000199028.76452.a9}.

\bibitem[Obeso et~al.(2011{\natexlab{a}})Obeso, Wilkinson, Casabona, Bringas,
  Alvarez, Alvarez, Pav\'{o}n, Rodr\'{\i}guez-Oroz, Mac\'{\i}as, Obeso, and
  Jahanshahi]{ObesoWilkinsonCasabonaEtAl11}
Ignacio Obeso, Leonora Wilkinson, Enrique Casabona, Maria~Luisa Bringas, Mario
  Alvarez, L\'{a}zaro Alvarez, Nancy Pav\'{o}n, Maria-Cruz Rodr\'{\i}guez-Oroz,
  Ra\'{u}l Mac\'{\i}as, Jose~a Obeso, and Marjan Jahanshahi.
\newblock {Deficits in inhibitory control and conflict resolution on cognitive
  and motor tasks in Parkinson's disease.}
\newblock \emph{Experimental brain research. Experimentelle Hirnforschung.
  Exp\'{e}rimentation c\'{e}r\'{e}brale}, 212\penalty0 (3):\penalty0 371--84,
  July 2011{\natexlab{a}}.
\newblock ISSN 1432-1106.
\newblock \doi{10.1007/s00221-011-2736-6}.
\newblock URL \url{http://www.ncbi.nlm.nih.gov/pubmed/21643718}.

\bibitem[Obeso et~al.(2011{\natexlab{b}})Obeso, Wilkinson, and
  Jahanshahi]{ObesoWilkinsonJahanshahi11}
Ignacio Obeso, Leonora Wilkinson, and Marjan Jahanshahi.
\newblock {Levodopa medication does not influence motor inhibition or conflict
  resolution in a conditional stop-signal task in Parkinson's disease.}
\newblock \emph{Experimental brain research. Experimentelle Hirnforschung.
  Experimentation cerebrale}, July 2011{\natexlab{b}}.
\newblock ISSN 1432-1106.
\newblock \doi{10.1007/s00221-011-2793-x}.
\newblock URL \url{http://www.ncbi.nlm.nih.gov/pubmed/21796541}.

\bibitem[Oosterlaan et~al.(1998)Oosterlaan, Logan, and
  Sergeant]{OosterlaanLoganSergeant98}
Jaap Oosterlaan, Gordon~D. Logan, and Joseph~A. Sergeant.
\newblock Response inhibition in ad/hd, cd, comorbid ad/hd+cd, anxious, and
  control children: A meta-analysis of studies with the stop task.
\newblock \emph{The Journal of Child Psychology and Psychiatry and Allied
  Disciplines}, 39\penalty0 (03):\penalty0 411--425, 1998.
\newblock URL
  \url{http://journals.cambridge.org/action/displayAbstract?fromPage=online\&a%
id=10427}.

\bibitem[O'Reilly and Frank(2006)]{OReillyFrank06}
Randall~C. O'Reilly and Michael~J. Frank.
\newblock Making working memory work: A computational model of learning in the
  prefrontal cortex and basal ganglia.
\newblock \emph{Neural Computation}, 18:\penalty0 283--328, 2006.
\newblock URL \url{http://www.ncbi.nlm.nih.gov/pubmed/16378516}.

\bibitem[O'Reilly and Munakata(2000)]{OReillyMunakata00}
Randall~C. O'Reilly and Yuko Munakata.
\newblock \emph{{Computational Explorations in Cognitive Neuroscience:
  Understanding the Mind by Simulating the Brain}}.
\newblock The MIT Press, Cambridge, MA, January 2000.

\bibitem[Osman et~al.(1986)Osman, Kornblum, and Meyer]{OsmanKornblumMeyer86}
A~Osman, S~Kornblum, and D~E Meyer.
\newblock The point of no return in choice reaction time: controlled and
  ballistic stages of response preparation.
\newblock \emph{Journal of experimental psychology. Human perception and
  performance}, 12\penalty0 (3):\penalty0 243--258, 09 1986.
\newblock URL \url{http://www.ncbi.nlm.nih.gov/pubmed/2943853}.

\bibitem[Palminteri et~al.(2009)Palminteri, Lebreton, Worbe, Grabli, Hartmann,
  and Pessiglione]{PalminteriLebretonWorbeEtAl09}
S.~Palminteri, M.~Lebreton, Y.~Worbe, D.~Grabli, A.~Hartmann, and
  M.~Pessiglione.
\newblock Pharmacological modulation of subliminal learning in parkinson's and
  tourette's syndromes.
\newblock \emph{Proceedings of the National Academy of Sciences}, 2009.

\bibitem[Pare and Hanes(2003)]{PareHanes03}
Martin Pare and Doug~P. Hanes.
\newblock Controlled movement processing: Superior colliculus activity
  associated with countermanded saccades.
\newblock \emph{J. Neurosci.}, 23\penalty0 (16):\penalty0 6480--6489, July
  2003.
\newblock URL \url{http://www.jneurosci.org/cgi/content/abstract/23/16/6480}.

\bibitem[Parent and Hazrati(1995)]{ParentHazrati95}
A.~Parent and L.~N. Hazrati.
\newblock Functional anatomy of the basal ganglia. ii. the place of subthalamic
  nucleus and external pallidum in basal ganglia circuitry.
\newblock \emph{Brain Research. Brain Research Reviews}, 20\penalty0
  (1):\penalty0 128--154, 05 1995.
\newblock URL \url{http://www.ncbi.nlm.nih.gov/pubmed/7711765}.

\bibitem[Parmentier et~al.(2008)Parmentier, Elford, Escera, Andr\'{e}s, and
  {San Miguel}]{ParmentierElfordEsceraEtAl08}
Fabrice B~R Parmentier, Gregory Elford, Carles Escera, Pilar Andr\'{e}s, and
  Iria {San Miguel}.
\newblock {The cognitive locus of distraction by acoustic novelty in the
  cross-modal oddball task.}
\newblock \emph{Cognition}, 106\penalty0 (1):\penalty0 408--32, January 2008.
\newblock ISSN 0010-0277.
\newblock \doi{10.1016/j.cognition.2007.03.008}.
\newblock URL \url{http://dx.doi.org/10.1016/j.cognition.2007.03.008}.

\bibitem[Penad\'{e}s et~al.(2007)Penad\'{e}s, Catal\'{a}n, Rubia, Andr\'{e}s,
  Salamero, and Gast\'{o}]{PenadesCatalanRubiaEtAl07}
R.~Penad\'{e}s, R.~Catal\'{a}n, K.~Rubia, S.~Andr\'{e}s, M.~Salamero, and
  C.~Gast\'{o}.
\newblock Impaired response inhibition in obsessive compulsive disorder.
\newblock \emph{European psychiatry : the journal of the Association of
  European Psychiatrists}, 22\penalty0 (6):\penalty0 404--410, September 2007.
\newblock ISSN 0924-9338.
\newblock \doi{10.1016/j.eurpsy.2006.05.001}.
\newblock URL \url{http://dx.doi.org/10.1016/j.eurpsy.2006.05.001}.

\bibitem[Pouget et~al.(2011)Pouget, Logan, Palmeri, Boucher, Pare, and
  Schall]{PougetLoganPalmeriEtAl11}
Pierre Pouget, Gordon~D. Logan, Thomas~J. Palmeri, Leanne Boucher, Martin Pare,
  and Jeffrey~D. Schall.
\newblock {Neural Basis of Adaptive Response Time Adjustment during Saccade
  Countermanding}.
\newblock \emph{Journal of Neuroscience}, 31\penalty0 (35):\penalty0
  12604--12612, August 2011.
\newblock ISSN 0270-6474.
\newblock \doi{10.1523/JNEUROSCI.1868-11.2011}.
\newblock URL
  \url{http://www.jneurosci.org/cgi/doi/10.1523/JNEUROSCI.1868-11.2011}.

\bibitem[Ramos and Arnsten(2007)]{RamosArnsten07}
B.~Ramos and A.~Arnsten.
\newblock Adrenergic pharmacology and cognition: Focus on the prefrontal
  cortex.
\newblock \emph{Pharmacology \& Therapeutics}, 113\penalty0 (3):\penalty0
  523--536, March 2007.
\newblock ISSN 01637258.
\newblock \doi{10.1016/j.pharmthera.2006.11.006}.
\newblock URL \url{http://dx.doi.org/10.1016/j.pharmthera.2006.11.006}.

\bibitem[Ratcliff and Frank(2012)]{RatcliffFrank12}
Roger Ratcliff and Michael~J. Frank.
\newblock Reinforcement-based decision making in corticostriatal circuits:
  mutual constraints by neurocomputational and diffusion models.
\newblock \emph{Neural Computation}, 24\penalty0 (5):\penalty0 1186--1229, May
  2012.
\newblock URL \url{http://www.ncbi.nlm.nih.gov/pubmed/22295983}.

\bibitem[Ray et~al.(2009)Ray, Jenkinson, Brittain, Holland, Joint, Nandi, Bain,
  Yousif, Green, Stein, and Aziz]{RayJenkinsonBrittainEtAl09}
N.~J. Ray, N.~Jenkinson, J.~Brittain, P.~Holland, C.~Joint, D.~Nandi, P.~G.
  Bain, N.~Yousif, A.~Green, J.~S. Stein, and T.~Z. Aziz.
\newblock The role of the subthalamic nucleus in response inhibition: evidence
  from deep brain stimulation for parkinson's disease.
\newblock \emph{Neuropsychologia}, 47\penalty0 (13):\penalty0 2828--2834, Nov
  2009.
\newblock URL \url{http://www.ncbi.nlm.nih.gov/pubmed/19540864}.

\bibitem[Reilly et~al.(2006)Reilly, Harris, Keshavan, and
  Sweeney]{ReillyHarrisKeshavanEtAl06}
J.~L. Reilly, M.~S. Harris, M.~S. Keshavan, and J.~A. Sweeney.
\newblock Adverse effects of risperidone on spatial working memory in
  first-episode schizophrenia.
\newblock \emph{Archives of general psychiatry}, 63\penalty0 (11):\penalty0
  1189--1197, November 2006.
\newblock ISSN 0003-990X.
\newblock \doi{10.1001/archpsyc.63.11.1189}.
\newblock URL \url{http://dx.doi.org/10.1001/archpsyc.63.11.1189}.

\bibitem[Reilly et~al.(2007)Reilly, Harris, Khine, Keshavan, and
  Sweeney]{ReillyHarrisKhineEtAl07}
J.~L. Reilly, M.~S. Harris, T.~T. Khine, M.~S. Keshavan, and J.~A. Sweeney.
\newblock Antipsychotic drugs exacerbate impairment on a working memory task in
  first-episode schizophrenia.
\newblock \emph{Biological psychiatry}, 62\penalty0 (7):\penalty0 818--821,
  October 2007.
\newblock ISSN 0006-3223.
\newblock \doi{10.1016/j.biopsych.2006.10.031}.
\newblock URL \url{http://dx.doi.org/10.1016/j.biopsych.2006.10.031}.

\bibitem[Reuter and Kathmann(2004)]{ReuterKathmann04}
Benedikt Reuter and Norbert Kathmann.
\newblock {Using saccade tasks as a tool to analyze executive dysfunctions in
  schizophrenia.}
\newblock \emph{Acta psychologica}, 115\penalty0 (2-3):\penalty0 255--69, 2004.
\newblock ISSN 0001-6918.
\newblock \doi{10.1016/j.actpsy.2003.12.009}.
\newblock URL \url{http://www.ncbi.nlm.nih.gov/pubmed/14962403}.

\bibitem[Ridderinkhof(2002)]{Ridderinkhof02}
K~Richard Ridderinkhof.
\newblock {Micro- and macro-adjustments of task set: activation and suppression
  in conflict tasks.}
\newblock \emph{Psychological research}, 66\penalty0 (4):\penalty0 312--23,
  November 2002.
\newblock ISSN 0340-0727.
\newblock \doi{10.1007/s00426-002-0104-7}.
\newblock URL \url{http://www.ncbi.nlm.nih.gov/pubmed/12466928
  http://academic.research.microsoft.com/Publication/2054125/activation-and-su%
ppression-in-conflict-tasks-empirical-clarification-through-distributional
  http://dare.uva.nl/record/122002}.

\bibitem[Ridderinkhof et~al.(2004)Ridderinkhof, van~den Wildenberg, Segalowitz,
  and Carter]{RidderinkhofWildenbergSegalowitzEtAl04}
K~Richard Ridderinkhof, Wery P~M van~den Wildenberg, Sidney~J Segalowitz, and
  Cameron~S Carter.
\newblock {Neurocognitive mechanisms of cognitive control: the role of
  prefrontal cortex in action selection, response inhibition, performance
  monitoring, and reward-based learning.}
\newblock \emph{Brain and cognition}, 56\penalty0 (2):\penalty0 129--40,
  November 2004.
\newblock ISSN 0278-2626.
\newblock \doi{10.1016/j.bandc.2004.09.016}.
\newblock URL \url{http://www.ncbi.nlm.nih.gov/pubmed/15518930}.

\bibitem[Ridderinkhof et~al.(2011)Ridderinkhof, Forstmann, Wylie, Burle,
  van~den Wildenberg, and {Richard
  Ridderinkhof}]{RidderinkhofForstmannWylieEtAl11}
K~Richard Ridderinkhof, Birte~U. Forstmann, Scott~a. Wylie, Bor\'{\i}s Burle,
  Wery P.~M. van~den Wildenberg, and K.~{Richard Ridderinkhof}.
\newblock {Neurocognitive mechanisms of action control: resisting the call of
  the Sirens}.
\newblock \emph{Wiley Interdisciplinary Reviews: Cognitive Science}, 2\penalty0
  (2):\penalty0 174--192, March 2011.
\newblock ISSN 19395078.
\newblock \doi{10.1002/wcs.99}.
\newblock URL \url{http://doi.wiley.com/10.1002/wcs.99}.

\bibitem[Roberts et~al.(1994)Roberts, Hager, and Heron]{RobertsHagerHeron94}
Ralph~J. Roberts, Lisa~D. Hager, and Christine Heron.
\newblock Prefrontal cognitive processes: Working memory and inhibition in the
  antisaccade task.
\newblock \emph{Journal of Experimental Psychology: General}, 123:\penalty0
  374, January 1994.

\bibitem[Rougier et~al.(2005)Rougier, Noelle, Braver, Cohen, and
  O'Reilly]{RougierNoelleBraverEtAl05}
N.~P. Rougier, D.~Noelle, T.~S. Braver, J.~D. Cohen, and R.~C. O'Reilly.
\newblock Prefrontal cortex and the flexibility of cognitive control: {R}ules
  without symbols.
\newblock \emph{Proceedings of the National Academy of Sciences}, 102\penalty0
  (20):\penalty0 7338--7343, January 2005.

\bibitem[Rowe et~al.(2002)Rowe, Friston, Frackowiak, and
  Passingham]{RoweFristonFrackowiakEtAl02}
J.~Rowe, K.~Friston, R.~Frackowiak, and R.~Passingham.
\newblock Attention to action: specific modulation of corticocortical
  interactions in humans.
\newblock \emph{NeuroImage}, 17\penalty0 (2):\penalty0 988--998, October 2002.
\newblock ISSN 1053-8119.
\newblock URL \url{http://view.ncbi.nlm.nih.gov/pubmed/12377172}.

\bibitem[Rubchinsky et~al.(2003)Rubchinsky, Kopell, and
  Sigvardt]{RubchinskyKopellSigvardt03}
Leonid~L Rubchinsky, Nancy Kopell, and Karen~A Sigvardt.
\newblock Modeling facilitation and inhibition of competing motor programs in
  basal ganglia subthalamic nucleus-pallidal circuits.
\newblock \emph{Proceedings of the National Academy of Sciences of the United
  States of America}, 100:\penalty0 14427--32, 12 2003.
\newblock URL \url{http://www.ncbi.nlm.nih.gov/pubmed/14612573}.

\bibitem[Sakagami et~al.(2001)Sakagami, K., Lauwereyns, Koizumi, Kobayashi, and
  Hikosaka]{SakagamiTsutsuiLauwereynsEtAl01}
M.~Sakagami, Tsutsui K., J.~Lauwereyns, M.~Koizumi, S.~Kobayashi, and
  O.~Hikosaka.
\newblock A code for behavioral inhibition on the basis of color, but not
  motion, in ventrolateral prefrontal cortex of macaque monkey.
\newblock \emph{The Journal of {N}euroscience}, 21\penalty0 (13):\penalty0
  4801--4808, Jul 2001.
\newblock URL \url{http://www.ncbi.nlm.nih.gov/pubmed/11425907}.

\bibitem[Schachar and Logan(1990)]{SchacharLogan90}
Russell Schachar and Gordon~D. Logan.
\newblock Impulsivity and inhibitory control in normal development and
  childhood psychopathology.
\newblock \emph{Developmental Psychology}, 26\penalty0 (5):\penalty0 710--720,
  1990.
\newblock ISSN 0012-1649.
\newblock \doi{10.1037/0012-1649.26.5.710}.
\newblock URL \url{http://dx.doi.org/10.1037/0012-1649.26.5.710}.

\bibitem[Schlag and Schlag-Rey(1987)]{SchlagSchlag-Rey87}
J.~Schlag and M.~Schlag-Rey.
\newblock {Evidence for a supplementary eye field}.
\newblock \emph{J Neurophysiol}, 57\penalty0 (1):\penalty0 179--200, January
  1987.
\newblock URL \url{http://jn.physiology.org/cgi/content/abstract/57/1/179}.

\bibitem[Schlag-Rey and Schlag(1984)]{Schlag-ReySchlag84}
M.~Schlag-Rey and J.~Schlag.
\newblock {Visuomotor functions of central thalamus in monkey. I. Unit activity
  related to spontaneous eye movements}.
\newblock \emph{J Neurophysiol}, 51\penalty0 (6):\penalty0 1149--1174, June
  1984.
\newblock URL \url{http://jn.physiology.org/cgi/content/abstract/51/6/1149}.

\bibitem[Schlag-Rey et~al.(1997)Schlag-Rey, Amador, Sanchez, and
  Schlag]{SchlagReySanchezSchlag97}
M~Schlag-Rey, N~Amador, H~Sanchez, and J~Schlag.
\newblock Antisaccade performance predicted by neuronal activity in the
  supplementary eye field.
\newblock \emph{Nature}, 390:\penalty0 398, 12 1997.
\newblock URL \url{http://www.ncbi.nlm.nih.gov/pubmed/9389478}.

\bibitem[Schmidt et~al.(2012)Schmidt, Leventhal, Pettibone, Case, and
  Berke]{SchmidtLeventhalPettiboneEtAl12}
Robert Schmidt, Daniel Leventhal, Jeff Pettibone, Alaina Case, and Joshua
  Berke.
\newblock {Suppressing Actions in the Basal Ganglia}.
\newblock In \emph{The 9th annual Computational and Systems Neuroscience
  meeting}, page 139, 2012.

\bibitem[Sharp et~al.(2010)Sharp, Bonnelle, De~Boissezon, Beckmann, James,
  Patel, and Mehta]{SharpBonnelleDeBoissezonEtAl10}
D~J Sharp, V~Bonnelle, X~De~Boissezon, C~F Beckmann, S~G James, M~C Patel, and
  M~A Mehta.
\newblock Distinct frontal systems for response inhibition, attentional
  capture, and error processing.
\newblock \emph{Proceedings of the National Academy of Sciences of the United
  States of America}, Mar 2010.
\newblock URL \url{http://www.ncbi.nlm.nih.gov/pubmed/20220100}.

\bibitem[Shen et~al.(2008)Shen, Flajolet, Greengard, and
  Surmeier]{ShenFlajoletGreengardEtAl08}
Weixing Shen, Marc Flajolet, Paul Greengard, and D~James Surmeier.
\newblock Dichotomous dopaminergic control of striatal synaptic plasticity.
\newblock \emph{Science (New York, N.Y.)}, 321\penalty0 (5890):\penalty0
  848--851, 08 2008.
\newblock URL \url{http://www.ncbi.nlm.nih.gov/pubmed/18687967}.

\bibitem[Simmonds et~al.(2008)Simmonds, Pekar, and
  Mostofsky]{SimmondsPekarMostofsky08}
D.~Simmonds, J.~Pekar, and S.~Mostofsky.
\newblock Meta-analysis of go/no-go tasks demonstrating that fmri activation
  associated with response inhibition is task-dependent.
\newblock \emph{Neuropsychologia}, 46\penalty0 (1):\penalty0 224--232, 2008.
\newblock ISSN 00283932.
\newblock \doi{10.1016/j.neuropsychologia.2007.07.015}.
\newblock URL \url{http://dx.doi.org/10.1016/j.neuropsychologia.2007.07.015}.

\bibitem[Simon(1969)]{Simon69}
J~R Simon.
\newblock Reactions toward the source of stimulation.
\newblock \emph{Journal of experimental psychology}, 81:\penalty0 174--176, 10
  1969.
\newblock URL \url{http://www.ncbi.nlm.nih.gov/pubmed/5812172}.

\bibitem[Sloman(1996)]{Sloman96}
S.~A. Sloman.
\newblock The empirical case for two systems of reasoning.
\newblock \emph{Pscyhological Bulletin}, 119:\penalty0 3--22, January 1996.

\bibitem[Sommer and Wurtz(2002)]{SommerWurtz02}
Marc~A Sommer and Robert~H Wurtz.
\newblock A pathway in primate brain for internal monitoring of movements.
\newblock \emph{Science (New York, N.Y.)}, 296:\penalty0 1480--1482, 05 2002.
\newblock URL \url{http://www.ncbi.nlm.nih.gov/pubmed/12029137}.

\bibitem[Sommer and Wurtz(2004{\natexlab{a}})]{SommerWurtz04}
Marc~A Sommer and Robert~H Wurtz.
\newblock What the brain stem tells the frontal cortex. i. oculomotor signals
  sent from superior colliculus to frontal eye field via mediodorsal thalamus.
\newblock \emph{Journal of neurophysiology}, 91\penalty0 (3):\penalty0
  1381--1402, Mar 2004{\natexlab{a}}.
\newblock URL \url{http://www.ncbi.nlm.nih.gov/pubmed/14573558}.

\bibitem[Sommer and Wurtz(2004{\natexlab{b}})]{SommerWurtz04a}
Marc~A Sommer and Robert~H Wurtz.
\newblock {What the brain stem tells the frontal cortex. II. Role of the
  SC-MD-FEF pathway in corollary discharge.}
\newblock \emph{Journal of neurophysiology}, 91\penalty0 (3):\penalty0
  1403--23, March 2004{\natexlab{b}}.
\newblock ISSN 0022-3077.
\newblock URL \url{http://www.ncbi.nlm.nih.gov/pubmed/14573557}.

\bibitem[Sommer and Wurtz(2006)]{SommerWurtz06}
Marc~A Sommer and Robert~H Wurtz.
\newblock {Influence of the thalamus on spatial visual processing in frontal
  cortex.}
\newblock \emph{Nature}, 444\penalty0 (7117):\penalty0 374--7, November 2006.
\newblock ISSN 1476-4687.
\newblock URL \url{http://dx.doi.org/10.1038/nature05279}.

\bibitem[Sparks(2002)]{Sparks02}
D.~L. Sparks.
\newblock The brainstem control of saccadic eye movements.
\newblock \emph{Nature Reviews Neuroscience}, 3:\penalty0 952--964, January
  2002.

\bibitem[Stevens(2000)]{Stevens00}
A~Stevens.
\newblock {Event-related fMRI of auditory and visual oddball tasks}.
\newblock \emph{Magnetic Resonance Imaging}, 18\penalty0 (5):\penalty0
  495--502, June 2000.
\newblock ISSN 0730725X.
\newblock \doi{10.1016/S0730-725X(00)00128-4}.
\newblock URL \url{http://dx.doi.org/10.1016/S0730-725X(00)00128-4
  http://linkinghub.elsevier.com/retrieve/pii/S0730725X00001284}.

\bibitem[Stuphorn et~al.(2000)Stuphorn, Taylor, and
  Schall]{StuphornTaylorSchall00}
V~Stuphorn, T~L Taylor, and J~D Schall.
\newblock Performance monitoring by the supplementary eye field.
\newblock \emph{Nature}, 408:\penalty0 857, 12 2000.
\newblock URL \url{http://www.ncbi.nlm.nih.gov/pubmed/11130724}.

\bibitem[Stuphorn and Schall(2006)]{StuphornSchall06}
Veit Stuphorn and Jeffrey~D. Schall.
\newblock Executive control of countermanding saccades by the supplementary eye
  field.
\newblock \emph{Nat Neurosci}, 9\penalty0 (7):\penalty0 925--931, July 2006.
\newblock ISSN 1097-6256.
\newblock \doi{10.1038/nn1714}.
\newblock URL \url{http://dx.doi.org/10.1038/nn1714}.

\bibitem[Swann et~al.(2011{\natexlab{a}})Swann, Poizner, Houser, Gould,
  Greenhouse, Cai, Strunk, George, and Aron]{SwannPoiznerHouserEtAl11}
Nicole Swann, Howard Poizner, Melissa Houser, Sherrie Gould, Ian Greenhouse,
  Weidong Cai, Jon Strunk, Jobi George, and Adam~R Aron.
\newblock {Deep brain stimulation of the subthalamic nucleus alters the
  cortical profile of response inhibition in the beta frequency band: a scalp
  EEG study in Parkinson's disease.}
\newblock \emph{The Journal of neuroscience : the official journal of the
  Society for Neuroscience}, 31\penalty0 (15):\penalty0 5721--9, April
  2011{\natexlab{a}}.
\newblock ISSN 1529-2401.
\newblock \doi{10.1523/JNEUROSCI.6135-10.2011}.
\newblock URL
  \url{http://www.pubmedcentral.nih.gov/articlerender.fcgi?artid=3086079\&tool%
=pmcentrez\&rendertype=abstract}.

\bibitem[Swann et~al.(2011{\natexlab{b}})Swann, Cai, Conner, Pieters, Claffey,
  George, Aron, and Tandon]{SwannCaiConnerEtAl11}
Nicole~C Swann, Weidong Cai, Christopher~R Conner, Thomas~A Pieters, Michael~P
  Claffey, Jobi~S George, Adam~R Aron, and Nitin Tandon.
\newblock {Roles for the pre-supplementary motor area and the right inferior
  frontal gyrus in stopping action : electrophysiological responses and
  functional and structural connectivity}.
\newblock \emph{NeuroImage}, September 2011{\natexlab{b}}.
\newblock ISSN 1095-9572.
\newblock \doi{10.1016/j.neuroimage.2011.09.049}.
\newblock URL
  \url{http://linkinghub.elsevier.com/retrieve/pii/S1053811911011141}.

\bibitem[Taverna et~al.(2008)Taverna, Ilijic, and
  Surmeier]{TavernaIlijicSurmeier08}
Stefano Taverna, Ema Ilijic, and D.~James Surmeier.
\newblock Recurrent collateral connections of striatal medium spiny neurons are
  disrupted in models of {P}arkinson's disease.
\newblock \emph{The Journal of Neuroscience}, 28\penalty0 (21):\penalty0
  5504--5512, 5 2008.
\newblock URL \url{http://www.jneurosci.org/cgi/content/abstract/28/21/5504}.

\bibitem[van Koningsbruggen et~al.(2009)van Koningsbruggen, Pender, Machado,
  and Rafal]{van_KoningsbruggenPenderMachadoEtAl09}
Martijn~G. van Koningsbruggen, Tom Pender, Liana Machado, and Robert~D. Rafal.
\newblock Impaired control of the oculomotor reflexes in parkinson's disease.
\newblock \emph{Neuropsychologia}, June 2009.
\newblock ISSN 1873-3514.
\newblock \doi{10.1016/j.neuropsychologia.2009.06.018}.
\newblock URL \url{http://dx.doi.org/10.1016/j.neuropsychologia.2009.06.018}.

\bibitem[Vazey and Aston-Jones(2012)]{VazeyAston-Jones12}
Elena~M Vazey and Gary Aston-Jones.
\newblock {The emerging role of norepinephrine in cognitive dysfunctions of
  Parkinson's disease.}
\newblock \emph{Frontiers in behavioral neuroscience}, 6:\penalty0 48, January
  2012.
\newblock ISSN 1662-5153.
\newblock \doi{10.3389/fnbeh.2012.00048}.
\newblock URL \url{http://www.ncbi.nlm.nih.gov/pubmed/22848194}.

\bibitem[Verbruggen and Logan(2008)]{VerbruggenLogan08}
Frederick Verbruggen and Gordon~D Logan.
\newblock Response inhibition in the stop-signal paradigm.
\newblock \emph{Trends in cognitive sciences}, 12:\penalty0 418--424, Nov 2008.
\newblock URL \url{http://www.ncbi.nlm.nih.gov/pubmed/18799345}.

\bibitem[Verbruggen and Logan(2009{\natexlab{a}})]{VerbruggenLogan08Rev}
Frederick Verbruggen and Gordon~D. Logan.
\newblock Models of response inhibition in the stop-signal and stop-change
  paradigms.
\newblock \emph{Neuroscience \& Biobehavioral Reviews}, 33\penalty0
  (5):\penalty0 647--661, May 2009{\natexlab{a}}.
\newblock ISSN 01497634.
\newblock \doi{10.1016/j.neubiorev.2008.08.014}.
\newblock URL \url{http://dx.doi.org/10.1016/j.neubiorev.2008.08.014}.

\bibitem[Verbruggen and Logan(2009{\natexlab{b}})]{VerbruggenLogan09}
Frederick Verbruggen and Gordon~D Logan.
\newblock Models of response inhibition in the stop-signal and stop-change
  paradigms.
\newblock \emph{Neuroscience and biobehavioral reviews}, 33:\penalty0 647--661,
  May 2009{\natexlab{b}}.
\newblock URL \url{http://www.ncbi.nlm.nih.gov/pubmed/18822313}.

\bibitem[Verbruggen et~al.(2010)Verbruggen, Aron, Stevens, and
  Chambers]{VerbruggenAronStevensEtAl10}
Frederick Verbruggen, Adam~R Aron, Micha\"{e}l~A Stevens, and Christopher~D
  Chambers.
\newblock {Theta burst stimulation dissociates attention and action updating in
  human inferior frontal cortex.}
\newblock \emph{Proceedings of the National Academy of Sciences of the United
  States of America}, 107\penalty0 (31):\penalty0 13966--71, August 2010.
\newblock ISSN 1091-6490.
\newblock \doi{10.1073/pnas.1001957107}.
\newblock URL
  \url{http://www.pubmedcentral.nih.gov/articlerender.fcgi?artid=2922216\&tool%
=pmcentrez\&rendertype=abstract}.

\bibitem[Voon et~al.(2010)Voon, Pessiglione, Brezing, Gallea, Fernandez, Dolan,
  and Hallett]{VoonPessiglioneBrezingEtAl10}
V.~Voon, M.~Pessiglione, C.~Brezing, C.~Gallea, H.~H. Fernandez, R.~J. Dolan,
  and M.~Hallett.
\newblock Mechanisms underlying dopamine-mediated reward bias in compulsive
  behaviors.
\newblock \emph{Neuron}, 65:\penalty0 135--142, 2010.

\bibitem[Wallis and Miller(2003)]{WallisMiller03a}
Jonathan~D Wallis and Earl~K Miller.
\newblock Neuronal activity in primate dorsolateral and orbital prefrontal
  cortex during performance of a reward preference task.
\newblock \emph{The European journal of neuroscience}, 18:\penalty0 2069--81,
  11 2003.
\newblock URL \url{http://www.ncbi.nlm.nih.gov/pubmed/14622240}.

\bibitem[Wang et~al.(2005)Wang, Isoda, Matsuzaka, Shima, and
  Tanji]{WangIsodaMatsuzakaEtAl05}
Yan Wang, Masaki Isoda, Yoshiya Matsuzaka, Keisetsu Shima, and Jun Tanji.
\newblock {Prefrontal cortical cells projecting to the supplementary eye field
  and presupplementary motor area in the monkey.}
\newblock \emph{Neuroscience research}, 53\penalty0 (1):\penalty0 1--7,
  September 2005.
\newblock ISSN 0168-0102.
\newblock \doi{10.1016/j.neures.2005.05.005}.
\newblock URL \url{http://www.ncbi.nlm.nih.gov/pubmed/15992955}.

\bibitem[Watanabe and Munoz(2009)]{WatanabeMunoz09}
Masayuki Watanabe and Douglas~P. Munoz.
\newblock Neural correlates of conflict resolution between automatic and
  volitional actions by basal ganglia.
\newblock \emph{European Journal of Neuroscience}, 30:\penalty0 2165--2176,
  2009.
\newblock URL \url{http://dx.doi.org/10.1111/j.1460-9568.2009.06998.x}.

\bibitem[Watanabe and Munoz(2010)]{WatanabeMunoz10}
Masayuki Watanabe and Douglas~P. Munoz.
\newblock Presetting basal ganglia for volitional actions.
\newblock \emph{The Journal of Neuroscience}, 30\penalty0 (2), Jul 2010.
\newblock URL \url{http://www.ncbi.nlm.nih.gov/pubmed/20668198}.

\bibitem[Watanabe and Munoz(2011)]{WatanabeMunoz11}
Masayuki Watanabe and Douglas~P Munoz.
\newblock {Probing basal ganglia functions by saccade eye movements.}
\newblock \emph{The European journal of neuroscience}, 33\penalty0
  (11):\penalty0 2070--90, June 2011.
\newblock ISSN 1460-9568.
\newblock \doi{10.1111/j.1460-9568.2011.07691.x}.
\newblock URL \url{http://www.ncbi.nlm.nih.gov/pubmed/21645102}.

\bibitem[Wegener et~al.(2008)Wegener, Johnston, and
  Everling]{WegenerJohnstonEverling08}
Stephen~P Wegener, Kevin Johnston, and Stefan Everling.
\newblock {Microstimulation of monkey dorsolateral prefrontal cortex impairs
  antisaccade performance.}
\newblock \emph{Experimental brain research. Experimentelle Hirnforschung.
  Exp\'{e}rimentation c\'{e}r\'{e}brale}, 190\penalty0 (4):\penalty0 463--73,
  October 2008.
\newblock ISSN 1432-1106.
\newblock \doi{10.1007/s00221-008-1488-4}.
\newblock URL \url{http://www.ncbi.nlm.nih.gov/pubmed/18641976}.

\bibitem[Wiecki and Frank(2010)]{WieckiFrank10}
T.V. Wiecki and M.J. Frank.
\newblock Neurocomputational models of motor and cognitive deficits in
  {P}arkinson's disease.
\newblock In Anders Bjorklund and M.~Angela Cenci, editors, \emph{{Progress in
  Brain Research: Recent Advances in Parkinson's Disease - Part I: Basic
  Research}}, volume 183, chapter~14, pages 275--297. Elsevier, 2010.
\newblock URL \url{http://www.ncbi.nlm.nih.gov/pubmed/20696325}.

\bibitem[Wylie et~al.(2010)Wylie, Ridderinkhof, Elias, Frysinger, Bashore,
  Downs, van Wouwe, and van~den Wildenberg]{WylieRidderinkhofEliasEtAl10}
Scott~A. Wylie, K.~Richard Ridderinkhof, William~J. Elias, Robert~C. Frysinger,
  Theodore~R. Bashore, Kara~E. Downs, Nelleke~C. van Wouwe, and Wery P.~M.
  van~den Wildenberg.
\newblock Subthalamic nucleus stimulation influences expression and suppression
  of impulsive behaviour in parkinson's disease.
\newblock \emph{Brain}, Sep 2010.
\newblock URL \url{http://www.ncbi.nlm.nih.gov/pubmed/20861152}.

\bibitem[Xue et~al.(2008)Xue, Aron, and Poldrack]{XueAronPoldrack08}
Gui Xue, Adam~R. Aron, and Russell~A. Poldrack.
\newblock Common neural substrates for inhibition of spoken and manual
  responses.
\newblock \emph{Cereb. Cortex}, 18\penalty0 (8):\penalty0 1923--1932, August
  2008.
\newblock \doi{10.1093/cercor/bhm220}.
\newblock URL \url{http://dx.doi.org/10.1093/cercor/bhm220}.

\bibitem[Yeung and Cohen(2006)]{YeungCohen06}
N.~Yeung and J.~D. Cohen.
\newblock The impact of cognitive deficits on conflict monitoring. predictable
  dissociations between the error-related negativity and n2.
\newblock \emph{Psychological science : a journal of the American Psychological
  Society / APS}, 17\penalty0 (2):\penalty0 164--171, February 2006.
\newblock ISSN 0956-7976.
\newblock \doi{10.1111/j.1467-9280.2006.01680.x}.
\newblock URL \url{http://dx.doi.org/10.1111/j.1467-9280.2006.01680.x}.

\bibitem[Yeung et~al.(2004{\natexlab{a}})Yeung, Botvinick, and
  Cohen]{YeungBotvinickCohen04}
Nick Yeung, Matthew~M Botvinick, and Jonathan~D Cohen.
\newblock The neural basis of error detection: conflict monitoring and the
  error-related negativity.
\newblock \emph{Psychological review}, 111\penalty0 (4):\penalty0 931--959, 10
  2004{\natexlab{a}}.
\newblock URL \url{http://www.ncbi.nlm.nih.gov/pubmed/15482068}.

\bibitem[Yeung et~al.(2004{\natexlab{b}})Yeung, Botvinick, and
  Cohen]{YeungCohenBotvinick04}
Nick Yeung, Matthew~M Botvinick, and Jonathan~D Cohen.
\newblock The neural basis of error detection: conflict monitoring and the
  error-related negativity.
\newblock \emph{Psychological review}, 111\penalty0 (4):\penalty0 931--959, 10
  2004{\natexlab{b}}.
\newblock URL \url{http://www.ncbi.nlm.nih.gov/pubmed/15482068}.

\bibitem[Zaghloul et~al.(2012)Zaghloul, Weidemann, Lega, Jaggi, Baltuch, and
  Kahana]{ZaghloulWeidemannLegaEtAl12}
K.A. Zaghloul, C.T. Weidemann, B.C. Lega, J.L. Jaggi, G.H. Baltuch, and M.J.
  Kahana.
\newblock Neuronal activity in the human subthalamic nucleus encodes decision
  conflict during action selection.
\newblock \emph{The Journal of Neuroscience}, 32\penalty0 (7):\penalty0
  2453--2460, 2012.

\bibitem[Zandbelt and Vink(2010)]{ZandbeltVink10}
Bram~B Zandbelt and Matthijs Vink.
\newblock {On the role of the striatum in response inhibition.}
\newblock \emph{PloS one}, 5\penalty0 (11):\penalty0 e13848, January 2010.
\newblock ISSN 1932-6203.
\newblock \doi{10.1371/journal.pone.0013848}.
\newblock URL
  \url{http://www.pubmedcentral.nih.gov/articlerender.fcgi?artid=2973972\&tool%
=pmcentrez\&rendertype=abstract}.

\end{thebibliography}
\end{document}